\documentclass[12pt]{article}
\usepackage{geometry}                
\usepackage{lscape}                
\usepackage{graphicx}
\usepackage{amssymb}
\usepackage{amsmath}
\usepackage{epstopdf}
\usepackage{natbib}
\usepackage{hyperref}
\usepackage{setspace}
\usepackage{afterpage}
\usepackage{authblk}
\hypersetup{colorlinks=true, linkcolor=blue, citecolor=blue,urlcolor=blue}
\usepackage[all]{hypcap}
\def\eqlbl#1{\label{eq:#1}}
\def\eqref#1{(\ref{eq:#1})}

\title{Scaling Laws for Convection with Temperature-dependent Viscosity and Grain-damage}
\author[1,2]{Bradford J. Foley} 
\author[1]{David Bercovici}

\affil[1]{Yale University, Department of Geology and Geophysics, New Haven, CT}
\affil[2]{Now at: Carnegie Institution for Science, Department of Terrestrial Magnetism, Washington, DC}

\date{}                                           

\begin{document}
\maketitle
\begin{abstract}
Numerical experiments of convection with grain-damage are used to develop scaling laws for convective heat flow, mantle velocity, and plate velocity across the stagnant lid and plate-tectonic regimes.  Three main cases are presented in order of increasing complexity: a simple case wherein viscosity is only dependent on grainsize, a case where viscosity depends on temperature and grainsize, and finally a case where viscosity is temperature and grainsize sensitive, and the grain-growth (or healing) is also temperature sensitive.  In all cases, convection with grain-damage scales differently than Newtonian convection; whereas the Nusselt number, $Nu$, typically scales with the reference Rayleigh number, $Ra_0$, to the $1/3$ power, for grain-damage this exponent is larger because increasing $Ra_0$ also enhances damage.  In addition, $Nu$, mantle velocity, and plate velocity are also functions of the damage to healing ratio, ($D/H$); increasing $D/H$ increases $Nu$ because more damage leads to more vigorous convection.  For the fully realistic case, numerical results show stagnant lid convection, fully mobilized convection that resembles the temperature-independent viscosity case, and partially mobile or transitional convection, depending on $D/H$, $Ra_0$, and the activation energies for viscosity and healing.  Applying our scaling laws for the fully realistic case to Earth and Venus we demonstrate that increasing surface temperature dramatically decreases plate speed and heat flow, essentially shutting down plate tectonics, due to increased healing in lithospheric shear zones, as proposed previously. Contrary to many previous studies, the transitional regime between the stagnant lid and fully mobilized regimes is large, and the transition from stagnant lid to mobile convection is gradual and continuous.  Thus planets could exhibit a full range of surface mobility, as opposed to the bimodal distribution of fully mobile lid planets and stagnant lid planets that is typically assumed.  
\end{abstract}

\section{Introduction}

Answering the major questions of geodynamics, from the thermal evolution of the Earth to the origin of plate tectonics, requires a detailed understanding of the physics of mantle convection.  Mantle convection theory is best summarized in the form of scaling laws, which relate the important aspects of mantle convection, such as the heat flow and plate speed, to the key quantities governing convective circulation such as the Rayleigh number.  There has been extensive work in developing scaling laws for simple cases like constant viscosity convection \citep[e.g.][]{turcotte1967,Mckenzie1974,Sotin1999}, convection with temperature-dependent viscosity \citep[e.g.][]{christensen1984a,Morris1984,Ogawa1991,Davaille1993,Moresi1995,slava1995,Grasset1998}, and weakly non-Newtonian convection \citep[e.g.][]{parm1976,parm1982,christensen1984c,Reese1998,Solomatov2000b}.  However, the rheology of the mantle is necessarily more complex than these simple cases; a strongly non-linear rheology is required in order to generate plate tectonics \citep[e.g.][]{tackley2000,Berco2003}.  While there have been many studies on plate generation with exotic rheologies \citep[e.g.][]{weinstein1992,trompert1998,tackley2000a,vanheck2008,Foley2009}, only a few attempt to develop detailed scaling laws \citep[e.g.][]{moresi1998,Korenaga2010}.  In particular, no study has developed scaling laws for convection with damage physics, a promising mechanism for generating plate tectonics \citep[e.g.][]{brs1,Landuyt2008,Landuyt2009a,br4,br6,br2012,br2013,br2014}.  Therefore, the purpose of this paper is to develop scaling laws for the heat flow, interior convective velocity, and plate velocity for convection with damage.       

The damage physics we employ is a grainsize feedback referred to as grain-damage \citep[e.g.][]{br6,Landuyt2008,br2012}.  Grain-damage is motivated by its effectiveness at causing weakening and shear localization in the lithosphere, its allowance of dormant weak zones, and by the geological observation that peridotitic mylonites are ubiquitous in lithospheric shear zones \citep{White1980}.  In particular, grain-damage's allowance for dormant weak zones means this mechanism has rheological memory, a property that other mechanisms, such as plasticity, lack \citep[e.g.][]{moresi1998}.  Such memory of past deformation is thought to be crucial for initiating subduction  \citep{gurnis2000}.    

Grain-damage relies on a feedback between grainsize reduction and a grainsize dependent viscosity; deformation reduces the grainsize, making the material weaker and hence causing more deformation.  Previous studies have demonstrated that grain-damage is an effective mechanism for shear localization \citep{Landuyt2008}, and for producing significant lithospheric weakening \citep{Foley2012}.  However, neither study develops scaling laws for the heat flow or plate velocity.  Scaling laws for convection with grain-damage could differ significantly from those for Newtonian or weakly non-Newtonian convection, due to the effects of damage throughout the lithosphere and mantle.  We therefore perform a large suite of numerical models and use the results to derive scaling laws for convection with grain-damage from boundary layer theory.  

Convection with grain-damage has not been studied extensively, so we start with the very simple case of a viscosity that depends solely on grainsize, progressively adding the complexities of temperature-dependent viscosity and finally temperature-dependent grain growth in later sections.  The paper is therefore organized in the following manner: grain-damage and the numerical methods employed are reviewed in \S \ref{sec:background}; scaling laws for temperature-independent viscosity are derived and compared to numerical experiments in \S \ref{sec:temp_indep}; scaling laws for the stagnant lid regime are derived and compared to numerical experiments with temperature and grainsize sensitive viscosity in \S \ref{sec:stag_H0}; temperature-dependent healing is added, and scaling laws for the the stagnant lid, transitional, and fully mobile lid regimes are derived and compared to numerical results in \S  \ref{sec:full_dam}; the boundaries between the stagnant lid, transitional, and fully mobile regimes, are derived in \S \ref{sec:boundaries}; a curve for the onset of convection is derived in \S \ref{sec:onset_convec}; an application of our scaling laws to Earth and Venus and other implications of this study for the thermal and tectonic evolution of planets are given in \S \ref{sec:discussion}, and concluding remarks in \S \ref{sec:conclusions}.  

\section{Background}
\label{sec:background}

Our grain-damage mechanism relies on the feedback between deformation induced grainsize reduction and grainsize dependent viscosity. This combination may seem problematic, because these processes are thought to occur via distinct microphysical mechanisms: grainsize reduction nominally occurs through the propogation of dislocations in the dislocation creep regime, whose flow law is independent of grainsize; meanwhile grainsize sensitive flow occurs in a diffusional regime, which does not implicitly involve dislocations and grain reduction.  Thus, the dislocation creep and diffusion creep deformation regimes occur in separate domains of deformation space (depending on differential stress, temperature and grainsize) and therefore do not necessarily interact in a way that would cause a grainsize feedback \citep{Karato2008}.  However, in a two-phase material like peridotite, deformation and damage to the interface between phases (e.g. olivine and pyroxene) combined with pinning effects allows damage, grain-reduction, and diffusion creep to co-exist \citep{br2012}; this leads to a state of small grain permanent diffusion creep, which is observed in natural peridotitic mylonites  \citep{Warren2006}.    

The full theory for grain-damage with Zener pinning involves a two-phase material with a composite rheology, taking into account both dislocation and diffusion creep.  To make the problem more numerically tractable, we use a simplified version by assuming that the permanent diffusion creep ``pinned" state prevails throughout the mantle.  In the pinned state, the grainsize of the primary phase is controlled by the curvature of the interface with the secondary phase, and thus damage to the interface leads directly to damage of the primary phase.  We can therefore assume that the bulk grainsize of the material is governed by the same equation as the evolution of the interface curvature (equation 4d of \cite{br2012}), reducing the problem to a single phase (i.e. we no longer need to track the evolution of both the interface curvature and grainsize, we only need to track the grainsize).  The assumption that the pinned state prevails throughout the mantle also allows us to assume grainsize sensitive diffusion creep throughout the domain, and neglect non-Newtonian dislocation creep.  Thus the composite rheology is reduced to a simple grainsize sensitive rheology.  However, in reality the rheology will be controlled by whichever mechanism allows for the easiest deformation \citep[e.g.][]{Rozel2010}; i.e. when grains are large dislocation creep should predominate.  We discuss how the transition to dislocation creep would affect our scaling laws, and under what conditions this transition should occur, in Appendix \ref{sec:disl_creep}.  

\subsection{Damage Formulation}

The viscosity is sensitive to grainsize and temperature as expected for diffusion creep or grain boundary sliding \citep{hirth2003}:   
\begin{equation}
\eqlbl{eqv}
\mu = \mu_n \exp \left(\frac{E_v}{RT} \right) \left(\frac{A}{A_0} \right)^{-m}
\end{equation}
where $\mu_n$ is a constant, $E_v$ the diffusion creep activation energy ($E_v = 300$ kJ/mol \citep{karato1993}), $T$ the temperature, $R$ the universal gas constant, $A$ the fineness, or inverse grainsize \citep{br6, Landuyt2008,Foley2012}, and $A_0$ the reference fineness.  The constant $m$ is equal to 2 or 3 depending on the mechanism of diffusion creep \citep{hirth2003,karato1993}.  Specifically, diffusion along the grain boundary (Coble creep) gives $m=3$ while diffusion through the grain (Nabarro-Herring creep) results in $m=2$ \citep[e.g.][]{Evans1995}.  For numerical purposes, we focus our study on $m=2$, as $m=3$ produces higher degrees of localization and larger viscosity contrasts that can cause numerical convergence problems.  However, the scaling laws that we derive from boundary layer theory are general functions of $m$, and thus illustrate how different values of $m$ influence the behavior of convection with grain-damage.  We also perform some numerical experiments at $m=3$ to constrain the empirically derived scalings-laws for different values of $m$.    

In the pinned state, fineness is governed by the following evolution equation:
\begin{equation}
\eqlbl{eqd}
\frac{DA}{Dt} = \frac{f}{\gamma}\Psi - h A^p
\end{equation}
where $t$ is time, $f$ is the damage partitioning fraction, which can vary from zero to one, $\gamma$ is the surface free energy, $\Psi$ the deformational work, $h$ the healing rate, and $p$ a constant.  Deformational work is defined as $\Psi = \nabla \underline{v} : \underline{\underline{\tau}}$,  where $\underline{v}$ is the velocity and $\underline{\underline{\tau}}$ is the stress tensor \citep{br6,Landuyt2008}; see also \cite{austin2007}.  

The first term on the right side of \eqref{eqd} represents the partitioning of a fraction ($f$) of deformational work into surface free energy by reducing grainsize (increasing fineness).  The second term on the right side represents reduction of fineness due to normal grain growth.  The exponent $p$ ranges from 4-6 in the pinned state \citep{br2012}; we primarily use $p=4$ in this study to maintain a level of shear localization similar to that which would be achieved through $m=3$ and $p=5-6$, while still keeping the numerical models tractable over a large span of parameter space.  The healing rate constant, $h$, is a function of temperature with an Arrhenius form: 
\begin{equation}
\eqlbl{eqheal}
h = h_n \exp \left(\frac{-E_h}{RT} \right)     
\end{equation}
where $h_n$ is a constant and $E_h$ is the activation energy for grain growth.  The value of $E_h$ in the pinned state is poorly known, as most grain-growth experiments are performed with monomineralic samples, resulting in low values of $E_h \approx 200$ kJ/mol and rapid grain growth \citep[e.g.][]{Karato1989}.  In the pinned state, $E_h$ should be higher because grain-growth can only occur through diffusion of material from one grain to another \citep{br2012}.  The kinetics of this process are still uncertain, but some preliminary results indicate lager values of $E_h \approx 400-500$ kJ/mol \citep[e.g.][]{Faul2006,Hiraga2010}.           

\subsection{Governing Equations}

\label{sec:govern_eqs}

We study convection with grain-damage with a model of infinite Prandtl number, Boussinesq thermal convection heated from below.  The damage formulation, \eqref{eqd}, is non-dimensionalized using the following scales where primes denote non-dimensional variables: $\underline{x} = \underline{x}^{\prime}d$, where $d$ is the depth of the mantle; $t = t^{\prime}d^2/\kappa$, where $\kappa$ is the thermal diffusivity; $\underline{v} = \underline{v}^{\prime}\kappa/d$; $T = T^{\prime}  \Delta T + T_s$, where $\Delta T$ is the temperature difference across the mantle and $T_s$ is the surface temperature; $A = A^{\prime} A_0$; $\underline{\underline{\tau}} = \underline{\underline{\tau}}^{\prime} \mu_m \kappa/d^2$, where $\mu_m$ is the reference viscosity defined at $T_m = \Delta T + T_s$ in the absence of damage; and $E_v = E_v^{\prime} R \Delta T$ and $E_h = E_h^{\prime} R \Delta T$.  The resulting non-dimensional equation governing fineness evolution is: 
\begin{equation}
\eqlbl{eqdam}
\frac{DA'}{Dt'} = D \psi  \exp{\left(\frac{E_v'}{T'+T_s^*} - \frac{E_v'}{1+T_s^*}\right)} A'^{-m} - H\exp{\left(\frac{-E_h'}{T'+T_s^*} + \frac{E_h'}{1+T_s^*}\right)} A'^p 
\end{equation}
where $\psi = \nabla \underline{v'} : ( \nabla \underline{v'} +  (\nabla \underline{v'})^T)$, $T_s^* = T_s/\Delta T$, $D$ is the non-dimensional damage number and $H$ is the non-dimensional healing number.  These quantities are defined as $D = f\mu_m\kappa/(\gamma A_0 d^2)$ and $H = h_m A_0^{(p-1)}d^2/\kappa$ where $h_m = h(T_m)$.  

The equations for conservation of mass, momentum, and energy, respectively, expressed in terms of non-dimensional variables using the same scales as above are:  
\begin{equation}
\eqlbl{eqmass}
\nabla \cdot \underline{v'} = 0
\end{equation}
\begin{equation}
\eqlbl{eqmom}
0=-\nabla P' + \nabla \cdot (2 \mu' \dot{\underline{\underline{\varepsilon}'}}) + Ra_0T' \hat{\underline{z}}
\end{equation}
\begin{equation}
\eqlbl{eqenergy}
\frac{\partial{T'}}{\partial{t'}}+ \underline{v'} \cdot \nabla T' = \nabla^2 T'
\end{equation}
where $P'$ is the dynamic pressure, $\dot{\varepsilon_{ij}}' = (\partial{v'_i} / \partial{x'_j} + \partial{v'_j} / \partial{x'_i} )/2$ is the strain rate, $\hat{\underline{z}}$ is the unit vector in the vertical direction, and $Ra_0$ is the reference Rayleigh number; $Ra_0 = (\rho \alpha g \Delta T d^3)/(\kappa \mu_m)$ where $\rho$ is density, $\alpha$ is thermal expansivity, and $g$ is acceleration due to gravity.    

In addition, we define parameters to describe the variation of viscosity and healing across the mantle due to temperature dependence; $\mu_l' = \mu_l/\mu_m$, the viscosity ratio in the absence of damage, (with $A = A_{ref}$), and $h_l' = h_l/h_m$, the healing ratio, where the subscript $l$ denotes the value in the lithosphere (i.e. at $T' = 0$).  We also define the effective Frank-Kamenetskii parameter, $\theta = E_v'/(1+T_s^*)^2$, to describe the temperature dependence of viscosity \citep{Korenaga2009}.  The Frank-Kamenetskii parameter comes from the linear exponential viscosity law, an approximation to full Arrhenius viscosity law, and appears in scaling laws for stagnant lid convection.  Thus it is necessary to define $\theta$ for numerical experiments with the Arrhenius viscosity law in order to develop and fit scaling laws for stagnant lid convection \cite[e.g.][]{Korenaga2009}.    

\subsection{Numerical Methods}

We solve the coupled convection and damage equations using a 2-D Cartesian finite volume code, similar to that used in \cite{Foley2012}.  The code uses the SIMPLER algorithm to solve the momentum equations \citep{Patankar1980}, employing a multi-grid method for the diffusion terms in both the momentum and temperature equations.  The temperature equation uses a Crank-Nicholson time discretization and the non-oscillatory version of MPDATA for the advection term \citep{Smolark1984,Smolark1990}.  For the fineness evolution equation, we linearize the source terms using the technique laid out in \cite{Patankar1980}, and utilize a Crank-Nicholson time discretization and the non-oscillatory version of MPDATA for advection as in the temperature equation.  A non-oscillatory advection scheme for fineness is essential to the stability of the numerical solution.  Dispersive ripples in the fineness solution will cause ripples in the viscosity field, which can grow due to the feedback with the momentum equations, eventually causing the numerical solution to diverge. Most numerical experiments in this study were performed with a $4 \times 1$ aspect ratio domain, though some cases were run in larger aspect ratio domains, up to $16 \times 1$ (see Tables \ref{tab_simple}, \ref{tab_stag}, \ref{tab_full}, and \ref{tab_mp_data} for a compilation of all numerical results).  The typical resolution used was $512 \times 128$, though a higher resolution of $1024 \times 256$ was used for models with large $D/H$ and/or large $Ra_0$ (see Tables \ref{tab_simple}, \ref{tab_stag}, \ref{tab_full}, and \ref{tab_mp_data}).  Rerunning select cases at double the resolution typically only changes the results for the Nusselt number by $1-3$ \%, with a maximum change of 5.5 \%, and typically only changes the results for the plate speed by $3-5$ \%, with a maximum change of $6.5$ \% (see Appendix \ref{sec:res}); thus the resolution used for the numerical models does not significantly impact the results.    
    
\section{Temperature-Independent Viscosity}
\label{sec:temp_indep}

\begin{figure}
\includegraphics[scale = 0.75]{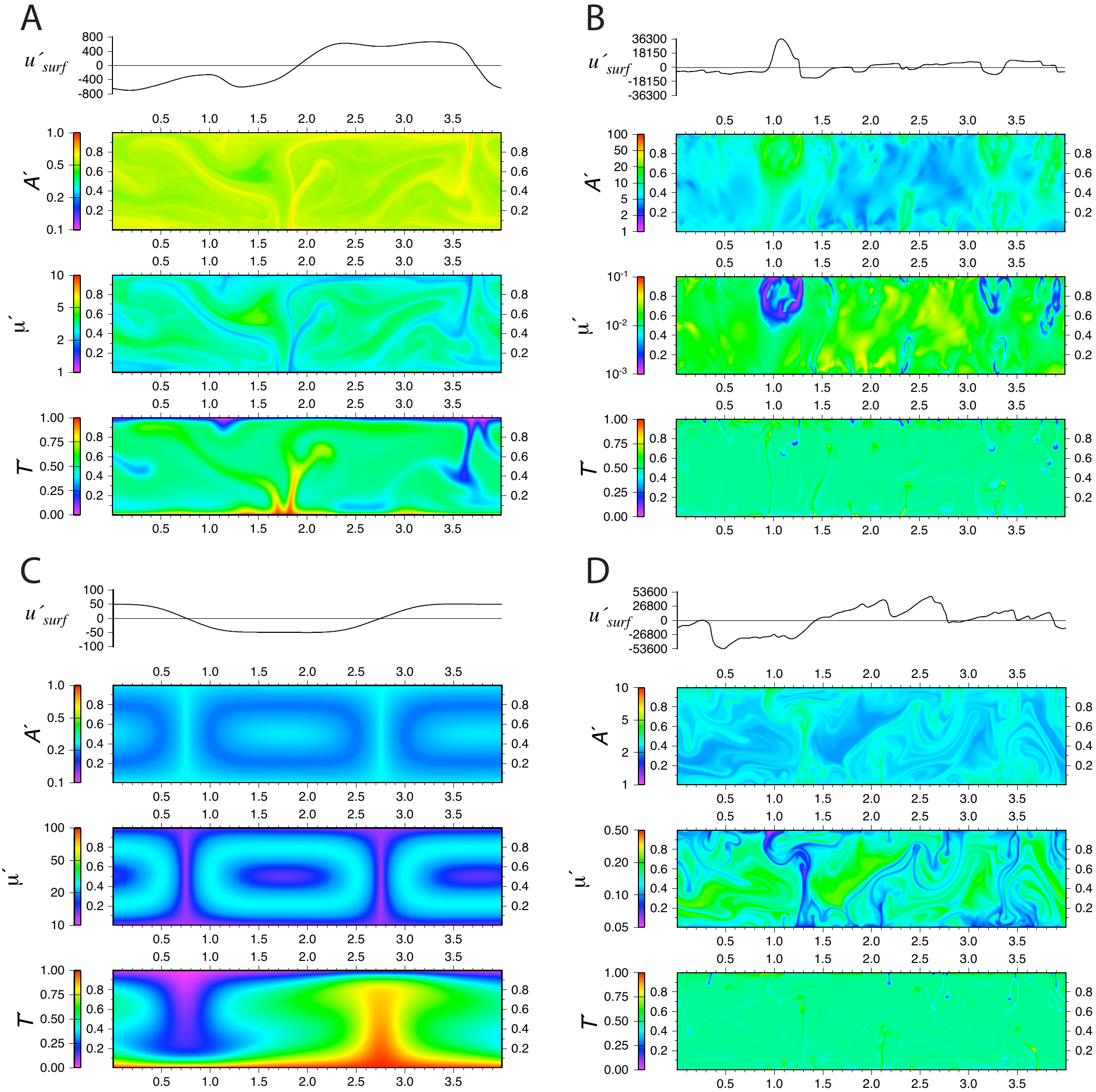}
\caption{\label {fig:pform_simple} Convection snapshots for a model with low $D/H$ ($D/H = 10^{-8}, Ra_0=10^6$) (a), high $D/H$ ($D/H = 10^{-4}, Ra_0=10^6$) (b), low $Ra_0$ ($D/H = 10^{-8}, Ra_0=10^5$) (c), and high $Ra_0$ ($D/H = 10^{-8}, Ra_0=5 \times 10^7$) (d).  All models use $m=2$ and $p=4$.  Each snapshot shows the surface velocity, $u'_{surf}$, the fineness field, $A'$, the viscosity field, $\mu'$, and the temperature field, $T'$.  } 
\end{figure}

We first explore the case where mantle viscosity is sensitive to grainsize only, and healing is constant (i.e. both viscosity and healing are insensitive to temperature).  This simple set-up is not a good approximation of mantle convection, but allows us to understand the effects of grain-damage in isolation; temperature-dependent viscosity and healing are added in later sections (\S \ref{sec:stag_H0} and \S \ref{sec:full_dam}).  The numerical models generally show a convective planform where upwellings and downwellings take the form of blob-like drips from the top and bottom thermal boundary layers, while the core of convection cells are dominated by a simple horizontal shear flow between the mobile boundary layers.  As seen in the scaling theory below (\S \ref{sec:scale_simple}), the fineness is controlled by the ratio of damage to healing, $D/H$, thus we discuss model results in terms of this quantity.  At low $D/H$, downwellings and upwellings are more sheet-like, becoming more drip-like as $D/H$ increases; a similar trend occurs for varying $Ra_0$ (Figure \ref{fig:pform_simple}).  As expected, both increasing $D/H$ or increasing $Ra_0$ enhances damage in the mantle interior (i.e. the isothermal core of convection cells) and boosts the vigor of convection.  

\subsection{Scaling Theory}
\label{sec:scale_simple}

\begin{figure}
\includegraphics[scale = 0.5]{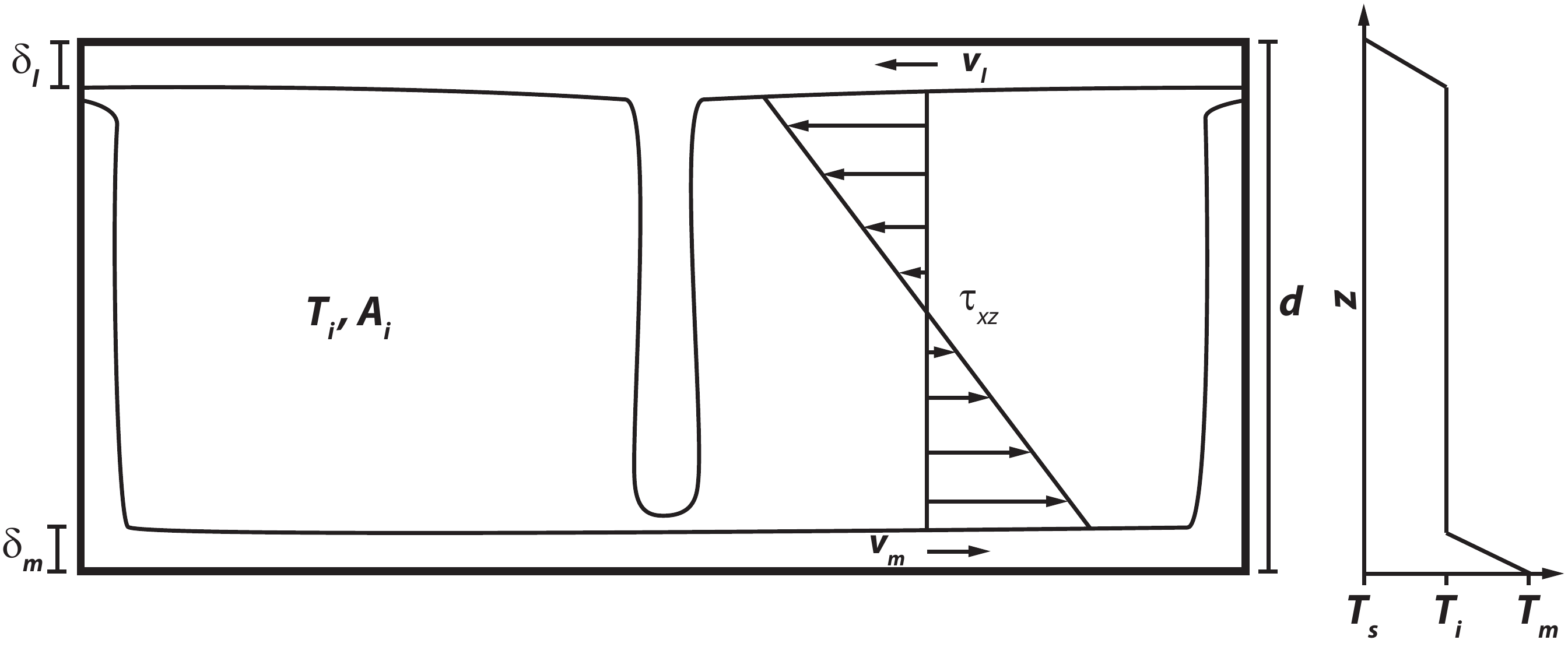}
\caption{\label {fig:scale_sketch_simple} Sketch of the idealized model for convection with grain-damage and temperature-independent viscosity and healing.  The assumed horizontal velocity profile and resulting mantle shear stress, $\tau_{xz}$, are shown (left), as well as the vertical temperature profile (right). The sketch also illustrates the definitions of $\delta_l$, $\delta_m$, $v_l$, $v_m$, $T_i$, and $A_i$.} 
\end{figure}

We derive scaling laws based on the idealized model in Figure \ref{fig:scale_sketch_simple}.  We assume that convection with grain-damage behaves like constant viscosity convection, with a viscosity set by the average interior fineness of the mantle, $A_i$; $A_i$ is determined by the typical stress scale in the mantle.  As seen by the numerical models, the shear stress, $\tau_{xz}$, which results from the horizontal shear flow between the top and bottom boundaries, is the dominant stress component driving damage in the mantle.  We note here that $\tau_{xz}$ is not the dominant stress component throughout the convecting layer; the normal stress, $\tau_{xx}$, ($\tau_{xx} = -\tau_{zz}$ due to mass conservation) is generally the largest stress component in the lithosphere and in upwelling or downwelling regions.  This difference is important when we develop scaling laws for plate-tectonic style convection, where we consider damage in the lithosphere (see \S \ref{sec:full_scale}). A scaling law for the convective shear stress with grain-damage is derived assuming $\tau_{xz} = (2\mu_{eff} v_l)/d$, where $v_l$ is the lithosphere (or plate) velocity, and the factor of two arises from the fact that the horizontal velocity profile goes from $v_l$ at the surface to zero at $z=d/2$ (see Figure  \ref{fig:scale_sketch_simple}).  The effective interior mantle viscosity, $\mu_{eff}$, is defined as $\mu_{eff} = \mu_i(A_i/A_0)^{-m}$, where $\mu_i$ is the undamaged viscosity at the average interior mantle temperature, $T_i$; with temperature-independent viscosity, $\mu_i = \mu_m$.  

We assume the plate velocity follows the classical scaling, $v_l = C_1(\kappa/d)Ra_{eff}^{2/3}$, where $C_1$ is a constant of order 0.1 \citep{turc1982,slava1995}, and the effective Rayleigh number ($Ra_{eff}$) is defined at the effective interior mantle viscosity, $\mu_{eff}$.  Combining these equations gives the scaling law for convective shear stress in the mantle: 
\begin{equation}
\eqlbl{tauxz1}
\tau_{xz} = 2 C_1 \frac{\kappa}{d^2} \mu_m Ra_0^{\frac{2}{3}} \left(\frac{A_i}{A_0}\right)^{-\frac{m}{3}}
\end{equation}
Non-dimensionalizing $\tau_{xz}$ by the stress scale $\mu_m (\kappa/d^2)$, and $A$ by the reference fineness $A_0$, yields the following simplified equation where, as before, primes denote non-dimensional variables: 
\begin{equation}
\eqlbl{tauxz}
\tau_{xz}' = 2 C_1 Ra_0^{\frac{2}{3}} A_i'^{-\frac{m}{3}}
\end{equation}
The average fineness of the mantle is derived using the grainsize evolution equation \eqref{eqdam} in steady-state, and assuming that $\tau_{xz} >> \tau_{xx} (\tau_{zz})$ in the mantle, 
\begin{equation}
\eqlbl{steadydam}
D\tau_{xz}'^2A_i'^{m} = HA_i'^p .
\end{equation}
Thus the steady-state fineness is a function of the shear stress: 
\begin{equation}
\eqlbl{steadyfine}
A_i' = \left(\frac{D}{H} \tau_{xz}'^2 \right)^{\frac{1}{p-m}} .
\end{equation}
Combining equations \eqref{tauxz} and \eqref{steadyfine} and solving for $\tau_{xz}'$ gives the final scaling law for convective shear stress with grain-damage: 
\begin{equation}
\eqlbl{tauxz_scaling}
\tau_{xz}' = (2C_1)^{\frac{3(p-m)}{3p-m}} \left(\frac{D}{H}\right)^{-\frac{m}{3p-m}} Ra_0^{\frac{2(p-m)}{3p-m}}.
\end{equation}
         
Equation \eqref{tauxz_scaling} demonstrates that increasing $Ra_0$ increases the shear stress while increasing $D/H$ decreases shear stress.  Comparing this scaling law to the shear stress scaling for uniform viscosity elucidates how grain-damage influences convection.  For isoviscous convection $\tau_{xz} \sim \mu (\kappa/d^2) Ra^{2/3}$; this shows that $\tau_{xz} \sim \mu^{1/3}$, and thus decreasing $\mu$ results in a net drop in shear stress.  Therefore increasing $D/H$ decreases the shear stress because higher damage reduces the effective interior mantle viscosity.  Increasing $Ra_0$ increases shear stress, but by a smaller amount than for isoviscous convection (note that the exponent for $Ra_0$ in \eqref{tauxz_scaling} is less than $2/3$).  The increase in strain-rate that larger $Ra_0$ causes by boosting velocity is somewhat offset by increased damage in the mantle.   

Scaling laws for the velocity and boundary layer thickness (and hence Nusselt number) are derived in similar fashion.  As explained above, we assume that the plate velocity, $v_l$ scales like $v_l' = C_1 Ra_{eff}^{2/3}$, where the non-dimensional velocity $v' = v (d/\kappa)$.  Therefore 
\begin{equation}
\eqlbl{velocity}
v_l' = C_1 Ra_0^{\frac{2}{3}} A_i'^{\frac{2m}{3}} .
\end{equation}                  
Combining \eqref{velocity} with \eqref{steadyfine} and \eqref{tauxz_scaling} we find that velocity scales as 
\begin{equation}
\eqlbl{velocity_scale}
v_l' = C_1 (2C_1)^{\frac{4m}{3p-m}} \left(\frac{D}{H} \right)^{\frac{2m}{3p-m}} Ra_0^{\frac{2(p+m)}{3p-m}}.
\end{equation} 
Equation \eqref{velocity_scale} shows that either increasing $D/H$ or $Ra_0$ causes higher velocities.  Larger $D/H$ enhances damage in the mantle and therefore decreases the interior viscosity; with less resistance to flow convective velocities increase. Equation \eqref{velocity_scale} also shows that increasing $Ra_0$ results in a larger increase in velocity than for isoviscous convection, where $v_l \sim Ra^{2/3}$.  With grain-damage the boost in velocity that comes from a higher Rayleigh number also increases the damage, thereby reducing the interior viscosity and causing velocities to increase even further.  Also, we note that for this case of temperature-independent viscosity and healing, convection is symmetric and thus the plate velocity $v_l$ is equal to the horizontal velocity at the base of the mantle, $v_m$.   

The thickness of the top thermal boundary layer, $\delta_l$, scales as $\delta_l = dC_2 Ra_{eff}^{-1/3}$ where $C_2$ is a constant or order 1 ($C_2 \approx 2-3$ \citep{turc1982,slava1995}). Non-dimensionalizing $\delta_l$ by mantle thickness $d$, the thickness of the top boundary layer scales as  
\begin{equation}
\eqlbl{delta}
\delta_l' = C_2 Ra_0^{-\frac{1}{3}} A_i'^{-\frac{m}{3}}
\end{equation}  
Combining \eqref{delta} with \eqref{steadyfine} and \eqref{tauxz_scaling} gives the final scaling law for $\delta_l$: 
\begin{equation}
\eqlbl{delta_scale}
\delta_l' = C_2 (2C_1)^{-\frac{2m}{3p-m}} \left (\frac{D}{H} \right)^{-\frac{m}{3p-m}} Ra_0^{-\frac{p+m}{3p-m}}.
\end{equation}
Equation \eqref{delta_scale} demonstrates that either increasing $D/H$ or $Ra_0$ decreases $\delta_l'$.  Larger $D/H$ reduces the interior mantle viscosity, allowing the boundary layer to go unstable more easily, and thus $\delta_l$ is thinner.  Compared to isoviscous convection, where $\delta_l' \sim Ra^{-1/3}$, $Ra_0$ has a stronger influence on $\delta_l$ for convection with grain-damage.  The reason is the same as the previous scalings: changing $Ra_0$ increases damage in the mantle, in addition to increasing the relative strength of buoyancy forces at reference conditions.  In addition, the thickness of the bottom boundary layer, $\delta_m$, is equal to $\delta_l$ due to symmetry for this temperature-independent viscosity case.  The Nusselt number is related to the boundary layer thickness through the convective heat flux, 
\begin{equation}
\eqlbl{Nu1}
Nu =  \frac{k\frac{T_i-T_s}{\delta_l}}{k\frac{\Delta T}{d}} = \frac{T_i - T_s}{\Delta T} \frac{d}{\delta_l}
\end{equation}
Non-dimensionalizing temperature by $\Delta T$ \eqref{Nu1} becomes 
\begin{equation} 
\eqlbl{Nu_nondim}
Nu = T_i'\delta_l'^{-1} .
\end{equation}
The average internal mantle temperature, $T_i'$, is equal to $1/2$ for convection with temperature-independent viscosity and healing, and the Nusselt number therefore scales as 
\begin{equation}
\eqlbl{Nu_scale}
Nu = \frac{(2C_1)^{\frac{2m}{3p-m}}}{2C_2} \left (\frac{D}{H} \right)^{\frac{m}{3p-m}} Ra_0^{\frac{p+m}{3p-m}}.
\end{equation}  
Finally, combining equations \eqref{steadyfine} and \eqref{tauxz_scaling} gives a scaling law for the average internal mantle fineness, $A_i'$: 
\begin{equation}
\eqlbl{A_scale}
A_i' = (2C_1)^{\frac{6}{3p-m}} \left (\frac{D}{H}\right)^{\frac{3}{3p-m}} Ra_0^{\frac{4}{3p-m}} .
\end{equation}

\subsection{Comparison to Numerical Experiments}
\label{sec:comp_simple}

\begin{figure}
\includegraphics[scale = 0.75]{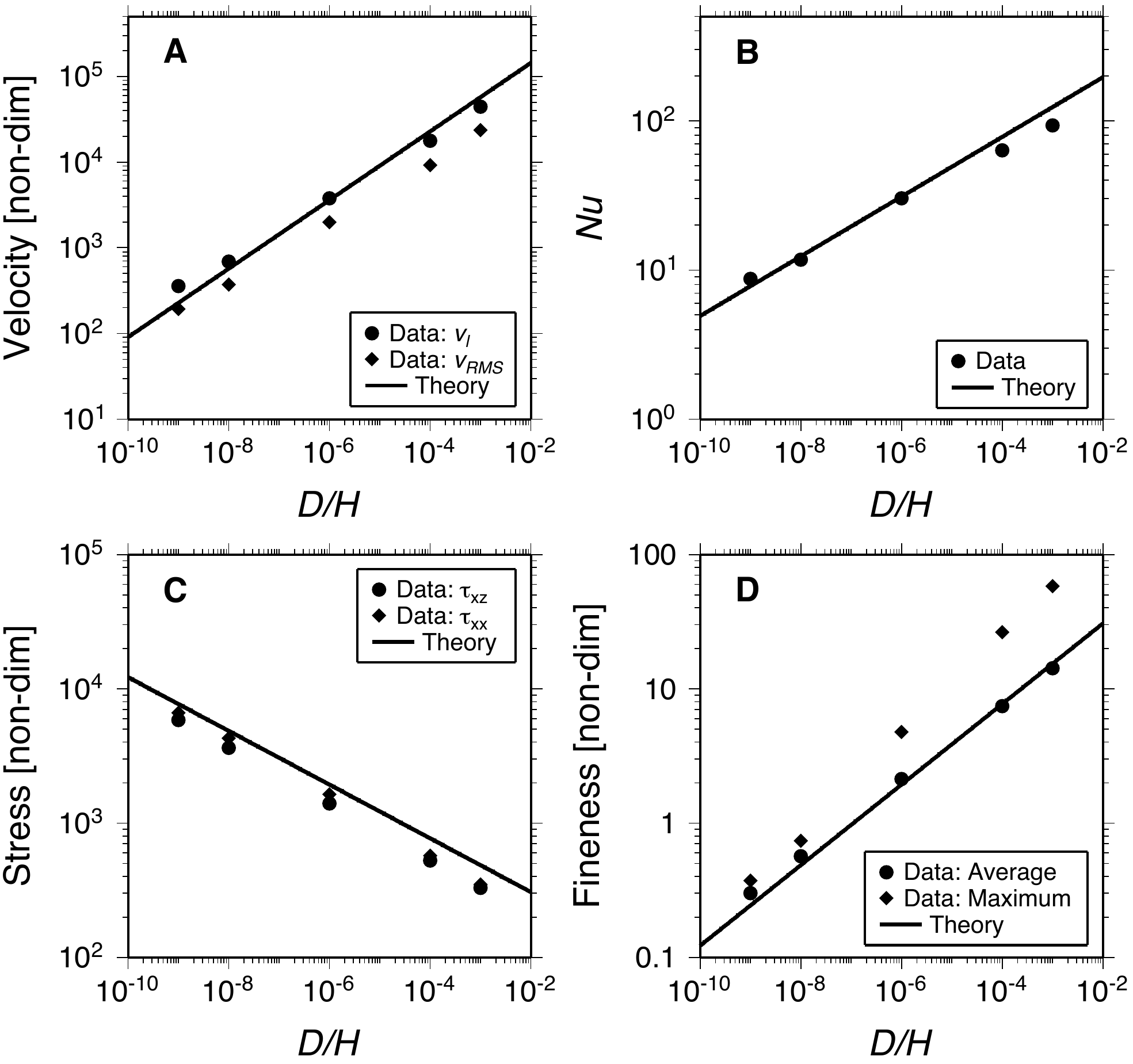}
\caption{\label {fig:dscale} Plots of velocity (a), Nusselt number (b), stress (c), and fineness (d) versus $D/H$ for temperature indepdent viscosity and healing (\S \ref{sec:temp_indep}).  Numerical results are plotted as symbols with theoretical scaling laws plotted as solid lines.  Figure (a) shows numerical data for $v_l'$ compared to the scaling law for $v_l'$ \eqref{velocity_scale}; data for $v_{RMS}'$ is also shown.  Figure (b) compares the numerical results for $Nu$ to the scaling law for $Nu$ \eqref{Nu_scale}.  Figure (c) compares the data for ${\tau_{xz}'}_{RMS}$ to the scaling law for $\tau_{xz}'$ \eqref{tauxz_scaling}; data for ${\tau_{xx}'}_{RMS}$ is also shown.  Figure (d) compares the data $\bar{A}'$ to the scaling law for $A_i'$ \eqref{A_scale} and also shows data for $A'_{max}$. } 
\end{figure}

We compare the theoretical scaling laws derived above to the results from numerical convection models.  To facilitate the comparison, we compute the following metrics from the numerical data:  the plate velocity,
\begin{equation} 
\eqlbl{max_plate}
v_l' = \frac{\mathrm{MAX}(u'(z'=1)) - \mathrm{MIN}(u'(z'=1))}{2}  ;
\end{equation}
the whole mantle RMS velocity, 
 \begin{equation}
 v_{RMS}' = \sqrt{\frac{1}{x_{max}'} \int_0^1 \int_0^{x_{max}'}( u'^2 + w'^2 ) \mathrm{d}x'\mathrm{d}z' } 
 \end{equation} 
 where we have used the fact that $0 \leq x' \leq x_{max}'$ and $0 \leq z' \leq 1$ in our numerical models; the Nusselt number, 
 \begin{equation}
 Nu =  \frac{1}{x_{max}'} \int_0^{x_{max}'} \left( \frac{\partial{T'}}{\partial{z'}} \right)_{z'=1} \mathrm{d}x'
 \end{equation}
 the whole mantle RMS stress, 
 \begin{equation}
 {\tau_{ij}'}_{RMS} = \sqrt{\frac{1}{x_{max}'} \int_0^1 \int_0^{x_{max}'} \tau_{ij}'^2 \mathrm{d}x'\mathrm{d}z' } 
 \end{equation}
 where $i$ and $j$ are replaced with $x$ or $z$ for the different components of the stress tensor; the volume averaged fineness, 
 \begin{equation}
 \bar{A}' = \frac{1}{x_{max}'} \int_0^1 \int_0^{x_{max}'} A' \mathrm{d}x'\mathrm{d}z'  ;
 \end{equation}
 and the maximum fineness, $A_{max}'$.  All metrics are then time-averaged after the numerical convection experiment has reached statistical steady-state.  

There is generally a good agreement between the numerical results where $D/H$ is varied (with constant $Ra_0 = 10^6$, $m=2$, and $p=4$), and the theoretical scaling laws with $C_1 = 0.15$ and $C_2 = 2.5$ (Figure \ref{fig:dscale}).  In particular the scaling laws for $v_l'$ \eqref{velocity_scale} and $Nu$ \eqref{Nu_scale} match the numerical results well (Figure \ref{fig:dscale}a,b).  In addition, $v_{RMS}'$ scales the same as $v_l'$, albeit smaller by a factor of $\approx 2$ due to averaging over areas where velocities are low (i.e. in the core of convection cells).      

\begin{figure}
\includegraphics[scale = 0.5]{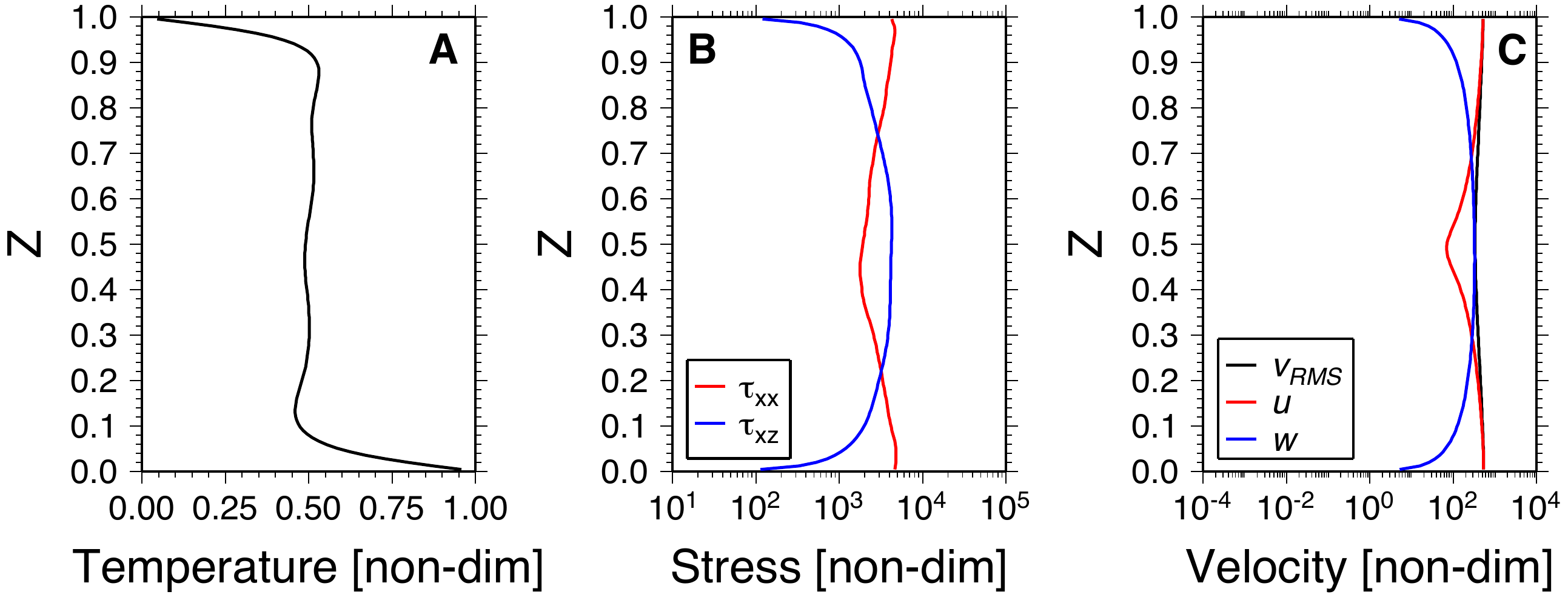}
\caption{\label {fig:z_profiles1} Vertical profiles of horizontally averaged temperature (a), stress (b), and velocity (c) for a representative numerical result with temperature-independent viscosity and healing in statistical steady-state.  The plot for stress (b) shows depth profiles of the shear stress, ${\tau'_{xz}}$, and normal stress, $\tau'_{xx}$.  The plot for velocity (C) shows depth profiles of the full velocity vector, $v'_{RMS}$, the horizontal velocity, $u'$, and vertical velocity, $w'$. All horizontal averages for stress and velocity are RMS averages. Parameters for model result shown here: $Ra_0 = 10^6$ and $D/H=10^{-8}$.  }  
\end{figure}

The scaling laws for $\tau_{xz}'$ \eqref{tauxz_scaling} and $A_i'$ \eqref{A_scale} fit the numerical results with a small offset (Figure \ref{fig:dscale}c,d).  Strictly speaking, the scaling laws for $\tau_{xz}'$ and $A_i'$ describe interior mantle values, or averages in the isothermal core of convection cells, and not whole mantle averages.  However, the difference between the whole mantle average and the interior value is likely to be small, as the well-mixed interior is by far the largest region of the convecting mantle.  This is confirmed by the numerical data which generally show a good agreement with the scaling laws for $\tau_{xz}'$ and $A_i'$ (Figure \ref{fig:dscale}c,d); there is only a small offset between the numerical data and the scaling laws for both quantities.  In addition, the numerical results show that ${\tau_{xx}'}_{RMS}$ scales in the same manner as the shear stress, and is typically somewhat larger in magnitude.  The ${\tau_{xx}'}_{RMS}$ average is likely dominated by the boundary layers and upwelling and downwelling regions in the mantle where normal stresses are highest.  Therefore the metric ${\tau_{xx}'}_{RMS}$ does not reflect the typical value of $\tau_{xx}$ in the core of convection cells; $\tau_{xz}$ is larger than $\tau_{xx}$ in the convecting interior (Figure \ref{fig:z_profiles1}b).  The maximum fineness, which is confined to the boundary layers and is likely driven by the large normal stresses where downwellings and upwellings first go unstable, appears to scale differently than $\bar{A}'$ (Figure \ref{fig:dscale}d).  As the driving force for $A_{max}'$ is different than for $\bar{A}'$, it is not unexpected that they scale differently.  

\begin{figure}
\includegraphics[scale = 0.75]{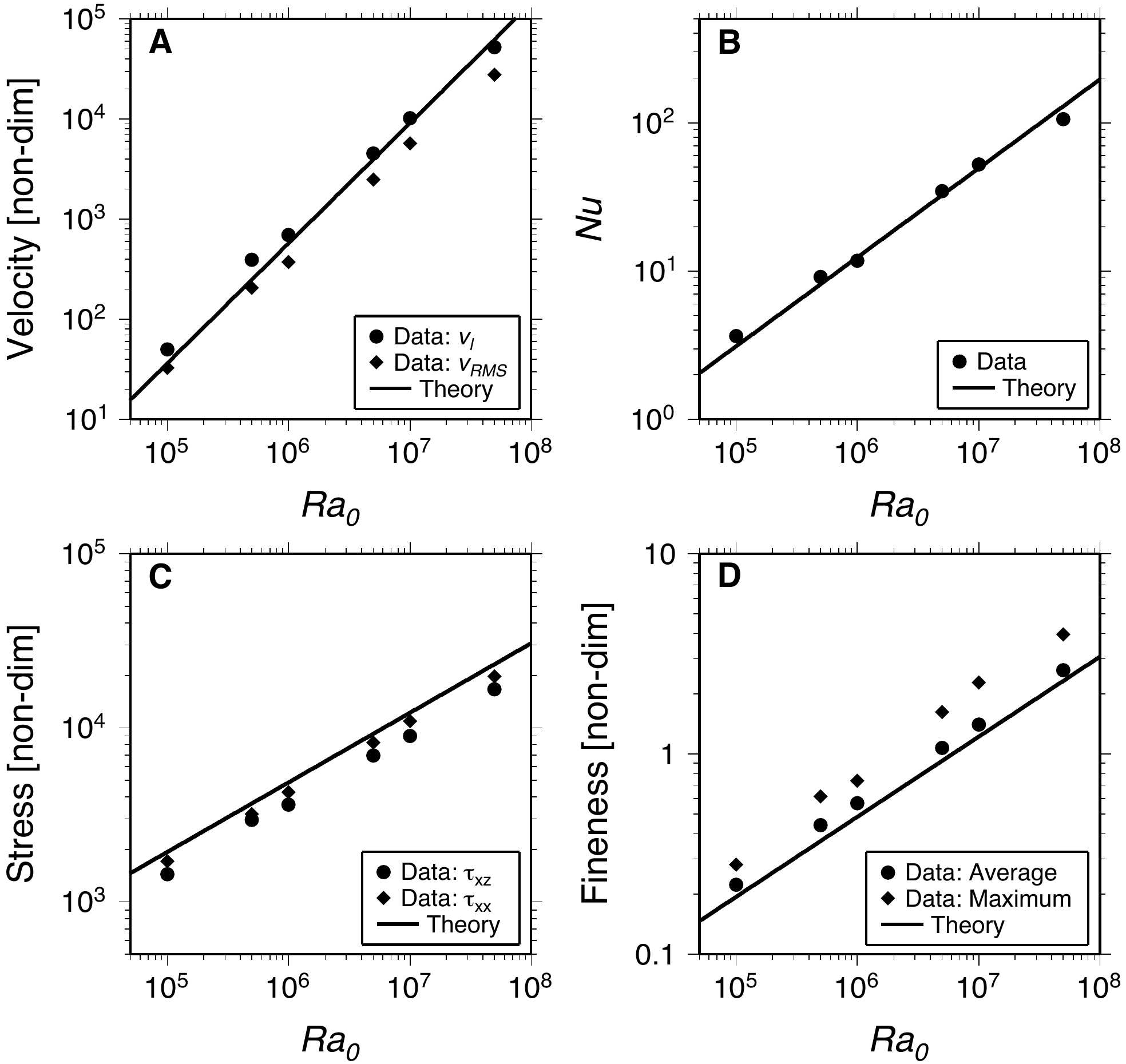}
\caption{\label {fig:Ra_scale} Plots of time averaged velocity (a), Nusselt number (b), stress (c), and fineness (d) versus Rayleigh number, $Ra_0$, as in Figure \ref{fig:dscale}.}  
\end{figure}

The numerical results for varying $Ra_0$ (with $D/H=10^{-8}$, $m=2$, and $p=4$) compare well to the theoretical scaling laws (Figure \ref{fig:Ra_scale}).  The results are similar to the experiments with varying $D/H$, in particular the plate velocities and Nusselt number, which are well fit by the theory.  As in Figure \ref{fig:dscale}, the convective stresses and mantle fineness show small offsets between the numerical data and the scaling theory due to the way the numerical data is calculated.  

\section{Temperature-Dependent Viscosity: Stagnant Lid Regime} 
\label{sec:stag_H0}

Adding a strongly temperature-dependent viscosity (with a constant healing rate) allows us to explore how grain-damage influences convection in the stagnant lid regime, the regime where the cold, high viscosity lithosphere no longer participates in convection \citep{Ogawa1991,Davaille1993,Moresi1995,slava1995}.  Developing scaling laws for convection with grain-damage in the stagnant lid regime is important because it allows us to establish an important baseline scenario, the fully stagnant lid end member, with which we can compare mobile and plate-like models to in later sections (see \S\ref{sec:full_dam}).  In addition, our results for stagnant lid convection may be applicable to planetary bodies that do not exhibit plate tectonics, such as Mars or rocky and icy satellites in the solar system.    
 
\begin{figure}
\includegraphics[scale = 0.75]{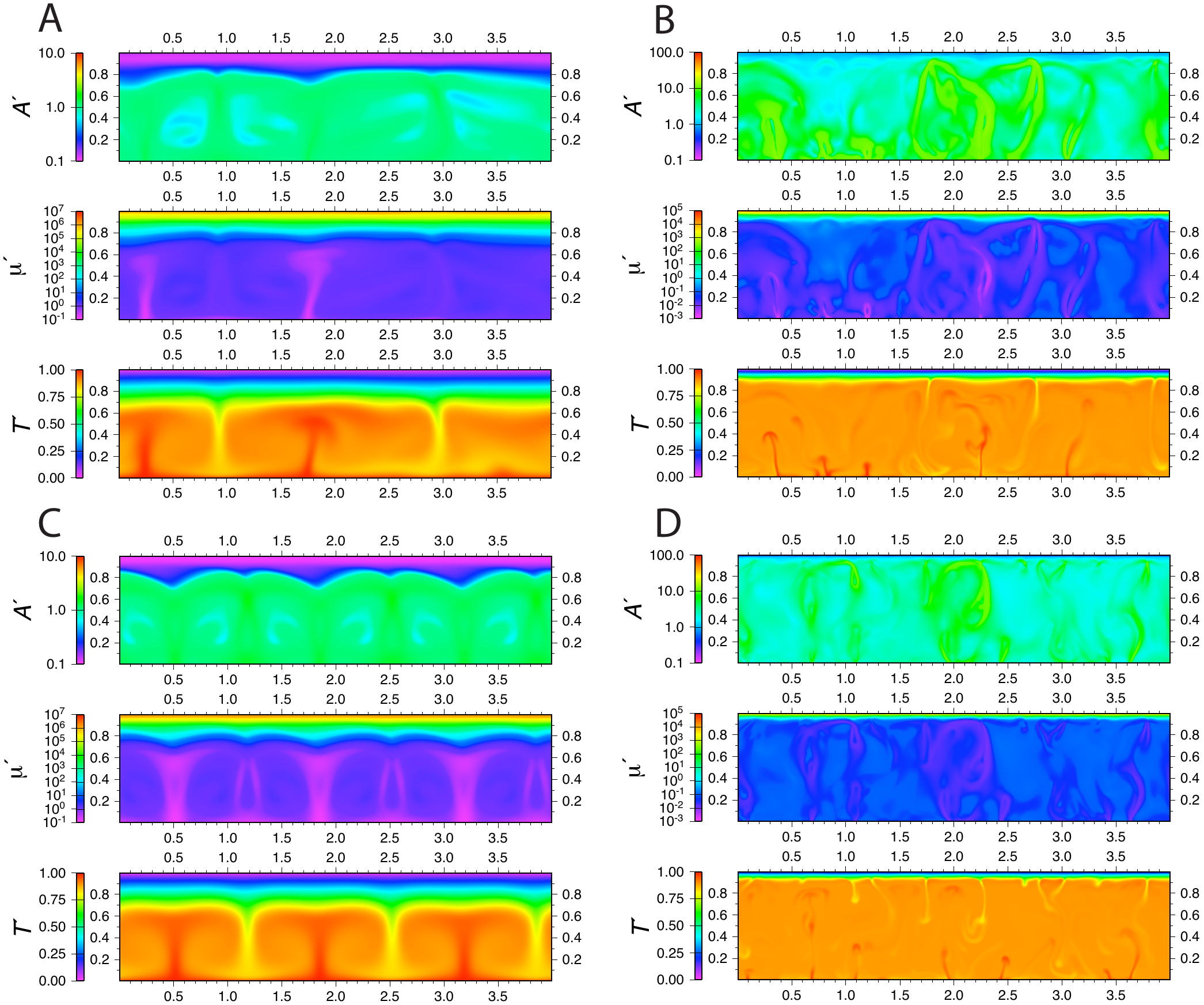}
\caption{\label {fig:pform_stag} Snapshots of stagnant lid convection (\S \ref{sec:stag_H0}) for a model with low $D/H$ ($D/H = 10^{-6}, Ra_0=10^6$ ) (a), high $D/H$ ($D/H = 10^{-3}, Ra_0=10^6$) (b), low $Ra_0$ ($D/H = 10^{-5}, Ra_0=3 \times 10^5$) (c), and high $Ra_0$ ($D/H = 10^{-5}, Ra_0=10^7$) (d).  All models use $m=2$, $p=4$, $E_v' = 23.03$, and $T_s^* = 1$. Each snapshot shows the fineness field, $A'$, the viscosity field, $\mu'$, and the temperature field, $T'$.  } 
\end{figure}

Stagnant lid convection occurs when the viscosity ratio across the top thermal boundary layer, $\mu_l/\mu_i$, reaches a critical value of about $3000$, which corresponds to a viscosity ratio across the mantle, $\mu_l/\mu_m$ of $\approx 10^4$ for bottom heated convection \citep{slava1995}.  We thus use viscosity ratios at or above $10^4$ in this section (where $\mu_l' = 10^5$) and in \S \ref{sec:full_dam}.  Numerical models show changes in the convective planform with $D/H$ or $Ra_0$ that are similar to the temperature-independent viscosity case (Figure \ref{fig:pform_stag}).  At low $D/H$ or $Ra_0$, convection beneath the lid is sluggish with sheet-like upwellings and downwellings; we even observe the cessation of convection entirely at very low $D/H$ or $Ra_0$ (\S\ref{sec:onset_convec}).  With increasing $D/H$ or $Ra_0$, upwellings and downwellings become drip-like, and the lid becomes flat and relatively thin.   

\subsection{Scaling Theory}
\label{sec:stag_H0_scale}

\begin{figure}
\includegraphics[scale = 0.55]{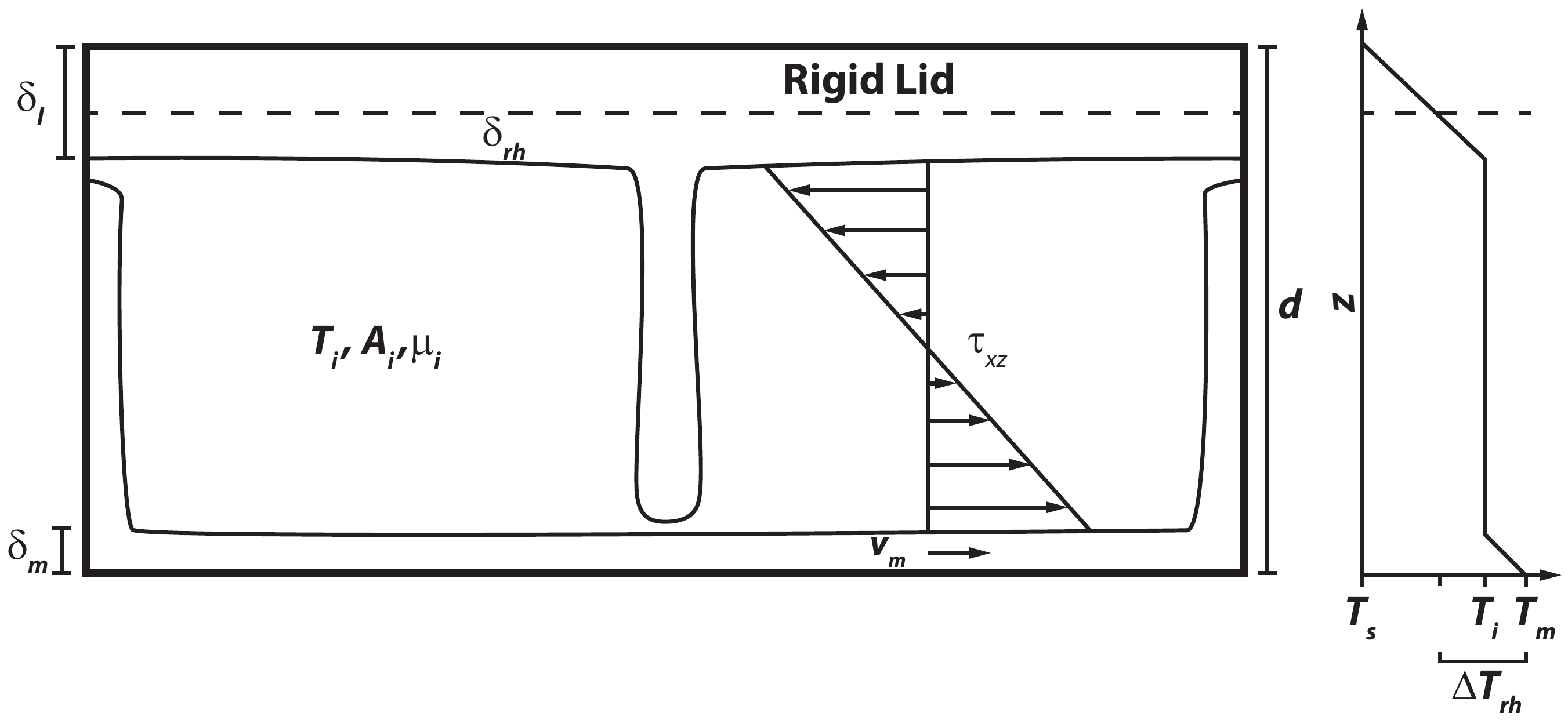}
\caption{\label{fig:scale_sketch_stag} Sketch of the idealized model for convection with grain-damage in the stagnant lid regime. The locations of the immobile lid and rheological boundary layer ($\delta_{rh}$) are shown in the sketch on the left, as well as the asymmetry between $\delta_l$ and $\delta_m$.  On the right, the temperature profile shows the rheological temperature scale, $\Delta T_{rh}$, and the definition of $T_i$ in the stagnant lid regime.  } 
\end{figure}

In the stagnant lid regime, convection only occurs within the region of warm temperatures beneath the lid where the viscosity variation is small.  As a result, the temperature difference actually involved in driving convection is significantly reduced; this reduced temperature difference is typically called the rheological temperature scale, $\Delta T_{rh}$ (Figure \ref{fig:scale_sketch_stag}) \citep[e.g.][]{slava1995,Solomatov2000b}.  $\Delta T_{rh}$ is determined by assuming that the viscosity in the convecting region is nearly constant \citep[e.g.][]{Reese1998}, and thus $\Delta T_{rh} = \Delta T a_{rh}/\theta$, where $\theta$ is the Frank-Kamenestkii parameter as defined in \S \ref{sec:govern_eqs}, and $a_{rh}$ is a constant that determines how large the viscosity variation is within the convecting region \citep{slava1995,Reese1998,Solomatov2000b}. Estimates typically give $a_{rh} \approx 2$ \citep{Solomatov2000b,Korenaga2009}, and this value also applies to our damage models that include temperature-dependent healing, which make up the bulk of results in this study (specifically we find that $a_{rh} \approx 1.8$ for this case; see \S  \ref{sec:comp_stag_H0}).  However, for the case considered in this section, where $E_h = 0$, faster healing leads to a reduced value of $a_{rh} \approx 1.3$, because less effective damage means less of the high viscosity lid can participate in convection.  The reduced value of $a_{rh}$ is calculated from our numerical results, and allows for a straight-forward fit of the scaling laws to the numerical data (see \S \ref{sec:comp_stag_H0}). \cite{slava1995} developed scaling laws for the stagnant lid regime by assuming that convection beneath the lid behaves like constant viscosity convection driven by the reduced temperature drop, $\Delta T_{rh}$.  In the limit of $Nu >> 1$ (i.e. the thickness of the rigid lid is negligible), convective velocity beneath the lid is thus
\begin{equation}
\eqlbl{stag_v}
v_m = \left (\frac{\kappa}{d} \right) C_4 \left ( \frac{Ra_{eff}a_{rh}}{\theta} \right)^{\frac{2}{3}} .
\end{equation}
As this scaling law describes the velocity in the convecting region, we associate this velocity with the basal mantle velocity, $v_m$.  The basal mantle velocity sets the shear stress scale across the convecting region of the mantle, and thus the damage in the mantle, just as $v_l$ does in the temperature-independent viscosity case.  

The top thermal boundary layer consists of the rigid lid, which does not participate in convection, and a thin region at the base of the lid that does participate in convection; this active region is called the rheological boundary layer, $\delta_{rh}$ (Figure \ref{fig:scale_sketch_stag}) \citep{slava1995}.  The thickness of $\delta_{rh}$ is related to the thickness of the entire top thermal boundary layer, $\delta_l$, through the heat flow: $k(\Delta T_{rh}/\delta_{rh}) = k (\Delta T/\delta_l)$, and thus $\delta_{rh} = \delta_l a_{rh}/\theta$.  Using this relationship, and again assuming that convection beneath the lid behaves like constant viscosity convection driven by $\Delta T_{rh}$, $\delta_l$ scales as 
\begin{equation}
\eqlbl{stag_delta}
\delta_l = d C_3 \left(\frac{\theta}{a_{rh}}\right)^{\frac{4}{3}} Ra_{eff}^{-\frac{1}{3}}, 
\end{equation}  
in the limit of $Nu >> 1$ \citep{slava1995}.
 
However, there is some disagreement over these scaling laws.  Boundary layer theories and steady-state numerical results find that $\delta_l \sim \theta Ra_{eff}^{1/5}$ \citep{Morris1984,Reese1998}; however, for time-dependent convection the steady-state boundary layer theory breaks down and \eqref{stag_delta} is the correct scaling \citep{Solomatov2000b,Korenaga2009}.  There is also disagreement over the scaling for velocity.  \cite{Solomatov2000b} find that $v \sim (Ra_{eff}/\theta)^{1/2}$ for their internally heated results.  However, they also state that \eqref{stag_v} fits the bottom heated experiments of \cite{Dumoulin1999}, and that the $1/2$ power-law scaling may be a transitional regime found at low Rayleigh numbers.  We find that \eqref{stag_v} is the correct scaling for our results.  

Scaling laws for convection in the stagnant lid regime with grain-damage are derived in a similar fashion to those for temperature-independent viscosity convection, using the stagnant lid scaling laws for $\delta_l$ \eqref{stag_delta} and $v_m$ \eqref{stag_v} and the grainsize evolution equation \eqref{eqdam} in steady-state. As before, we assume that convection is controlled by the interior mantle viscosity and therefore $\mu_{eff} = \mu_i (A_i/A_0)^{-m}$.  We also assume that $\tau_{xz}$ is the dominant stress component in the convecting interior, and that $\tau_{xz} = (2\mu_{eff} v_m)/d$.  Combining the equation for shear stress with \eqref{stag_v} gives
\begin{equation}
\eqlbl{tau_grainsize}
\tau_{xz}= 2 C_4 \mu_m \frac{\kappa}{d^2} \left(\frac{Ra_0 a_{rh}}{\theta}\right)^{\frac{2}{3}} \left(\frac{\mu_i}{\mu_m}\right)^{\frac{1}{3}} \left(\frac{A_i}{A_0}\right)^{-\frac{m}{3}} .
\end{equation} 
Non-dimesionalizing by the same scales used in section \S \ref{sec:scale_simple}, 
\begin{equation}
\eqlbl{stress_stag}
\tau_{xz}'= 2 C_4 \left(\frac{Ra_0 a_{rh}}{\theta}\right)^{\frac{2}{3}}  \mu_i'^{\frac{1}{3}} A_i'^{-\frac{m}{3}} .
\end{equation}  
With temperature-dependent viscosity, the fineness evolution equation in steady-state (again assuming that $\tau_{xz} >> \tau_{xx}$) is 
\begin{equation}
\frac{D}{\mu_i'}\tau_{xz}'^2 A_i'^{m} = HA_i'^p
\end{equation}
and therefore the average fineness in the convecting interior is 
\begin{equation}
\eqlbl{steady_fine_stag}
A_i' = \left(\frac{D}{H\mu_i'} \tau_{xz}'^2 \right)^{\frac{1}{p-m}} .
\end{equation}
Combining equations \eqref{stress_stag} and \eqref{steady_fine_stag} gives the final scaling law for shear stress in the convecting region, 
\begin{equation}
\eqlbl{scale_stress_stag}
\tau_{xz}'= (2 C_4)^{\frac{3(p-m)}{3p-m}} \mu_i'^{\frac{p}{3p-m}} \left(\frac{D}{H}\right)^{-\frac{m}{3p-m}} \left(\frac{Ra_0 a_{rh}}{\theta}\right)^{\frac{2(p-m)}{3p-m}}  .
\end{equation}
Equation \eqref{scale_stress_stag} is the same is \eqref{tauxz_scaling} with the additional factors of $\theta$ and $\mu_i'$ that arise due to temperature-dependent viscosity.  Increasing the Frank-Kamenestkii parameter decreases shear stress because it lowers $\Delta T_{rh}$, the effective temperature difference driving convection beneath the lid.  The interior mantle viscosity does not play a major role in how shear stress scales, as $\mu_i'$ is of order 1.  However, correctly accounting for the deviation of $\mu_i$ from the reference viscosity, $\mu_m$, is important for fitting the data from the numerical experiments.  Combining equations \eqref{steady_fine_stag} and \eqref{scale_stress_stag} gives an equation for the average interior mantle fineness: 
\begin{equation}
\eqlbl{scale_a_stag}
A_i' = (2C_4)^{\frac{6}{3p-m}}  \mu_i'^{-\frac{1}{3p-m}} \left(\frac{D}{H}\right)^{\frac{3}{3p-m}} \left(\frac{Ra_0 a_{rh}}{\theta} \right)^{\frac{4}{3p-m}} .
\end{equation}  

The scaling laws for velocity and Nusselt number are derived in similar fashion.  From \eqref{stag_v}, the non-dimensional basal mantle velocity is 
\begin{equation}
\eqlbl{stag_v2}
v_m' = C_4 \left (\frac{Ra_0 a_{rh}}{\theta} \right)^{\frac{2}{3}} \mu_i'^{-\frac{2}{3}} A_i'^{\frac{2m}{3}} 
\end{equation}
which gives the final scaling law for velocity when \eqref{scale_a_stag} is substituted into \eqref{stag_v2}, 
\begin{equation}
\eqlbl{scale_v_stag}
v_m' = C_4(2C_4)^{\frac{4m}{3p-m}}\mu_i'^{-\frac{2p}{3p-m}} \left(\frac{D}{H} \right)^{\frac{2m}{3p-m}} \left(\frac{Ra_0 a_{rh}}{\theta} \right)^{\frac{2(p+m)}{3p-m}} .
\end{equation} 
From \eqref{Nu_nondim} and \eqref{stag_delta} the Nusselt number is 
\begin{equation}
\eqlbl{Nu_stag1}
Nu = \frac{T_i'}{C_3} \left(\frac{\theta}{a_{rh}}\right)^{-\frac{4}{3}} Ra_0^{\frac{1}{3}} \mu_i'^{-\frac{1}{3}} A_i'^{\frac{m}{3}}
\end{equation}
and the final scaling law for Nu is obtained by substituting \eqref{scale_a_stag} into \eqref{Nu_stag1}, 
\begin{equation}
\eqlbl{scale_Nu_stag}
Nu = \frac{T_i'}{C_3} (2C_4)^{\frac{2m}{3p-m}} \mu_i'^{-\frac{p}{3p-m}} \left(\frac{\theta}{a_{rh}}\right)^{-\frac{4p}{3p-m}} \left(\frac{D}{H} \right)^{\frac{m}{3p-m}}  Ra_0^{\frac{p+m}{3p-m}} .
\end{equation}
The scaling laws for the stagnant lid regime derived above only differ from the scaling laws for temperature-independent viscosity by the factors of $\theta$ and $\mu_i$ that come into the stagnant lid scaling laws.  

\subsection{Comparison to Numerical Experiments}
\label{sec:comp_stag_H0}

\begin{figure}
\includegraphics[scale = 0.75]{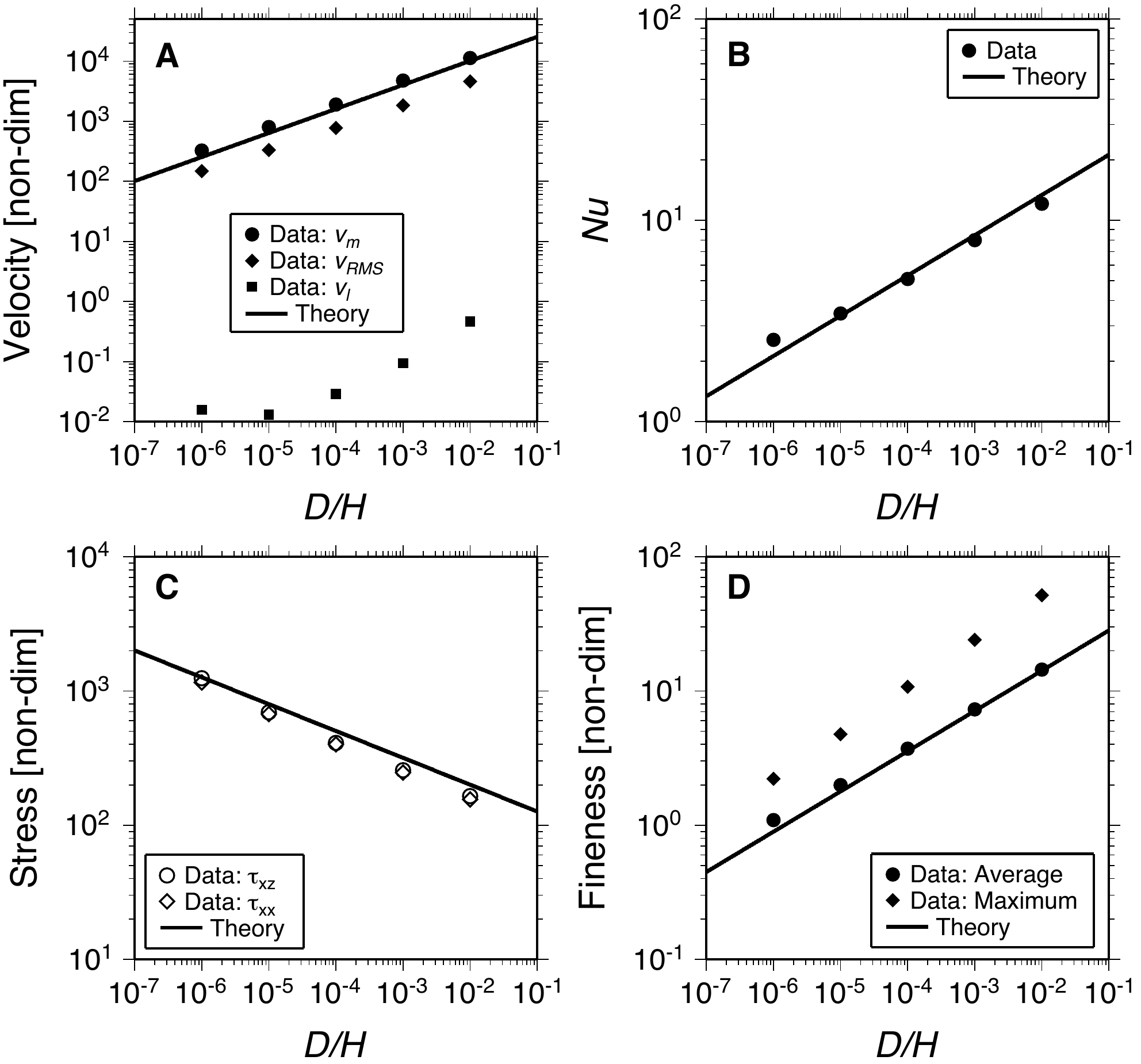}
\caption{\label {fig:dscale_stag} Plots of velocity (a), Nusselt number (b), stress (c), and fineness (d) versus $D/H$ for convection with temperature-dependent viscosity and constant healing (\S \ref{sec:stag_H0}).  Numerical results are plotted as symbols with theoretical scaling laws plotted as solid lines.  Figure (a) shows numerical data for $v_m'$ compared to the scaling law for $v_m'$ \eqref{scale_v_stag}; data for $v_{RMS}'$ and $v_l'$ are also shown.  Figure (b) compares the numerical results for $Nu$ to the scaling law for $Nu$ \eqref{scale_Nu_stag}.  Figure (c) compares the data for the sub-lid ${\tau_{xz}'}_{RMS}$ to the scaling law for $\tau_{xz}'$ \eqref{scale_stress_stag}; data for the sub-lid  ${\tau_{xx}'}_{RMS}$ is also shown.  Figure (d) compares the data for sub-lid $\bar{A}'$ to the scaling law for $A_i'$ \eqref{scale_a_stag} and also shows data for $A'_{max}$. }  
\end{figure}

\begin{figure}
\includegraphics[scale = 0.75]{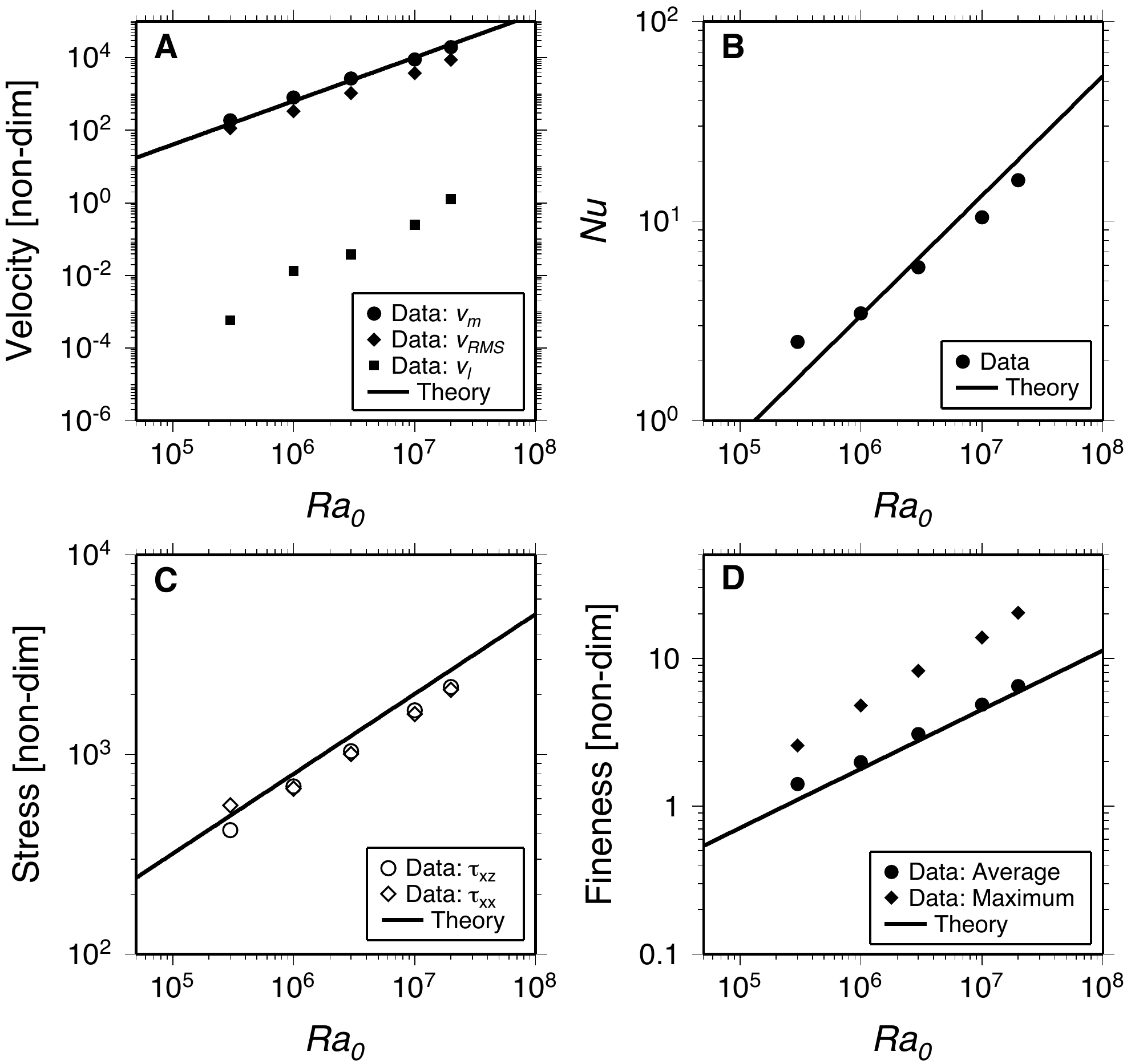}
\caption{\label {fig:Ra_scale_stag} Plots of time averaged velocity (a), Nusselt number (b), stress (c), and fineness (d) versus Rayleigh number, $Ra_0$, as in Figure \ref{fig:dscale_stag}.}  
\end{figure}

\begin{figure}
\includegraphics[scale = 0.5]{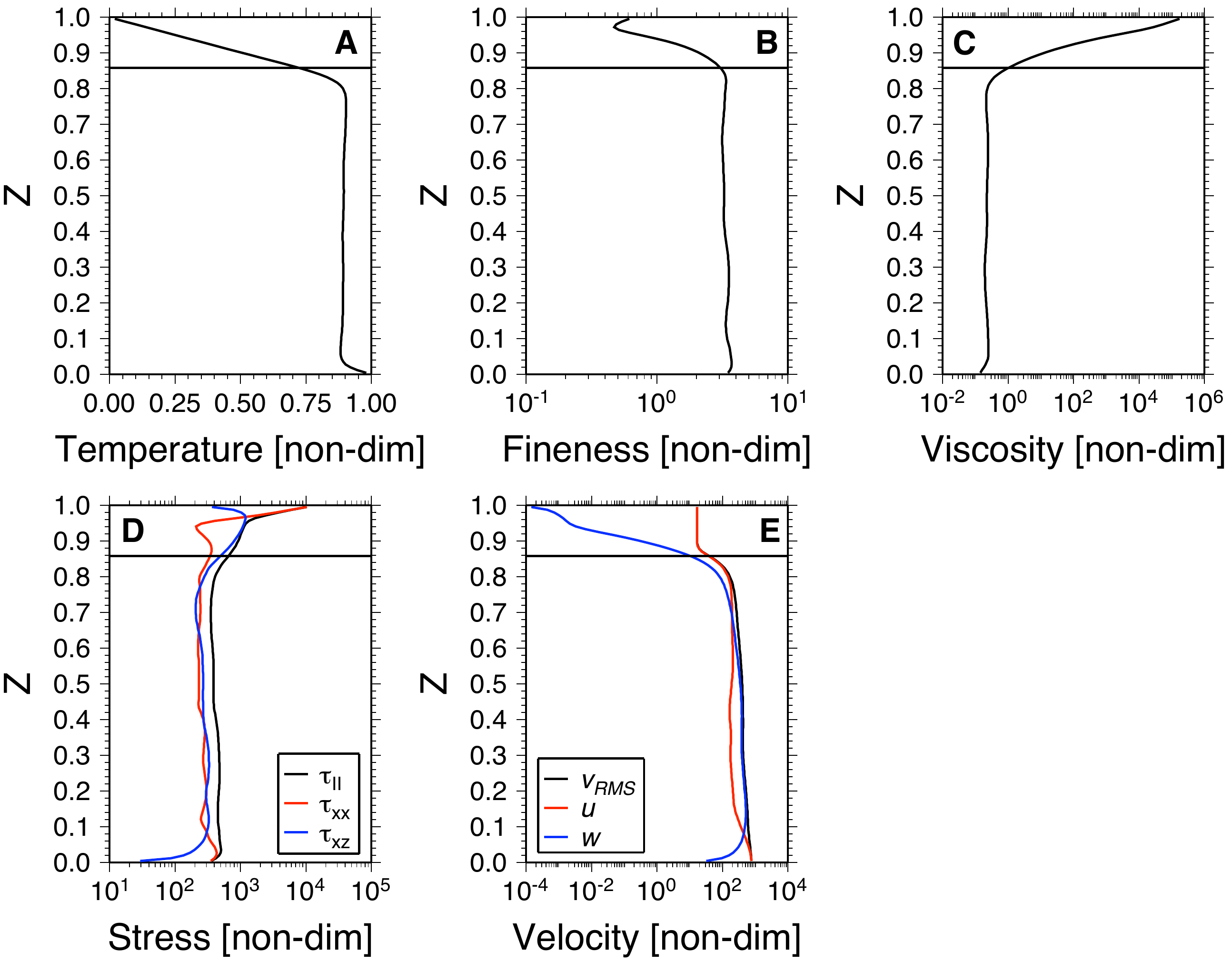}
\caption{\label {fig:z_profiles} Vertical profiles of horizontally averaged temperature (a), fineness (b), viscosity (c), stress (d), and velocity (e) for a representative stagnant lid case in statistical steady-state.  The plot for stress (d) shows depth profiles of the second invariant of the stress tensor, $\langle \tau'_{II} \rangle$, shear stress, $\langle {\tau'_{xz}}\rangle_{RMS}$, and normal stress, $\langle \tau'_{xx} \rangle_{RMS}$.  The plot for velocity (e) shows depth profiles of the full velocity vector, $\langle v'\rangle_{RMS}$, the horizontal velocity, $\langle u' \rangle_{RMS}$, and vertical velocity, $\langle w' \rangle_{RMS}$.  All plots show the base of the stagnant lid, defined by $\langle w'\rangle_{RMS} = 10$, as a horizontal black line.  Parameters for model result shown here: $Ra_0 = 10^6$, $D/H=10^{-4}$, $E_v' = 23.03$, and $T_s^* = 1$.  }  
\end{figure}

We compare our theory to numerical results in the stagnant lid regime where $D/H$ and $Ra_0$ are varied separately (Figures \ref{fig:dscale_stag} and \ref{fig:Ra_scale_stag}).  The theoretical curves use $C_3 = 4.24$, $C_4 = 0.125$, and $a_{rh}=1.3$, while the numerical models varying $D/H$ use $Ra_0 = 10^6$, $m=2$, $p=4$, $E_v' = 23.03$, and $T_s^* = 1$; those varying $Ra_0$ use $D/H = 10^{-5}$ and the same parameters otherwise.  The temperature-dependent viscosity relation used here results in an effective Frank-Kamenetskii parameter of $\theta = 5.76$ and a viscosity ratio of $\mu_l' = 10^5$. The constant $a_{rh}$ is determined by calculating $\Delta T'_{rh}$ from our numerical models, and using $a_{rh} = \Delta T'_{rh} \theta$ (see \S \ref{sec:stag_H0_scale}). Assuming symmetry between $\delta'_m$ and $\delta'_{rh}$, $\Delta T'_{rh} = 2(1-T_i')$ (see Figure \ref{fig:scale_sketch_stag}), and the internal temperature, $T_i'$, is calculated from the numerical results by taking the sub-lid volume average of the temperature field.  The numerical results consistently yield a value of $a_{rh} \approx 1.3$. The sub-lid average, also used for fineness and stress, is defined as a volume average only within the convecting region (i.e. a volume average beneath the base of the lid).  Computing averages only within the convecting region is important because the rigid lid has a significant effect on whole mantle averages, especially for stress which is large in the lid. We define the base of the stagnant lid as the depth where the horizontally averaged vertical velocity profile reaches a non-dimensional value of 10.  As illustrated by depth profiles of horizontally averaged temperature, velocity, and stress, the definition of $\langle w' \rangle_{RMS} = 10$ as the base of the stagnant lid is, if anything, a conservative choice (Figure \ref{fig:z_profiles}); i.e. some of the lid is included in the average.  Using a different value of $\langle w' \rangle_{RMS}$ to define the start of the convecting region does not change how the quantities considered here scale with $D/H$ and $Ra_0$.  Finally, we calculate $v_m'$ from the numerical results using \eqref{max_plate} with $u'$ evaluated at $z'=0$ instead of $z'=1$.  

The numerical results for both the experiments varying $D/H$ and those varying $Ra_0$ fit the data well, similar to the results for temperature-independent viscosity.  The basal mantle velocity, the sub-lid average shear stress, and the sub-lid average fineness all match the theory (Figures \ref{fig:dscale_stag}a,c,d and \ref{fig:Ra_scale_stag}a,c,d).  The Nusselt number deviates from the theory more than any of the other observables (Figures \ref{fig:dscale_stag}b and \ref{fig:Ra_scale_stag}b), but the data appears to asymptote to the theoretical curves at high $D/H$ and $Ra_0$.  

\section{Temperature-Dependent Viscosity and Healing}
\label{sec:full_dam}

With both temperature-dependent viscosity and healing, numerical convection results display a wide range of behavior from stagnant lid convection to convection that resembles the temperature-independent viscosity case.  We also observe a parameter range where convection ceases entirely; the cessation of convection is more readily observed with temperature-dependent viscosity because it occurs at higher $D/H$ or $Ra_0$ (i.e. the region of parameter space where convection will not occur is larger).  As shown by \cite{Landuyt2009a} and \cite{Foley2012}, stagnant lid convection occurs when either the damage to healing ratio in the lithosphere is low, or when the Rayleigh number is low, while the cessation of convection occurs when the damage to healing ratio in the mantle is very low, or $Ra_0$ is very low, as is explained further in \S \ref{sec:onset_convec}.  The numerical models show that with increasing $D/H$, the lid becomes steadily more mobilized over a broad region of parameter space (Figure \ref{fig:pform_full1}).  At low $D/H$, convection shows weak mobility with a ``mushy lid" as damage is just able to soften the high viscosity lithosphere.  At higher $D/H$, more coherent downwellings can form leading to faster plate velocities; however, the plate velocity is still slower than the basal mantle velocity due to the resistance provided by the viscosity of damaged lithospheric shear zones.  In addition, the wavelength of instability for the top boundary layer is large compared to the bottom boundary layer; this is a result of the long horizontal length scale required for the top boundary layer to go unstable.  Finally, at large $D/H$, models resembles temperature-independent viscosity convection because damage is so effective at weakening the lithosphere.  Convection becomes symmetric: the plate velocity converges to the basal mantle velocity and the wavelength of instability for the top boundary layer becomes equal to that of the bottom boundary layer.  Similar trends occur for varying $Ra_0$.       

\afterpage{
\begin{landscape}
\begin{figure}
\includegraphics[scale = 0.75]{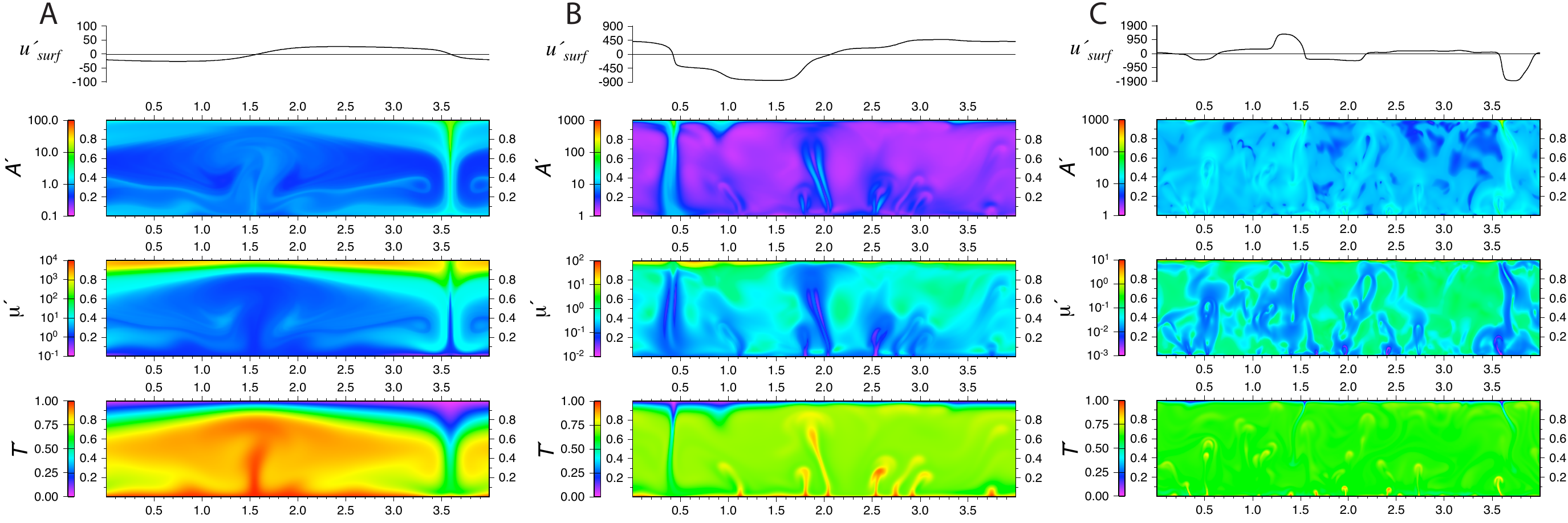}
\caption{\label {fig:pform_full1} Snapshots of convection with temperature-dependent viscosity and healing (\S\ref{sec:full_dam}) in the transitional regime.  Models use three different $D/H$ values: $D/H = 10^{-7}$ (a), $D/H = 10^{-5}$ (b), and $D/H = 10^{-3}$ (c).  All models use $Ra_0 = 10^6$, $m=2$, $p=4$, $E_v' = E_h'=23.03$, and $T_s^* = 1$.  Each snapshot shows the surface velocity, $u'_{surf}$, the fineness field, $A'$, the viscosity field, $\mu'$, and the temperature field, $T'$.  } 
\end{figure}   
\end{landscape}}

\subsection{Scaling Theory} 
\label{sec:full_scale}

The numerical results show three main regimes of convection: the fully stagnant lid regime, the transitional regime, and the fully mobile regime, with most models shown here falling in the transitional regime. The fully stagnant lid regime is defined by the assumption that convection takes place beneath a rigid lid, and that the effective viscosity of the mantle interior controls instability of drips off the base of this lid; thus the important characteristics of convection, such as $\delta_l'$ or $v_m'$, are set by the interior mantle viscosity (\S \ref{sec:stag_H0_scale}). The viscosity that effectively controls convection changes in the transitional regime, where the viscosity of damaged lithospheric shear zones determines instability of the top thermal boundary layer and thus the plate motion.  In this regime, plates show a wide degree of mobility relative to the interior, from near stagnant lid behavior when convection first enters the transitional regime, to near isoviscous convection, or full mobilization, when convection enters the fully mobile regime.  Finally, the fully mobile regime is defined as the point when damage in the  lithosphere is so effective that the viscosity of lithospheric shear zones no longer provides any significant resistance to flow.  Thus the interior mantle viscosity again controls boundary layer thickness and convective velocity.  As a result, the scaling laws for the fully stagnant lid and fully mobile regimes derived below provide end member limits to the numerical data.  For the example case of varying $D/H$, convection follows the fully stagnant lid scaling law at low $D/H$, deviates from this limit toward the transitional regime scaling law at moderate $D/H$, and finally converges to the fully mobile limit at high $D/H$. 

We derived scaling laws for the fully stagnant lid regime in \S \ref{sec:stag_H0_scale} for temperature-insensitive healing.  Here we briefly show how to extend the stagnant lid regime scaling laws to incorporate temperature-dependent healing.  As in \S \ref{sec:stag_H0_scale}, shear stress follows     
\begin{equation}
\eqlbl{stress_stag2}
\tau_{xz}'= 2 C_4 \left(\frac{Ra_0 a_{rh}}{\theta}\right)^{\frac{2}{3}} \mu_i'^{\frac{1}{3}} A_i'^{-\frac{m}{3}} .
\end{equation}  
The average mantle fineness is found using the fineness evolution equation in steady-state,
\begin{equation}
\frac{D}{\mu_i'}\tau_{xz}'^2 A_i'^{m} = H h_i' A_i'^p
\end{equation}
where $h_i'$ is the non-dimensional healing rate evaluated at the average interior mantle temperature, $T_i$.  Thus 
\begin{equation}
\eqlbl{steady_fine_stag2}
A_i' = \left(\frac{D}{H h_i' \mu_i'} \tau_{xz}'^2 \right)^{\frac{1}{p-m}} 
\end{equation}
and
\begin{equation}
\eqlbl{scale_stress_stag2}
\tau_{xz}'= (2 C_4)^{\frac{3(p-m)}{3p-m}} \mu_i'^{\frac{p}{3p-m}} \left(\frac{D}{H h_i'}\right)^{-\frac{m}{3p-m}} \left(\frac{Ra_0 a_{rh}}{\theta}\right)^{\frac{2(p-m)}{3p-m}}  .
\end{equation}
The remaining scaling laws can be derived in similar fashion and are as follows: 
\begin{equation}
\eqlbl{scale_a_stag2}
A_i' = (2C_4)^{\frac{6}{3p-m}} \mu_i'^{-\frac{1}{3p-m}}  \left(\frac{D}{H h_i'} \right)^{\frac{3}{3p-m}}  \left(\frac{Ra_0 a_{rh}}{\theta} \right)^{\frac{4}{3p-m}}
\end{equation}  
\begin{equation}
\eqlbl{scale_v_stag2}
v_m' = C_4(2C_4)^{\frac{4m}{3p-m}} \mu_i'^{-\frac{2p}{3p-m}} \left(\frac{D}{H h_i'} \right)^{\frac{2m}{3p-m}} \left(\frac{Ra_0 a_{rh}}{\theta} \right)^{\frac{2(p+m)}{3p-m}} 
\end{equation} 
\begin{equation}
\eqlbl{scale_Nu_stag2}
Nu = \frac{T_i'}{C_3} (2C_4)^{\frac{2m}{3p-m}} \mu_i'^{-\frac{p}{3p-m}} \left(\frac{\theta}{a_{rh}}\right)^{-\frac{4p}{3p-m}} \left(\frac{D}{H h_i'} \right)^{\frac{m}{3p-m}}  Ra_0^{\frac{p+m}{3p-m}} .
\end{equation}

In the fully mobile regime, convection resembles the temperature-independent viscosity case: the top and bottom boundary layers are symmetric, and the internal temperature $T_i' = 0.5$.  Scaling laws for the fully mobile limit have the same form as equations \eqref{scale_stress_stag2} - \eqref{scale_Nu_stag2} (with $T_i' = 0.5$), except the Frank-Kamenetskii parameter $\theta$ drops out of the equations, and the constants $C_3$ and $C_4$ are replaced by $C_2$ and $C_1$, respectively.  Scaling laws in the fully mobile limit no longer depend on $\theta$ because convection now involves the entire temperature drop across the mantle, $\Delta T$.  The scaling laws are thus, 
\begin{equation}
\eqlbl{scale_stress_mob2}
\tau_{xz}'= (2 C_1)^{\frac{3(p-m)}{3p-m}} \mu_i'^{\frac{p}{3p-m}} \left(\frac{D}{Hh_i'}\right)^{-\frac{m}{3p-m}} Ra_0^{\frac{2(p-m)}{3p-m}}  
\end{equation}
\begin{equation}
\eqlbl{scale_a_mob2}
A_i' = (2C_1)^{\frac{6}{3p-m}} \mu_i'^{-\frac{1}{3p-m}}  \left(\frac{D}{H h_i'}\right)^{\frac{3}{3p-m}}  Ra_0 ^{\frac{4}{3p-m}}
\end{equation}  
\begin{equation}
\eqlbl{scale_v_mob2}
v_m' = C_1(2C_1)^{\frac{4m}{3p-m}}\mu_i'^{-\frac{2p}{3p-m}} \left(\frac{D}{Hh_i'} \right)^{\frac{2m}{3p-m}} Ra_0^{\frac{2(p+m)}{3p-m}} 
\end{equation} 
\begin{equation}
\eqlbl{scale_Nu_mob2}
Nu = \frac{T_i'}{C_2} (2C_1)^{\frac{2m}{3p-m}} \mu_i'^{-\frac{p}{3p-m}} \left(\frac{D}{Hh_i'} \right)^{\frac{m}{3p-m}}  Ra_0^{\frac{p+m}{3p-m}} .
\end{equation}   

We derive new scaling laws for convection in the transitional regime, as there are no previous studies directly constraining this style of convection.  \cite{slava1995} explored convection in a transitional regime at low viscosity ratios (e.g. $\mu_l' \approx 10^2 - 10^3$) that lies between constant viscosity convection and stagnant lid convection, however, it is unclear if this scaling theory is applicable to the transitional regime we observe, which sits between stagnant lid convection and fully mobile convection at high viscosity ratios.  

\begin{figure}
\includegraphics[scale = 0.55]{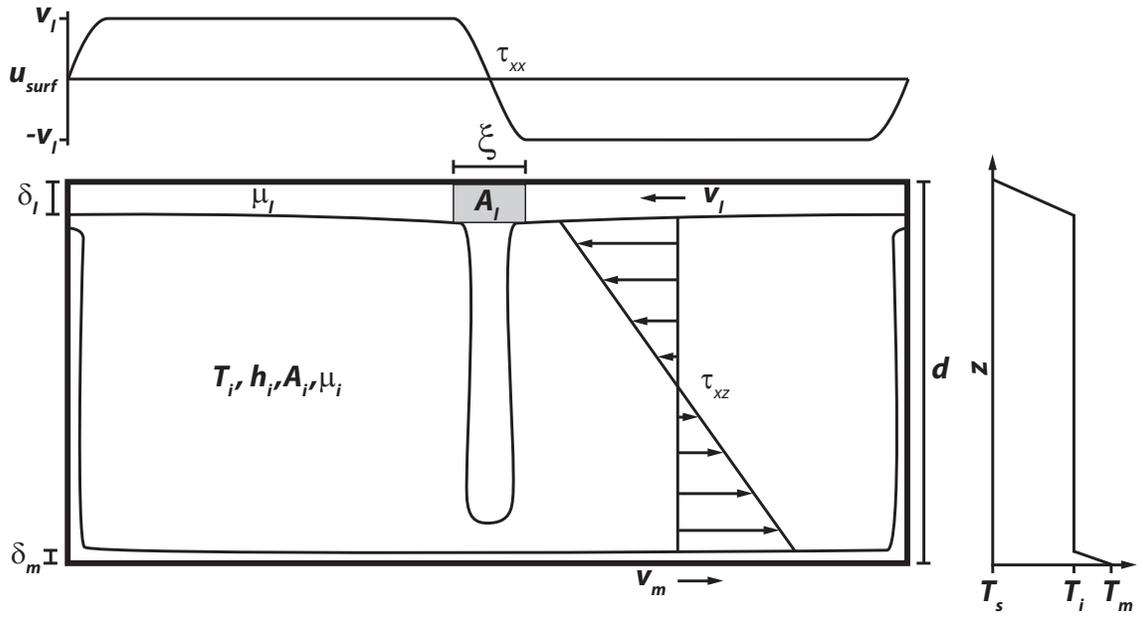}
\caption{\label {fig:scale_sketch_trans} Sketch of the idealized model for convection with grain-damage in the transitional regime. Sketch of the convective planform (bottom left) shows the lithospheric thickness $\delta_l$ is controlled by the viscosity of the damaged shear zone, with a fineness $A_l$.  The vertical temperature profile (right) denotes $T_i$, and shows the asymmetry between the top and bottom boundary layers.  The surface velocity (top) shows how the velocity goes from $v_l$ to $-v_l$ over the lithospheric shear zone of thickness $\xi$.  This sets the lithospheric normal stress, $\tau_{xx}$, which drives damage in the lithosphere.}  
\end{figure}
 
We first derive an equation for the thickness of the top thermal boundary layer, $\delta_l$, as scaling laws for $Nu$ and the plate velocity are easily obtained from those for $\delta_l$ (see the conceptual sketch, Figure \ref{fig:scale_sketch_trans}).  The scaling law for $\delta_l$ is determined semi-empirically; we use boundary layer stability analysis to constrain which non-dimensional parameters should appear in the scaling law, and empirically determine the scaling law exponents from our numerical experiments. We choose to use empirical fits because the simple boundary layer theory used in previous sections (\S \ref{sec:scale_simple} and \S\ref{sec:stag_H0_scale}) fails for transitional regime convection, where plate motion is dominated by the viscosity within the boundary layer rather than in the mantle interior.  

Assuming that the top boundary layer, $\delta_l$, is approximately at the margin of convective instability, we write 
\begin{equation}
Ra_c \approx \frac{\rho g \alpha (T_i - T_s) \delta_l^3}{\kappa \mu_l (\frac{A_l}{A_0})^{-m}}
\end{equation}
where $Ra_c$ is the critical Rayleigh number for convection with free-slip boundaries ($Ra_c \approx 700$) and $A_l$ is the characteristic fineness of lithospheric shear zones. Strictly speaking, the top boundary layer may not be at the margin of convective instability for bottom heated convection, due to the influence of plumes from the bottom boundary layer \citep{Moore2008}.  However, as we will later employ empirical fits to determine the final scaling law, the small errors introduced by this assumption are accounted for.  Non-dimensionalizing and solving for $\delta_l$, and also using $T_i - T_s = \Delta T T_i'$, 
\begin{equation}
\eqlbl{lith_trans}
\delta_l' \approx \left(\frac{Ra_c \mu_l'}{Ra_0 T_i'} \right)^{\frac{1}{3}} A_l'^{-\frac{m}{3}}.
\end{equation}
As per the scaling law derivations in previous sections, an equation for the lithospheric fineness is derived from the fineness evolution equation in steady-state. However, the dominant stress scale responsible for damage in lithospheric shear zones is different from the stress scale that dominates in the mantle interior.  Lithospheric motion is primarily horizontal flow towards convergent regions (above downwellings), and away from divergent regions (above upwellings), and thus the dominant stresses at nascent plate boundaries are the normal stresses associated with these convergent or divergent flows; this dominance of normal stresses at lithospheric shear zones is a near universal feature of the numerical models (see also \cite{Foley2012}). Therefore, given that $\tau_{xx} >> \tau_{xz}$ in the lithosphere, the deformational work term is 
\begin{equation}
\underline{\underline{\dot{\varepsilon}'}} : \underline{\underline{\tau'}} = \frac{\underline{\underline{\tau'}} : \underline{\underline{\tau'}}}{2\mu_l' A_l'^{-m}} = \frac{\tau_{xx}'^2}{\mu_l' A_l'^{-m}}
\end{equation}
where we have used the fact that $\tau_{xx} = \tau_{zz}$ due to mass conservation. The steady-state fineness equation is thus
\begin{equation}
\frac{D}{\mu_l'} \tau_{xx}'^2 A_l'^m = Hh_l'A_l'^p,
\end{equation}
and the lithospheric fineness is
\begin{equation}
\eqlbl{steady_fine_lith}
A_l' = \left(\frac{D\tau_{xx}'^2}{Hh_l'\mu_l'} \right)^{\frac{1}{p-m}} .
\end{equation}

The lithospheric normal stress, $\tau_{xx}'$, can be written in terms of the strain-rate in a lithospheric shear zone as $\tau_{xx}' \sim (\mu_l' A'^{-m} v_l')/\xi$, where $\xi$ is the characteristic width of the shear zone (Figure \ref{fig:scale_sketch_trans}); there is no known scaling law for $\xi$.  Since we assume that the lithospheric shear zone viscosity controls the dynamics of the lithosphere, the shear zone thickness must be a function of lithospheric quantities and properties, namely the lithospheric damage to healing ratio, $D/(Hh_l')$, the plate velocity, $v_l'$, and the lithospheric viscosity, $\mu_l'$. Therefore $A_l'$ is some unknown function of $\mu_l'$, $v_l'$, and $D/(Hh_l')$.  

The plate velocity, $v_l$, is related to $\delta_l$ via thermal diffusion. The top boundary layer grows diffusively for a time $L/v_l$, where $L$ is the horizontal distance the boundary layer must travel before going unstable (i.e. the plate length).  Thus the boundary layer thickness $\delta_l = (\kappa L/v_l)^{1/2}$, which, when non-dimensionalized (where the non-dimensional plate-length, $L'$, is defined as $L' = L/d$), gives $\delta_l' = (L'/v_l')^{1/2}$.  Using this relationship between $\delta_l'$ and $v_l'$, assuming that the unknown function for $A_l'$ has power-law dependencies on $L'$, $\mu_l'$, $v_l'$, and $D/(Hh_l')$, and using \eqref{lith_trans}, $\delta_l'$ scales as    
\begin{equation}
\eqlbl{scale_lith_trans}
\delta_l' = C_5 L'^{\beta_L} \mu_l'^{\beta_{\mu}} \left(\frac{D}{Hh_l'} \right)^{\beta_{D}} ({Ra_0 T_i'}) ^{\beta_{Ra}}  . 
\end{equation}   
where $C_5$, $\beta_L$, $\beta_{\mu}$, $\beta_D$, and $\beta_{Ra}$ are constants determined from the numerical results. 

The plate length, $L'$, arises from the convecting system, and can be calculated from boundary layer theory \citep[e.g.][]{turc1982}; however, as an alternative, we choose to exploit our numerical results and calculate $L'$ directly from the models.  We therefore treat the plate length, $L'$, as an unknown in \eqref{scale_lith_trans}, similar to $D/(Hh_l')$ or $Ra_0$, and determine the influence of varying $L'$ empirically from the numerical results. We compute $L'$ from the numerical models using the relationship between plate speed and boundary layer thickness; this gives $L' = v_l'\delta_l'^2$, where $\delta_l' = T_i'/Nu$ from the definition of the Nusselt number, and $v_l'$, $T_i'$, and $Nu$ are output from the numerical models (see \S \ref{sec:comp_simple} and \S \ref{sec:comp_stag_H0}).  The numerical results show that for a fixed aspect ratio numerical domain, $L'$ is nearly constant in the transitional regime ($L' \approx 1.2 - 1.8$, with an average value of $L' \approx 1.5$ for the models with a $4 \times 1$ aspect ratio).  We therefore assume that $L'$ can be approximated as independent of damage to healing ratio, Rayleigh number, and viscosity ratio, and that the influence of $L'$ on the top boundary layer thickness can be constrained using a set of models where the aspect ratio of the numerical domain is changed (increasing the aspect ratio of the numerical model from $4 \times 1$ to $16 \times 1$ causes $L'$ to increase from $\approx 1.5$ to $\approx 4$).  Determining the influence of plate length in this manner is analogous to how the roles of $D/(Hh_l')$ or $Ra_0$ are assessed using models where these parameters are varied.  

\begin{figure}
\includegraphics[scale = 0.65]{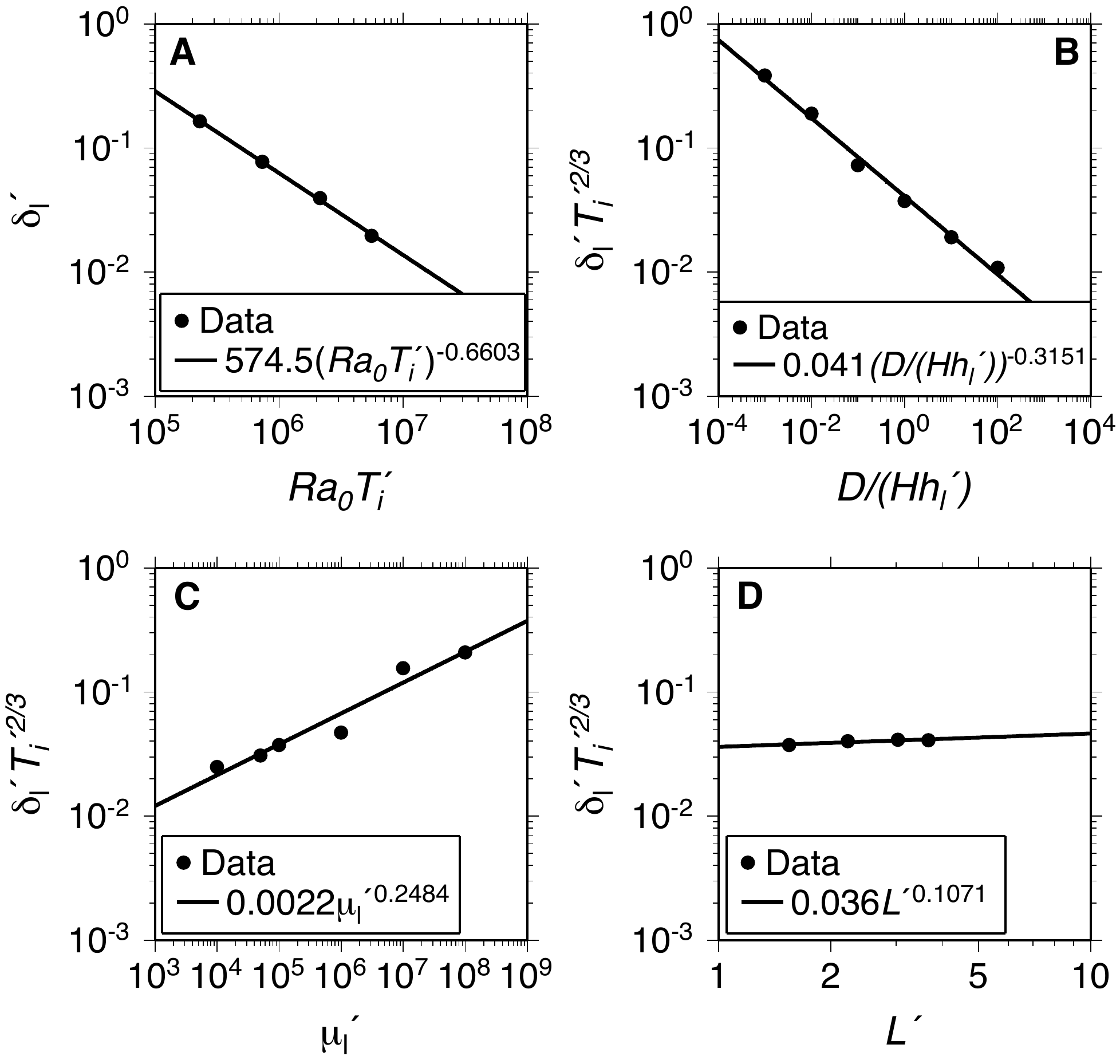}
\caption{\label {fig:fitplots} Plots of numerical data with least squares fits for varying $Ra_0$ (a), $D/(Hh_l')$ (b), $\mu_l'$ (c), and $L'$ (d).  The internal temperature, $T_i'$, varies as $Ra_0$, $D/(Hh_l')$, $\mu_l'$, or $L'$ are varied.  We therefore take the variation of $T_i'$ into account by fitting $Ra_0T_i'$ to the data for $\delta_l'$ (a), and fitting $D/(Hh_l')$, $\mu_l'$, and $L'$ to $\delta_l'T_i'^{2/3}$.  The results shown here use $m=2$, $p=4$, and: (a) $D/H=10^{-5}$, $\mu_l'=10^6$, $h_l' = 10^{-5}$, $L'=1.5$ (for a $4 \times 1$ aspect ratio numerical domain); (b) $Ra_0=10^{6}$, $\mu_l'=10^5$, $L' = 1.5$; (c) $Ra_0=10^{6}$, $D/H=10^{-5}$, $h_l' = 10^{-5}$, $L'=1.5$; (d) $Ra_0=10^{6}$, $D/H=10^{-5}$, $h_l' = 10^{-5}$, $\mu_l' = 10^5$.  The results in (d) use numerical domains with aspect ratios ranging from $4 \times 1$ to $16 \times 1$.}  
\end{figure}

We determine the scaling law exponents empirically.  We first perform a least squares fit to numerical experiments where $Ra_0$ is varied, and find $\beta_{Ra} \approx -2/3$ for $m=2$ and $p=4$.  Using $\beta_{Ra} = -2/3$, we perform least squares fits to determine $\beta_D \approx -1/3$, $\beta_{\mu} \approx 1/4$, and $\beta_L \approx 1/10$ (see Table \ref{tab_mp} for the scaling law constants with different combinations of $m$ and $p$); all three fits give $C_5 \approx 20$ for $m=2$ and $p=4$ (Figure \ref{fig:fitplots}).  Our constraint on $\beta_L$ is only based on a limited number of numerical experiments.  However, given that the influence of $L'$ is significantly less than the influence of damage to healing ratio, Rayleigh number, and viscosity ratio, and that the plausible range of variation in $L'$ for planetary mantle convection is also significantly less than these other parameters, which can vary many orders of magnitude, our simple model to incorporate the effect of varying plate length is sufficient for constraining the first order scaling laws. 

\begin{table}
\caption{Transitional Regime Scaling Law Constants for Various $m$ and $p$}
\label{tab_mp}
\begin{tabular}{c c c c c c c}
\hline
$m$ & $p$ & $C_5$ & $\beta_{Ra}$ & $\beta_D$ & $\beta_{\mu}$ & $\beta_L$ \\ \hline
2 & 4 & 20 & -0.6603  & -0.3151 & 0.2484 & 0.1071 \\
3 & 4 & 86 & -0.8515  & -0.4919 & 0.2598 & -0.0218 \\
3 & 5 & 11 & -0.6232  & -0.4342 & 0.1931 & 0.0582 \\
\end{tabular}      
\end{table}

To close our scaling analysis of transitional regime convection, we need a relation for the internal temperature, $T_i'$; for this we require a scaling law for the bottom boundary layer, $\delta_m$.  This can be derived fully from boundary layer stability theory because the viscosity of the bottom boundary layer is low, and we can assume that the interior mantle viscosity controls the boundary layer thickness, as in the scaling laws for stagnant lid and fully mobile convection.  Assuming marginal stability of the boundary layer, 
\begin{equation}
Ra_c \sim \frac{\rho g \alpha (T_m-T_i) \delta_m^3}{\kappa \mu_i (\frac{A_i}{A_0})^{-m}} ;
\end{equation} 
non-dimensionalizing and using $T_m - T_i = \Delta T (1-T_i')$, the bottom boundary layer thickness is 
\begin{equation}
\eqlbl{deltam_marginal}
\delta_m' \sim \left(\frac{Ra_c \mu_i'}{Ra_0 (1-T_i')} \right)^{\frac{1}{3}} A_i'^{-\frac{m}{3}}.
\end{equation}
The steady-state fineness equation is given by \eqref{steady_fine_stag2}, where $\tau_{xz}' = 2\mu_i'A_i'^{-m} v_m'$. As is the case for the top thermal boundary layer, $v_m$ is related to $\delta_m$ via thermal diffusion: $\delta_m = (\kappa L _m/ v_m)^{1/2}$, where $L_m$ is the wavelength of convection for the bottom boundary layer. When non-dimensionalized, $\delta_m' = (L_m'/v_m')^{1/2}$.  We calculate $L_m'$ from the numerical models using a procedure analogous to the calculation of $L'$ (see text below equation \eqref{scale_lith_trans}): $L_m' = v_m'\delta_m'^2$, where $\delta_m'=(1-T_i')/Nu$ and $Nu$, $v_m'$, and $T_i'$ are output from the models. We find that $L_m' \approx 1$ for nearly all cases, and thus we assume $\delta_m' = v_m'^{-1/2}$.  The steady-state mantle fineness is then
\begin{equation}
\eqlbl{internal_A_trans}
A_i' = \left(\frac{4D\mu_i'}{Hh_i' \delta_m'^4} \right)^{\frac{1}{p+m}} .
\end{equation}  
Eliminating $A_i'$ between and \eqref{deltam_marginal} and \eqref{internal_A_trans}, and introducing a proportionality constant $C_6$ into \eqref{deltam_marginal}, the bottom boundary layer thickness is 
\begin{equation}
\eqlbl{deltam_trans_scale}
\delta_m' = C_6 \mu_i'^{\frac{p}{3p-m}} \left(\frac{4D}{Hh_i'} \right)^{-\frac{m}{3p-m}} \left(\frac{Ra_c}{Ra_0(1-T_i')} \right)^{\frac{p+m}{3p-m}}  .
\end{equation}
The constant $C_6$ is determined by the fact that \eqref{deltam_trans_scale} should converge to \eqref{delta_scale} when viscosity and healing are temperature-independent and $T_i' = 0.5$; this gives $C_6 \approx 0.07$.  

We close the problem by solving for the average internal mantle temperature, $T_i'$, using a simple energy balance.  The heat flux out of the mantle through the top boundary layer must match the heat flux into the mantle through the bottom boundary layer, and thus $T_i'/\delta_l' = (1-T_i')/\delta_m'$.  Solving for $T_i'$, 
\begin{equation}
T_i' = \frac{\delta_l'}{\delta_l' + \delta_m'} .
\end{equation}

Scaling laws for $Nu$, $v_l'$, $v_m'$, and $A_i'$, are easily obtained from \eqref{scale_lith_trans} and \eqref{deltam_trans_scale}.  Combining \eqref{scale_lith_trans} and \eqref{Nu_nondim}, the Nusselt number is
\begin{equation}
\eqlbl{Nu_trans_scale}
Nu = \frac{T_i'}{C_5} L'^{-\frac{1}{10}} \mu_l'^{-\frac{1}{4}} \left(\frac{D}{Hh_l'} \right)^{\frac{1}{3}} \left(Ra_0 T_i' \right)^{\frac{2}{3}} . 
\end{equation}
Using $v_l' = L' \delta_l^{-2}$, the plate velocity is 
\begin{equation}
\eqlbl{vl_trans_scale}
v_l' = C_5^{-2}  L'^{\frac{4}{5}} \mu_l'^{-\frac{1}{2}} \left(\frac{D}{Hh_l'} \right)^{\frac{2}{3}} \left(Ra_0 T_i'\right)^{\frac{4}{3}} ,
\end{equation} 
and using $v_m' = \delta_m'^{-2}$, the basal mantle velocity is
\begin{equation}
\eqlbl{vm_trans_scale}
v_m' = C_6^{-2} \mu_i'^{-\frac{2p}{3p-m}} \left(\frac{4D}{Hh_i'} \right)^{\frac{2m}{3p-m}} \left(\frac{Ra_0(1-T_i')}{Ra_c} \right)^{\frac{2(p+m)}{3p-m}}  .
\end{equation} 
Finally, the average internal fineness is obtained from \eqref{internal_A_trans} and \eqref{deltam_trans_scale}, 
\begin{equation}
\eqlbl{Ai_trans_scale}
A_i' = C_6^{-\frac{4}{p+m}} \mu_i'^{-\frac{1}{3p-m}} \left(\frac{4D}{Hh_i'} \right)^{\frac{3}{3p-m}}
\left(\frac{Ra_0(1-T_i')}{Ra_c} \right)^{\frac{4}{3p-m}} . 
\end{equation}

The scaling laws for $v_m'$ and $A_i'$ have the same form as the fully stagnant lid and fully mobile laws, because they are controlled by the viscosity of the interior mantle.  They only differ from the fully stagnant lid and fully mobile scaling laws by the internal temperature, which determines both the buoyancy of the bottom boundary layer and the viscosity and healing rates in the mantle.  The scaling laws for $Nu$ and $v_l'$ have entirely different forms than the stagnant lid and fully mobile laws, as the driving forces for damage in lithospheric shear zones are dominated by in-plane normal stresses, which are different from those in the interior of convection cells where shear stresses dominate.  The scaling laws for these lithospheric quantities are stronger functions of $D/H$ and $Ra_0$ than either end member limit, and will thus make the transition from the fully stagnant lid limit to the fully mobile limit as either $D/H$ or $Ra_0$ increase.  

\subsection{Comparison to Numerical Experiments}
\label{sec:comp_full_dam}

\begin{figure}
\includegraphics[scale = 0.75]{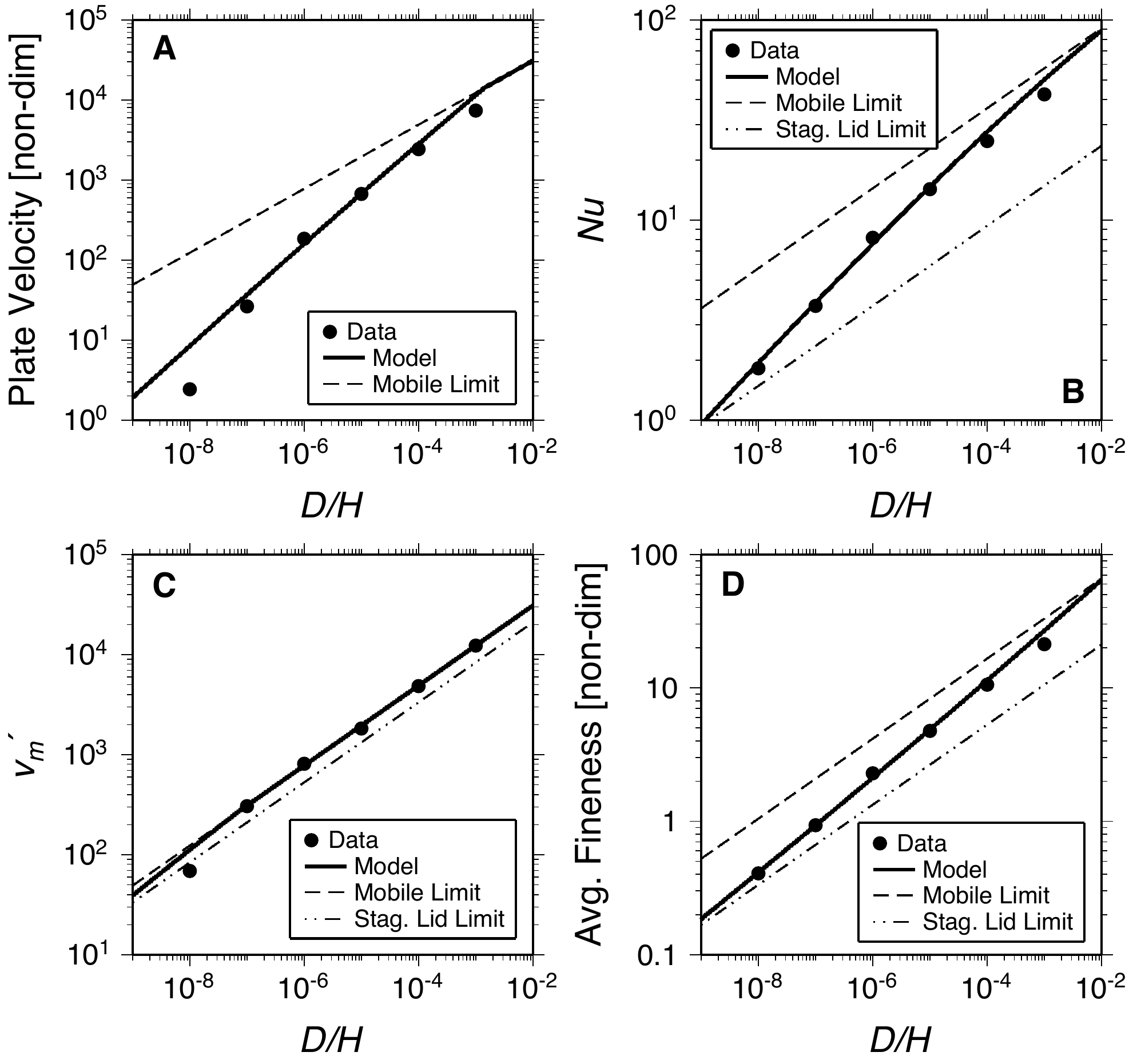}
\caption{\label {fig:dscale_trans} Plots of plate velocity (a), Nusselt number (b), basal mantle velocity (c), and volume averaged fineness (d) versus $D/H$ for convection with temperature-dependent viscosity and healing (\S \ref{sec:full_dam}).  Numerical results are plotted as symbols with scaling laws plotted as lines.  Panel (a) shows numerical data for $v_l'$ compared to the scaling law for $v_l'$ (\eqref{vl_trans_scale}); the mobile lid limit for plate velocity is also shown (equation \eqref{scale_v_mob2}; no scaling law for $v_l'$ in the stagnant lid regime exists).  Panel (b) compares the numerical results for $Nu$ to the scaling law for $Nu$ \eqref{Nu_trans_scale}, with the stagnant lid \eqref{scale_Nu_stag2} and fully mobile limits also plotted \eqref{scale_Nu_mob2}.  Panel (c) compares the data for $v_m'$ to the scaling law for $v_m'$ \eqref{vm_trans_scale}, and also shows the stagnant lid \eqref{scale_v_stag2} and fully mobile limits \eqref{scale_v_mob2}.  Panel (d) compares the data for $\bar{A}'$ to the scaling law for $A_i'$ \eqref{Ai_trans_scale} and also shows the stagnant lid \eqref{scale_a_stag2} and fully mobile limits \eqref{scale_a_mob2}. All models use $Ra_0 = 10^6$, $\mu_l' = 10^5$, $h_l' = 10^{-5}$, $m=2$, $p=4$, and $L'=1.5$.  }  
\end{figure}

\begin{figure}
\includegraphics[scale = 0.75]{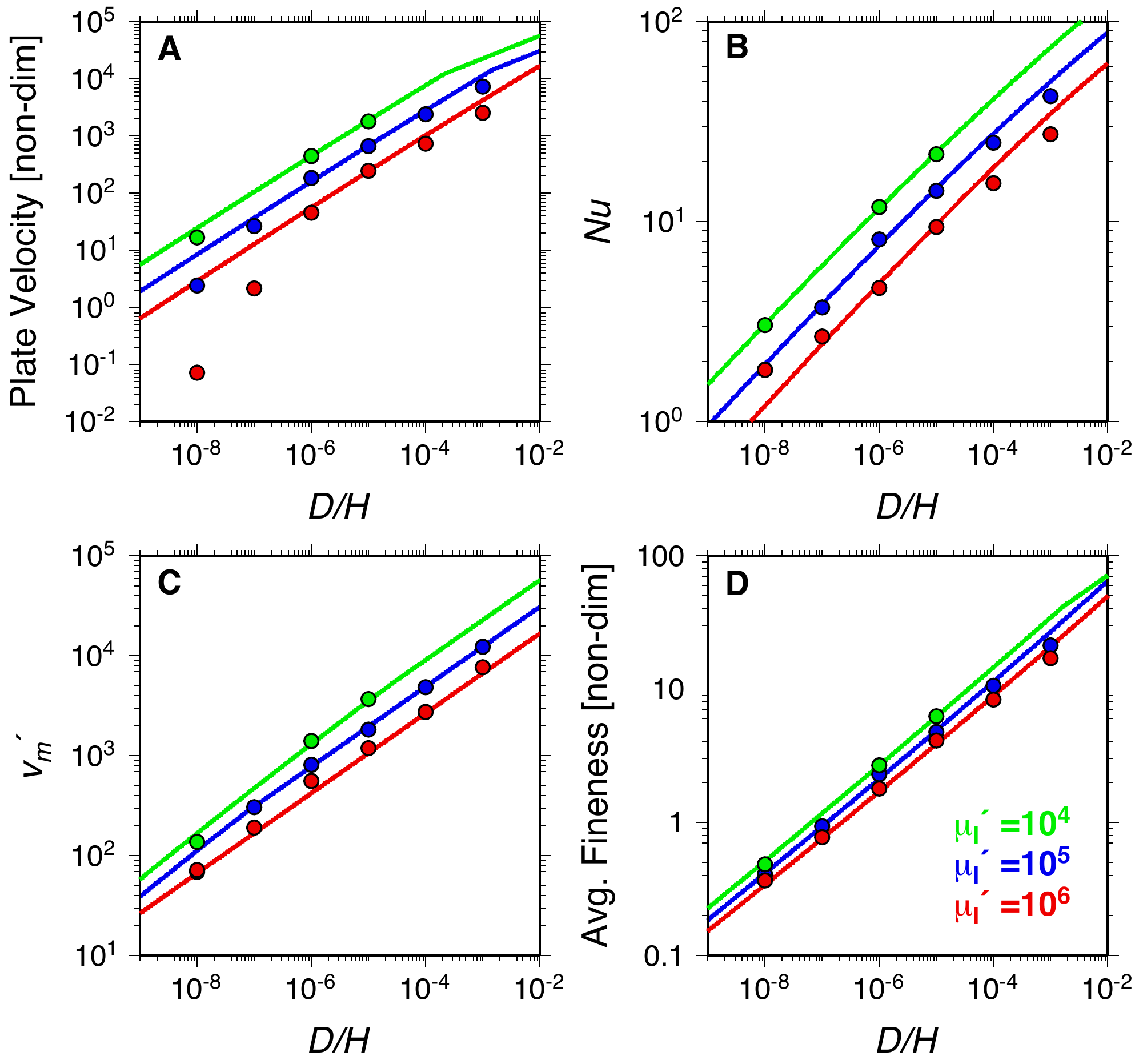}
\caption{\label {fig:dscale_trans_mul} Comparison of numerical data (circles) to scaling laws (solid lines), as in Figure \ref{fig:dscale_trans}, at three different viscosity ratios, $\mu_l' = 10^4$ (green), $\mu_l' = 10^5$ (blue), and $\mu_l' = 10^6$ (red). All models use $Ra_0 = 10^{6}$, $m=2$, $p=4$, $h_l' = 10^{-5}$, and $L'=1.5$.}  
\end{figure}

\begin{figure}
\includegraphics[scale = 0.75]{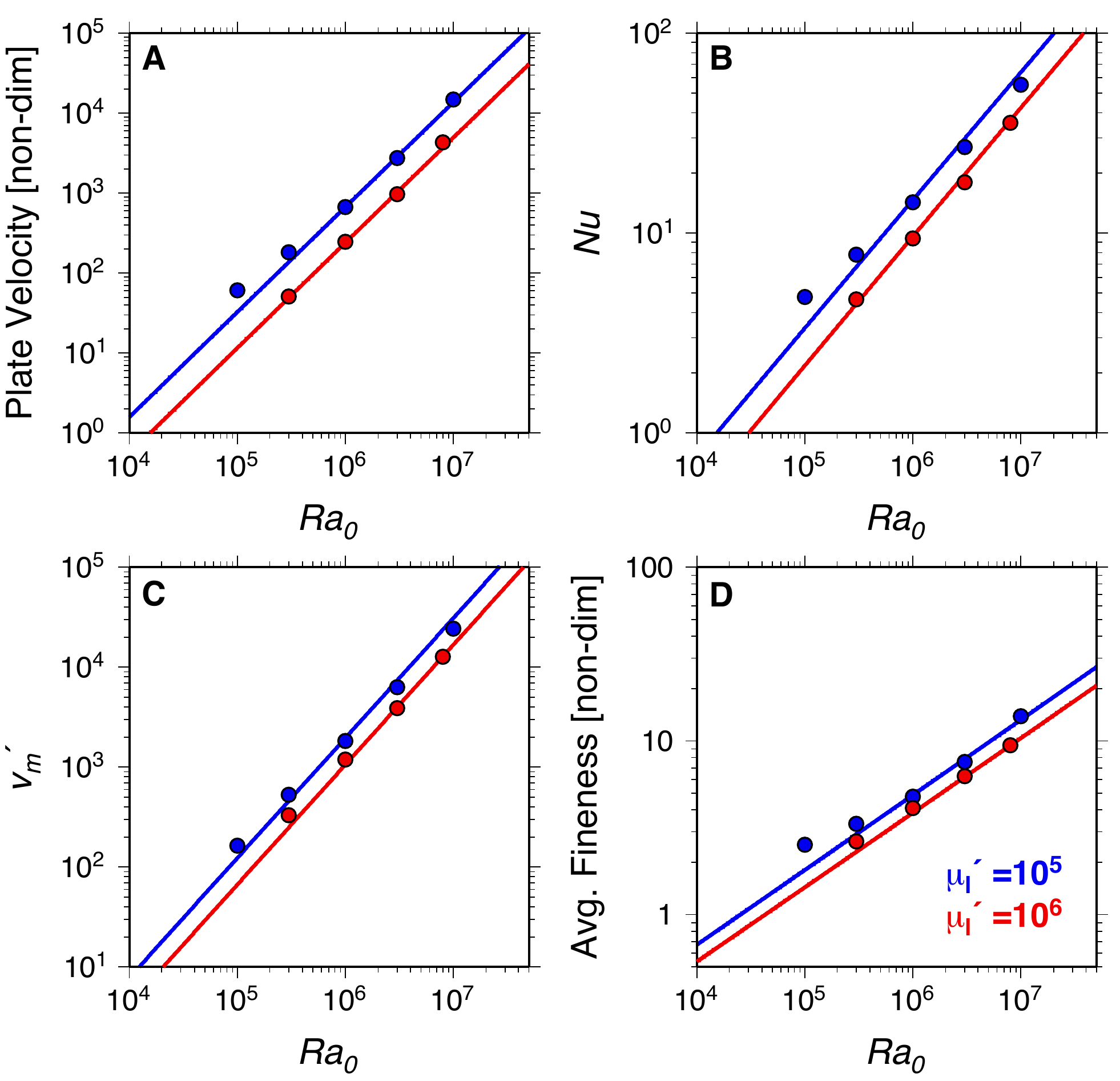}
\caption{\label {fig:Rascale_trans_mul} Comparison of numerical data with varying $Ra_0$ (circles) to scaling laws (solid lines), as in Figure \ref{fig:dscale_trans_mul}, at two different viscosity ratios, $\mu_l' = 10^5$ (blue), and $\mu_l' = 10^6$ (red). Models shown here use $D/H = 10^{-5}$, $m=2$, $p=4$, $h_l' = 10^{-5}$, and $L'=1.5$.}  
\end{figure}

\begin{figure}
\includegraphics[scale = 0.75]{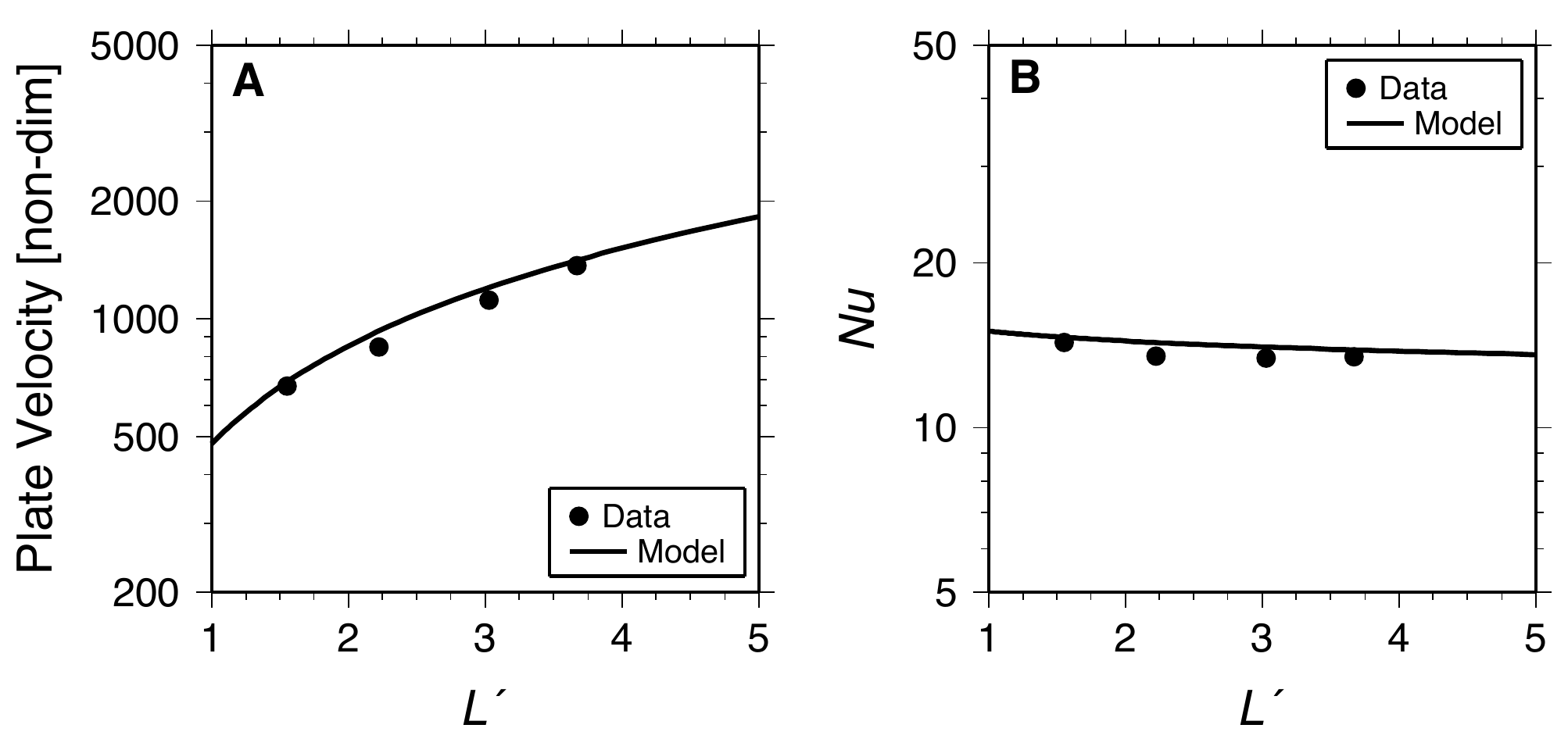}
\caption{\label {fig:L_trans_mul} Comparison of numerical data with varying $L'$ (circles) to scaling laws (solid lines), for (a) plate velocity (scaling law is given by \eqref{vl_trans_scale}) and (b) Nusselt number (scaling law is given by \eqref{Nu_trans_scale}). Models shown here use $D/H = 10^{-5}$, $m=2$, $p=4$, $h_l' = 10^{-5}$, $Ra_0=10^6$, and $\mu_l' = 10^5$.} 
\end{figure}

The theoretical scaling laws for convection with temperature-dependent viscosity and healing compare well to the numerical data for $v_l'$, $Nu$, $v_m'$, and $A'_i$ (see Figures \ref{fig:dscale_trans}, \ref{fig:dscale_trans_mul}, \ref{fig:Rascale_trans_mul}, and \ref{fig:L_trans_mul}).  In addition, the fully stagnant lid and fully mobile limits (shown in Figure \ref{fig:dscale_trans}) illustrate how convection results evolve through the transitional regime and converge to the fully mobile limit as $D/H$ increases. To plot the fully stagnant lid scaling laws, we calculate $a_{rh}$ from stagnant lid numerical results, in the same manner as outlined in \S \ref{sec:comp_stag_H0}, and find $a_{rh} \approx 1.82$. All results shown here use $m=2$, $p=4$, $E_h' = 23.03$ which results in $h_l' = 10^{-5}$, and, save for those shown in Figure \ref{fig:L_trans_mul}, a $4 \times 1$ aspect ratio domain resulting in $L'\approx1.5$.      

There is some deviation of the numerical results from the theory at low $D/H$ and low $Ra_0$ when convection is sluggish or approaching the fully stagnant lid regime.  In particular, the plate velocity deviates from the scaling laws near the transition to stagnant lid convection (e.g. the data for low $D/H$ with $\mu_l' = 10^6$).  This is not unexpected given that there is no scaling law for the plate velocity in the stagnant lid regime.  There may be an intermediate mode as convection switches from instability of the whole top boundary layer to instability of a sub-layer beneath a rigid lid that our scaling laws do not capture.  Nevertheless, our scaling laws capture the asymptotic behavior, and are able to match the experiments over a wide parameter range.  

There is, also, some minor discrepancy between where the transitional regime scaling laws intersect the fully mobile limit.  The laws for $Nu$ and $A_i'$ converge to the fully mobile limit at approximately the same value of $D/H$, while the transitional regime scaling law for $v_l'$ converges to the fully mobile limit at a slightly lower value. The disagreement in the end of the transitional regime between $v_l'$ and $Nu$ or $A_i'$ is likely due to the small errors implicit in boundary layer theories because of their simplifying assumptions; in this case the constant plate-length premise may break down near the boundary between fully mobile and transitional regime convection. The clearest definition of the regime boundaries is based on the scaling law for $\delta_l'$, as discussed further in \S \ref{sec:boundaries}.  The scaling law for $v_m'$ converges to the fully mobile limit at a much lower $D/H$ than the other scaling laws; this occurs because there is little difference in deeper mantle circulation between the transitional and fully mobile regimes, and thus $v_m'$ reaches its maximum value with increasing $D/H$ well before the fully mobile limit.  The only factor in the $v_m'$ scaling laws that varies between the different regimes is the internal temperature, which evolves as convection progresses through the transitional regime.  The influence of $T_i'$ is small because it has competing effects on the mobility of the bottom boundary layer: lower $T_i'$ increases the buoyancy of the boundary layer and decreases the internal mantle healing rate, $h_i'$, thus increasing $v_m'$; however, it also increases the internal mantle viscosity, $\mu_i'$, which acts to decrease $v_m'$.    

Our scaling laws for $Nu$ and $v_l'$ also provide a good fit to the numerical results where the plate length, $L'$, is varied via the use of larger numerical model domains (Figure \ref{fig:L_trans_mul}); this indicates that our scaling laws are able to successfully incorporate the influence of longer wavelength flow.  Furthermore, the fact that the numerical results from the $4 \times 1$ aspect ratio domain are well fit by our scaling laws with a constant $L' = 1.5$ throughout most of the transitional regime (Figures \ref{fig:dscale_trans}, \ref{fig:dscale_trans_mul}, and \ref{fig:Rascale_trans_mul}), justifies our approximation of $L'$ as independent of damage to healing ratio, viscosity ratio, and Rayleigh number.  The velocity at the base of mantle, $v_m'$, is assumed to be nearly independent of plate length in the scaling laws (the only influence of $L'$ is indirect; changing $L'$ changes the internal temperature, which in turn affects $v_m'$).  However, in practice $v_m'$ does increase as $L'$ increases ($v_m'$ increases by a factor of $\approx 1.5$ as $L'$ increases by a factor of $\approx 2$), because our calculation of $v_m'$ becomes biased by a rapid, localized divergent flow that occurs where downwellings impinge upon the base of the mantle.  This flow has significantly higher velocities than the typical flow along the bottom boundary layer.  With a larger plate length downwellings are stronger, due to a thicker lithosphere and more rapid plate speed, and our method for calculating $v_m'$ from the numerical models (see \S\ref{sec:comp_stag_H0}) becomes dominated by this impingement induced flow. 

\section{Regime Boundaries}
\label{sec:boundaries}

As described in the previous section, convection with grain-damage shows three regimes of behavior: the fully stagnant lid regime, the transitional regime, and the fully mobile regime.  Here we demonstrate how to define the boundaries between these regimes, and their physical meaning.  The different regimes result from different dynamics governing the size of the top thermal boundary layer, thus we define the regime boundaries based on the scaling laws for $\delta_l'$.  Specifically, the three regimes result from the instability of the top thermal boundary layer involving different viscosity scales; i.e., the effective interior mantle viscosity governs convection in both the fully mobile and fully stagnant lid regimes, while the effective viscosity of lithospheric shear zones governs convection in the transitional regime.  

Convection will switch from the fully stagnant lid regime to the transitional regime when transitional regime convection can transport heat more efficiently (i.e. have a thinner top thermal boundary layer) than fully stagnant lid convection.  Therefore the boundary between these two regimes is defined by the intersection of the $Nu$ scaling law for the fully stagnant lid regime \eqref{scale_Nu_stag2} with the $Nu$ scaling for the transitional regime \eqref{Nu_trans_scale}.  As the regime boundary is at the margin of the stagnant lid limit, the internal temperature, $T_i$, is given by the rheological temperature scale for stagnant lid convection: $T_i = 1-a_{rh}/(2\theta)$, where $a_{rh} \approx 1.82$.  

The onset of transitional regime convection can be physically interpreted as a competition between whether whole lithosphere instability or instability of a sub-layer can produce a thinner lithosphere and more efficient heat transport.  In the fully stagnant lid regime, instability of a sub-layer off the base of the rigid lid transports heat more efficiently than foundering of the whole lithosphere, because the lithosphere would have to grow to an enormous thickness in order to have sufficient negative buoyancy to overcome its own internal viscous resistance.  In the transitional regime, whole lithosphere instability dominates over sub-layer instability, because damage is able to weaken the lithosphere to such a degree that it can go unstable at a thickness less than what would occur with sub-layer instability. 

\begin{figure}
\includegraphics[scale = 0.8]{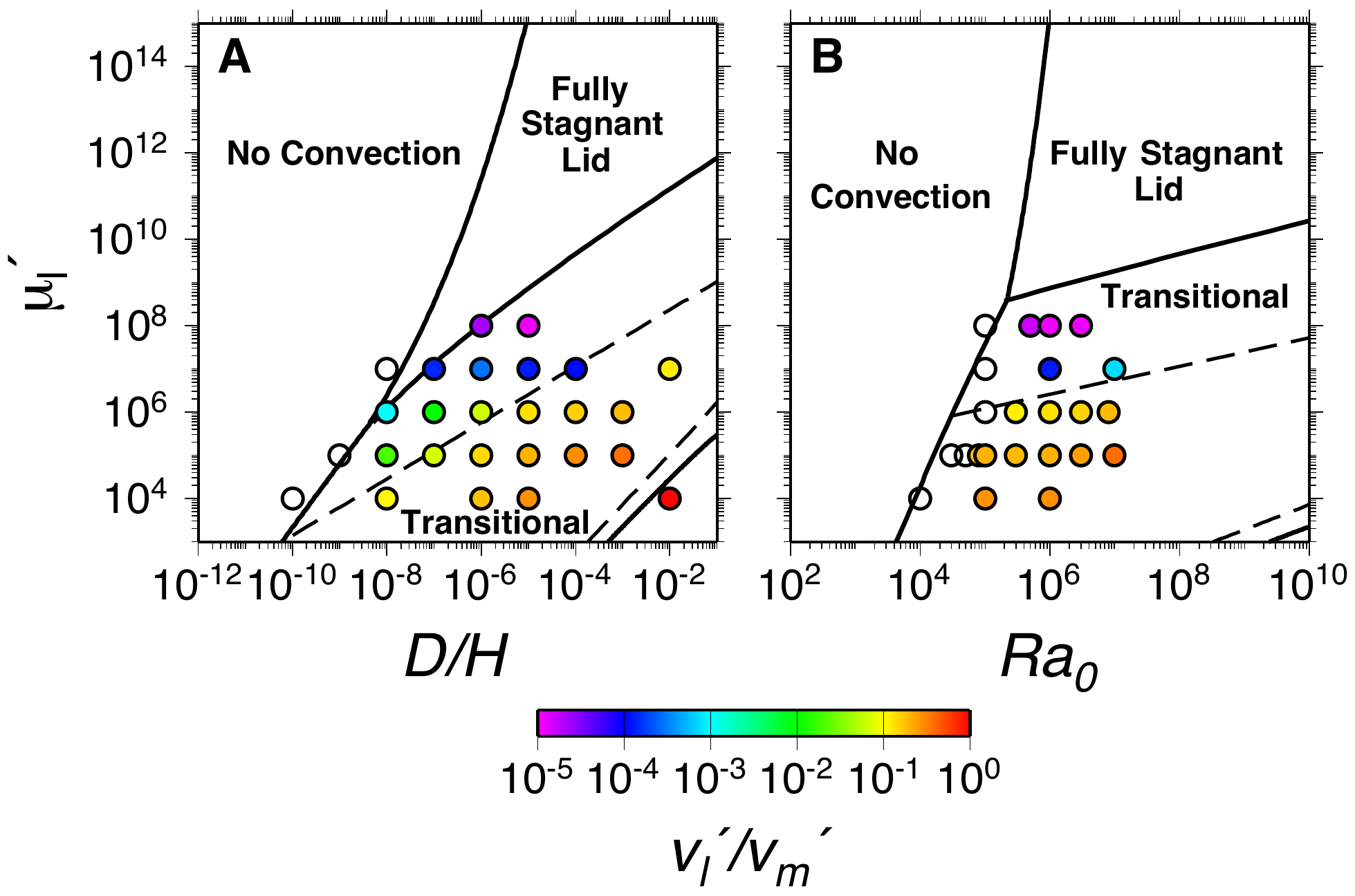}
\caption{\label {fig:regime} Regime diagrams showing the three regimes of convection and the onset of convection, with boundaries denoted by solid lines, for varying $D/H$ (a) and $Ra_0$ (b) (the fully mobile regime is in the bottom right corner, unlabeled).  Regime boundaries are derived from theory (as explained in \S \ref{sec:boundaries}) and data is shown as circles, colored by the value of $v_l'/v_m'$ (results with no convection are white circles). The dashed lines show the boundaries for the nominal plate-tectonic regime as described in the main text; the boundary nearer to the fully stagnant lid regime is determined from \eqref{boundary1}, and the boundary nearer to the fully mobile regime is determined from \eqref{boundary2}.  Theoretical curves and numerical results assume $m=2$, $p=4$, and $E_h' =23.03$; (a) fixes $Ra_0 = 10^6$ while (b) fixes $D/H = 10^{-5}$. All numerical results presented in this figure used a $4\times1$ aspect ratio domain and therefore the theoretical curves assume $L'=1.5$.}  
\end{figure}
  
Convection will enter the fully mobile regime when the heat transport from transitional regime convection matches the heat transport from fully mobile convection; thus the regime boundary is defined by equating $Nu$ in the transitional regime \eqref{Nu_trans_scale} to $Nu$ in the fully mobile regime \eqref{scale_Nu_mob2}.  This regime boundary corresponds to the point where damage is so effective in the lithosphere that the viscous resistance of lithospheric shear zones is no longer significant compared to the viscous resistance of the underlying mantle.  Because instability of the top thermal boundary layer must be controlled by the largest viscosity resisting flow, convection will switch to being governed by the effective interior mantle viscosity.  As a result, an alternative, and perhaps more direct, definition of the boundary between the transitional and fully mobile regimes would be when the lithospheric shear zone viscosity equals the effective viscosity of the mantle interior \citep[e.g.][]{Foley2012}.  However, unlike \cite{Foley2012}, we do not have a scaling law for the shear zone viscosity, and developing one is beyond the scope of this study (\cite{Foley2012} was able to develop a scaling law for the shear zone viscosity by assuming the mantle was effectively Newtonian). Our simple approach using the scaling laws for $\delta_l'$ appears to be a good approximation based on the numerical data. 

Two regime diagrams are shown in Figure \ref{fig:regime}; one in $\mu_l' -D/H$ space and one in $\mu_l'-Ra_0$ space.  The boundary between the fully stagnant lid regime and the transitional regime, and the boundary between the transitional regime and the fully mobile regime, roughly parallel each other.  The gap between the two boundaries is large, showing the importance of the transitional regime in the parameter space; this regime dominates because the fully stagnant lid and fully mobile regimes are extreme, end-member scenarios. Intuitively, increasing $D/H$ or $Ra_0$ pushes either boundary to higher $\mu_l'$, as either more damage or more vigorous convection makes it easier to mobilize the lithosphere.  However, $D/H$ has an apparently larger influence on the regime boundaries, as the slopes are steeper with increasing $D/H$ than with increasing $Ra_0$. The numerical data show that the surface becomes increasingly mobilized as one progresses through the transitional regime towards the fully mobile regime, going from low $v_l'/v_m'$ at the boundary with the fully stagnant lid regime to $v_l'/v_m' \approx 1$ at the fully mobile regime.  While our regime boundaries are based on the thermal boundary layer thickness instead of the plate velocity, the two are linked (e.g. \S \ref{sec:full_scale}); as $\delta_l'$ shrinks across the transitional regime, $v_l'$ increases in tandem.  Furthermore because $\delta_l'$ decreases more rapidly than $\delta_m'$ in the transitional regime, the plate velocity relative to the mantle velocity, $v_l'/v_m'$, also increases. In addition, the stability curve for the onset of convection (derived below in \S \ref{sec:onset_convec}) is also shown.  At moderate to low viscosity ratios, the transitional regime boundary intersects the stability curve, and thus convection will skip the fully stagnant lid regime and begin in the transitional regime straight away.  

We approximate a nominal plate-tectonic regime using two simple definitions based on our scaling laws. Our numerical results indicate that plate-tectonic style convection, i.e. convection with relatively rigid, mobile plates and slab-like downwellings, occurs within the transitional regime, with varying degrees of surface mobility. Thus there are two boundaries for plate-tectonic style convection within the transitional regime: the boundary nearer to the stagnant lid regime where surface mobility becomes so sluggish that downwellings become drip-like, and the boundary nearer to the fully mobile regime where the lithosphere is so thoroughly weakened by damage that convection resembles constant viscosity convection.  Based on inspection of the numerical results, we define the first boundary, where convection goes from sluggish subduction to more fully developed, slab-like downwellings, as the point where $v_l'/v_m' = 0.1$.  For $m=2$ and $p=4$, this boundary is determined by solving the following equation that combines equations \eqref{vl_trans_scale} and \eqref{vm_trans_scale},   
\begin{equation}
\eqlbl{boundary1}
C_5^{-2} L'^{\frac{4}{5}} \mu_l'^{-\frac{1}{2}} \left(\frac{D}{Hh_l'} \right)^{\frac{2}{3}} \left(Ra_0 T_i'\right)^{\frac{4}{3}} = 0.1 C_6^{-2} \mu_i'^{-\frac{4}{5}} \left(\frac{4D}{Hh_i'} \right)^{\frac{2}{5}} \left(\frac{Ra_0(1-T_i')}{Ra_c} \right)^{\frac{6}{5}} .
\end{equation}
We define the second boundary, where convection begins to resemble isoviscous convection, when $\delta_l'$ is only 10 \% larger than it would be in the fully mobile regime, and, for $m=2$ and $p=4$, this boundary is found by solving 
\begin{equation}
\eqlbl{boundary2}
1.1 C_5 L'^{\frac{1}{10}} \mu_l'^{\frac{1}{4}} \left(\frac{D}{Hh_l'} \right)^{-\frac{1}{3}} ({Ra_0 T_i'}) ^{-\frac{2}{3}} = C_3 (2C_4)^{-\frac{2}{5}} \mu_i'^{\frac{2}{5}} \left(\frac{D}{Hh_i'} \right)^{-\frac{1}{5}}  Ra_0^{-\frac{3}{5}} ,
\end{equation} 
which is derived by combining equations \eqref{Nu_trans_scale} and \eqref{scale_Nu_mob2}.  Plate-tectonic convection takes up most of the transitional regime, with a relatively large region of non-plate-tectonic convection characterized by sluggish, drip-like subduction within the transitional regime near the fully stagnant lid regime, and a narrow region of overly fragmented non-plate-tectonic style convection adjacent to the boundary with the fully mobile regime.  This indicates that planets can exist in a non-plate-tectonic state while still having some form of sluggish surface mobility and lithospheric dripping, while a planet would have to practically be in the fully mobile regime before excessive damage in the lithosphere eliminates plate-tectonic style convection.       

\section{Onset of Convection}
\label{sec:onset_convec}

As shown in \S\ref{sec:stag_H0} and \S\ref{sec:full_dam}, convection can be completely stopped by healing when either damage is weak or Rayleigh number is low.  Here we provide a simple approximation for the boundary between the convective and non-convective states.  Although a scaling law for how healing suppresses convection is useful for interpreting our numerical results and the general behavior of convection with grain-damage, it is not necessarily applicable to convection in planetary mantles; this is because at the large grainsizes associated with high healing, the rheology will be dominated by the dislocation creep mechanism and the viscosity will no longer be grainsize sensitive.  We describe the transition to convection dominated by dislocation creep in Appendix \ref{sec:disl_creep}.

The boundary between the convective and non-convective states can be approximated by setting the Nusselt number equal to 1, and thus can be determined from either equations \eqref{Nu_trans_scale} or \eqref{scale_Nu_stag2}, depending on the regime.  Figure \ref{fig:regime} shows the stability curve in addition to the boundaries of the regimes.  At low to moderate viscosity ratios the stability curve is defined by the transitional regime scaling law (e.g. setting \eqref{Nu_trans_scale} equal to 1); when the stability curve intersects the boundary with the fully stagnant lid regime, it is then defined by the fully stagnant lid scaling law (e.g. setting \eqref{scale_Nu_stag2} equal to 1).  The theory fits the numerical data well for the regime diagram in $\mu_l'-D/H$ space, and appears to have the right slope but is off by a factor of $\approx 10$, for the regime diagram in $\mu_l'-Ra_0$ space.  Our scaling laws tend to deviate from the numerical results at low $Ra_0$, possibly due to deviation from the assumption of thin boundary layers implicit in boundary layer theory, so the stability curve based on our scaling laws naturally has the same error.   

Defining the boundary between convective and non-convective states using scaling laws derived from boundary layer theory may seem counterintuitive, as boundary layer theories are typically not applicable to the onset of convection.  However, our approach is necessary for two reasons: 1) the steady-state grainsize is undefined in the static state, and thus there is no general background state to perturb for a linear stability analysis; 2) we want to know the long term, steady-state behavior, and not just whether convection would begin under a set of initial conditions.  Our approach accomplishes this by essentially asking whether the interior mantle fineness that would result from finite amplitude convection is sufficient to allow convection to continue.  In addition, there would likely be some hysteresis with a linear stability theory, that is not present using finite amplitude scaling laws.  For example, with parameters that will lead to a non-convective state (e.g. low $D/H$), an experiment started from a static, conductive temperature profile can initially go unstable while grains are small, only to see convection shut off later as grain growth dominates.  

Finally, we note that our approach for defining the stability curve differs from that of \cite{Stengel1982} and \cite{slava1995}, where the effective Rayleigh number of a sub-layer is maximized and compared to the critical Rayleigh number.  We found this theory problematic when grain-damage is added because the sub-layer optimization can give unphysical results with even moderate values of $E_h$, and because it does not accurately capture the thermal structure, and therefore healing rates, of the interior of convection cells.  Despite these differences, our approach of setting $Nu = 1$ to define the stability curve closely approximates the results of \cite{slava1995} for the case where viscosity is only sensitive to temperature (the case \cite{slava1995} considers).         

\section{Discussion}
\label{sec:discussion}

\subsection{Plate Speed and Heat Flow on Earth and Venus}
\label{sec:application}

To demonstrate an application of our scaling laws to mantle convection on terrestrial planets, we investigate how plate speed and heat flow are influenced by surface temperature.  One possible explanation for the lack of plate tectonics on Venus is that the extremely high surface temperature leads to weak lithospheric buoyancy stresses \citep{Lenardic2008}, or rapid lithospheric healing \citep{Landuyt2009a,Foley2012}.  We can provide another test of this hypothesis by looking directly at how plate speed is coupled to surface temperatures using our scaling laws.  

We assume that the Earth lies within the transitional regime, because this is where plate-tectonic style convection occurs (see \S \ref{sec:boundaries}), and we fix the damage partitioning coefficient, $f$, to match Earth's current day plate speed and heat flow; this results in a value of $f$ that compares well with estimates based on experimental results and field observations, as discussed below (see text below \eqref{healing_constant}).  As Earth is in the transitional regime, we do not need to specify whether the mantle is dominated by dislocation or diffusion creep; as explained in Appendix \ref{sec:disl_creep}, the scaling laws for plate speed and heat flow are not affected by the dominant creep mechanism in the mantle because they are controlled by the dynamics of the lithosphere. We further assume that planets are dominated by internal heating, including both secular cooling and radiogenic heat production, and thus $T_m \approx T_i$.  This means that $\mu_i' = 1$, and $Ra_0$ and $D/H$ are defined at the given interior mantle temperature for a planet. Applying scaling laws developed for bottom heating to internally heated convection can lead to small errors due to the lack of plumes impinging on the lithosphere in the internally heated case \citep{Moore2008}.  However, this error is small (e.g. the $Nu \sim Ra$ scaling law exponent only changes by $\sim 10$ \%) \citep{Moore2008}, and would not change the overall results presented in this section.  We therefore assume that our scaling laws developed from bottom heated simulations are able to capture the first order physics of convection with grain-damage in planetary mantles. The dimensional scaling law for plate velocity in the transitional regime \eqref{vl_trans_scale}, is then 
\begin{equation}
v_l = \left(\frac{\kappa}{d C_5^2} \right) L'^{\frac{4}{5}} \left(\frac{\mu_l}{\mu_m} \right)^{-\frac{1}{2}} \left(\frac{D h_m}{H h_l} \right)^{\frac{2}{3}} Ra_0^{\frac{4}{3}} ,
\end{equation} 
where $\mu_l$ and $h_l$ are the viscosity and healing defined at the lithosphere temperature, $T_l$, and we set the plate length to $L'=1.5$.  A plate length of $L'=1.5$ is the average value from our numerical models performed in a $4 \times 1$ domain, the geometry used for the bulk of this study; we choose this value because the scaling laws are the most well constrained for this case.  As described earlier (see text below \eqref{scale_lith_trans}), different geometries, as well as rheological effects not included in our model, such as depth or pressure-dependent viscosity, can change the plate length \citep[e.g.][]{bung1996b,tack1996b,Lenardic2006,Zhong2007,Hoink2008}.  A different plate length would change the exact values of plate speed and heat flow calculated here, but not the overall trend with increasing surface temperature; i.e. the basic physics presented in this section still hold. From \eqref{Nu_trans_scale}, the scaling law for heat flow is 
\begin{equation}
q = \left(\frac{k (T_m - T_s)}{d C_5} \right) L'^{-\frac{1}{10}} \left(\frac{\mu_l}{\mu_m} \right)^{-\frac{1}{4}} \left(\frac{D h_m}{H h_l} \right)^{\frac{1}{3}} Ra_0^{\frac{2}{3}} . 
\end{equation}

In scaling the heat flux and plate speed with $T_s$, we fix all parameters aside from $T_m$, and the viscosities and healing rates in both the lithosphere and mantle, as these are most strongly affected by temperature.  There is no clear relationship for how mantle temperature scales with surface temperature, as this requires a full thermal evolution model and thus depends on the age of the planet and amount of radiogenic heating, among other factors.  We therefore use a simple model wherein we assume that the difference in the mantle temperature between Earth and Venus will be approximately 1/2 the difference in surface temperature, as suggested by \cite{Lenardic2008}.  This gives 
\begin{equation}
T_m = T_{m,0} + \frac{T_s - T_{s,0}}{2}
\end{equation} 
where $T_{m,0} = 1650$ K is the Earth's mantle potential temperature, and $T_{s,0} = 273$ K is Earth's surface temperature. This can be interpreted in terms of how the thermal evolution of Venus would differ from that of Earth due to a hotter surface temperature.  The hot climate on Venus lowers its heat loss relative to Earth, and thus Venus will have cooled less and will have a higher mantle temperature at the present day.  This effect should apply regardless of whether the mantle is dominated by secular-cooling or dominated by radiogenic heating. Furthermore, we find that our results are not strongly sensitive to the assumed mantle temperature scaling, and thus our results are robust to uncertainties in the thermal evolution of Earth and Venus. To define the lithosphere temperature, we simply set $T_l = (T_m + T_s)/2$, as this provides a good approximation to the definition given in \cite{Foley2012} where $T_l$ is determined by the transition from deformation by frictional sliding to semi-brittle/semi-viscous flow.  In general, our definition of $T_l$ is meant to represent the mid-lithosphere, the region of peak strength.  

To calculate the healing rate, we assume $E_h = 500$ kJ/mol, consistent with grain-growth experiments that take into account secondary phases and pinning \citep[e.g.][]{Evans2001,Faul2006,Hiraga2010}, and determine the constant $h_n$ (see \eqref{eqheal}) based on grain-growth experiments.  As shown by \cite{br2013}, the growth rate for the interface between phases in the pinned state (for a general $p$), can be related to experimentally determined grain-growth rates (with $p=3$), as 
\begin{equation}
\eqlbl{healing_constant}
h_n = \frac{h_{n,0}(p-1)}{500} \tilde{r}^{p-3} 
\end{equation}  
where $h_{n,0}$ is the experimentally determined grain-growth constant, and $\tilde{r}$ is the grainsize where the pinned state is reached; \cite{br2012} find $\tilde{r} \approx 1$ $\mu$m.  Fitting the data from \cite{Hiraga2010} to the grain-damage healing law, \cite{br2012} find a growth rate of $\approx 3 \times 10^{-15}$ m$^2$ s$^{-1}$ at a temperature of $\approx 1630$ K, and thus $h_{n,0} \approx 30$ m$^2$ s$^{-1}$ with our choice of $E_h = 500$ kJ/mol.  Using \eqref{healing_constant}, we find $h_n \approx 2 \times 10^{-7}$ m$^3$ s$^{-1}$.  We use $f \approx10^{-5}$ to match Earth's present day plate speed and heat flow.  This value of $f$ is in line with estimates based on experimental results and geological observations \citep{austin2007,Rozel2010}.  

 \begin{figure}
\includegraphics[scale = 0.5]{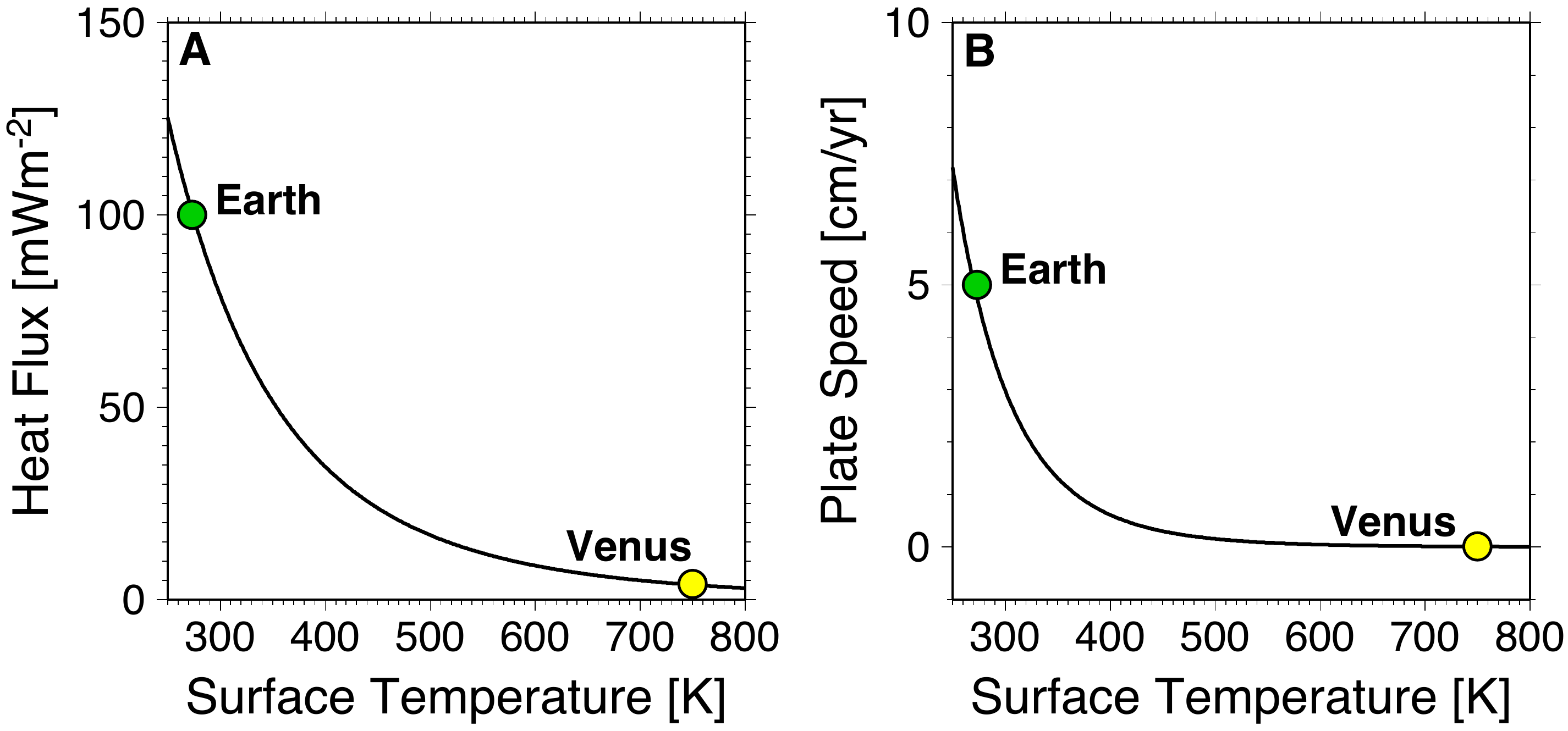}
\caption{\label {fig:app} Heat flux (a) and plate velocity (b) as a function of surface temperature, with Earth (green) and Venus (yellow) represented.  }  
\end{figure}

Figure \ref{fig:app} shows a large increase in plate speed from Venusian conditions to Earth-like conditions; this is due primarily to cooler temperatures suppressing lithospheric grain-growth, and enhancing the efficacy of damage.  A similar trend is observed for the heat flux.  At a Venusian surface temperature of 750 K we obtain a plate speed of $\approx 0.01$ cm/yr and a heat flux of $\approx 4$ mWm$^{-2}$.  This estimate is of the same order as others made from mantle convection scaling laws \citep{Reese1998}. Plate speed and heat flux also both increase sharply as surface temperature decreases from Earth's present day value.  If this trend continued, Earth would eventually enter the fully mobile regime, where plate speed and heat flow would become significantly less sensitive to surface temperature (i.e. the curves would flatten out with decreasing $T_s$). This highlights an important physical concept about convection in the transitional regime versus the end-member regimes.  The dependence of healing on surface temperature only comes into play in the transitional regime, where lithospheric damage controls convection; in the end-member regimes the only role surface temperature plays is in determining lithospheric buoyancy and mantle temperature (and viscosity), because the surface is either completely broken up, and is effectively insensitive to lithospheric viscosity and damage, or completely rigid and again effectively insensitive to the lithospheric viscosity. 

\subsection{Comparison to Other Models and Implications for Earth}

In addition to confirming that increasing surface temperature can shut down plate tectonics due to the effects of increased healing in the lithosphere, our scaling laws have further important implications for the Earth and other planets.  In particular, with grain-damage, the transition between stagnant lid convection and fully mobile convection is gradual and takes place over a large transitional regime, with plate-tectonics lying within the transitional regime. Most previous studies on plate generation, which utilize the pseudoplastic rheology, find this transition to be bifurcation-like, and thus have a very narrow or nonexistent transitional regime \citep[e.g.][]{moresi1998,tackley2000a,Korenaga2010}. The difference between our results and those of many previous studies points out an important distinction between grain-damage and the pseudoplastic yield stress rheology: in the simplest case, e.g. where viscosity is only a function of temperature and the yielding criterion, the yield stress rheology does not allow mobile lid convection with lithospheric shear zones that are stronger (or provide more viscous resistance to plate motion) than the mantle, and thus all plate-tectonic style convection is fully mobilized using this mechanism.  Using the yield stress rheology, the effective viscosity of lithospheric shear zones is $\mu_{eff} = (1/\mu_T + 1/\mu_y)^{-1}$, where $\mu_T$ is the unyielded temperature-dependent viscosity, $\mu_y = \tau_y/(2\dot{\varepsilon})$ is the yielded viscosity, $\tau_y$ is the yield stress, and $\dot{\varepsilon}$ is the second invariant of the strain-rate tensor. Solving for the strain-rate as a function of $\tau$, the second invariant of the stress tensor, gives
\begin{equation}
\dot{\varepsilon}= \frac{\tau}{2 \mu_T} \left(1- \frac{\tau}{\tau_y} \right)^{-1}
\end{equation}          
which becomes unbounded as $\tau$ approaches $\tau_y$ (when the stress is below the yield stress, the rheology is effectively Newtonian, with viscosity $\mu_T$).  Thus when the stress reaches the yield stress, lithospheric shear zones can be weakened without bound; this means that instability of the top thermal boundary layer, and mobility of plates, will be determined by the mantle viscosity (i.e. the mantle provides the most significant resistance to flow as compared to the lithosphere), and convection will enter the fully mobile regime when $\tau = \tau_y$ \citep[e.g.][]{moresi1998}.  There is no, or at least a very small, transitional regime because convection switches abruptly from a mantle controlled stagnant lid regime to a mantle controlled mobile regime as a function of the yield stress. 

Grain-damage has an effectively non-Newtonian power law rheology with a large $n$ (Appendix \ref{sec:disl_creep}) and does not become unbounded at high stress.  Therefore intermediate states where the damaged viscosity of the lithosphere determines the dynamics of the top thermal boundary layer can exist. We note that similar intermediate states can be found with the pseudoplastic rheology when viscosity layering within the mantle, such as a low viscosity asthenosphere, is introduced \citep{Hoink2010,Crowley2012}. In particular, \cite{Hoink2010} and \cite{Crowley2012} find a large transitional regime, defined as a regime where flow velocities in the asthenosphere exceed the plate velocity, because the low viscosity of the asthenosphere allows rapid channel flow, while the higher viscosity sub-asthenospheric mantle dictates the plate speed. Recognizing the possibility for intermediate states between stagnant lid convection and fully mobile convection, including the non-plate-tectonic sluggish subduction style of convection, has profound implications for our understanding of the tectonic modes of other terrestrial planets and exo-planets.  Venus is often interpreted as exhibiting stagnant lid convection \citep[e.g.][]{Phillips1981,Kaula1990,Reese1998}, possibly with episodic overturns of the lithosphere \citep[e.g.][]{Turcotte1993,moresi1998,Lenardic2008,Landuyt2009a}.  However, Venus has surface features that are strikingly similar to subduction zones on Earth \citep[e.g.][]{Sandwell1992}.  As demonstrated in \S \ref{sec:application}, Venus can be explained by convection in the transitional regime, close to the fully-stagnant lid regime, with a very slow ``plate" speed because the viscosity of shear zones in the Venusian lithosphere is high due to increased healing.  This gives a possible alternative interpretation of Venus as a planet with ``sluggish subduction" that could explain both the trench-like surface features and the lack of rapid, plate-tectonics style lithospheric recycling \citep{br2014}.   

Our work also has important implications for the thermal and tectonic evolution of the Earth.  Various authors have suggested that the early Earth had either more sluggish or intermittent plate tectonics than today due to stiffening of the lithosphere through melting \citep{korenaga2006}, increased crustal buoyancy due to melting \citep{Sleep1982,davies1992b}, a higher interior temperature causing a drop in convective stress \citep{ONeill2007b}, or closing of oceanic basins temporarily halting plate tectonics \citep{Silver2008}.  One motivation for the hypothesis of sluggish or intermittent early plate tectonics is that it may reconcile thermal evolution models based on scaling laws for mantle convection with geochemical and cosmochemical estimates of Earth's heat budget.  Grain-damage may produce a thermal history similar to that of \cite{korenaga2006}, where plate speed and heat flux decrease with increasing mantle temperature in the Archaean, due to the influence of mantle temperature on lithospheric healing.  A full thermal evolution model using the scaling laws presented here is outside the scope of this paper, but is an important future step in understanding the thermal evolution of the Earth.  

\section{Conclusions}
\label{sec:conclusions}

Scaling laws developed from boundary layer theory match numerical experiments of mantle convection with grain damage over a wide region of parameter space.  Two simplified cases, the temperature-independent viscosity case and constant healing case, demonstrate that our approach of scaling for the effective mantle rheology based on the fineness evolution equation in steady-state accurately describes convection in both the mobile lid and stagnant lid regimes.  A third, more realistic case incorporating both temperature-dependent viscosity and temperature-dependent healing shows three regimes with fundamentally different scaling behavior.  In the fully stagnant lid regime, grain-damage in the lithosphere is ineffective, and the heat flow and mantle velocity are determined by the viscous resistance of the interior mantle viscosity to drips off the base of the rigid lid.  In the transitional regime, damage in the lithosphere is effective enough to allow sinking and mobilization of the whole top thermal boundary layer.  The viscosity of lithospheric shear zones provides the primary viscous resistance to foundering and mobility of the lithosphere.  In the fully mobile regime, damage in the lithosphere is so effective that lithospheric shear zones no longer provide a significant resistance to flow; the main source of viscous resistance is again the viscosity of the mantle interior.  The scaling laws in all three regimes differ significantly from the traditional scaling law where $Nu \sim Ra_0^{1/3}$, because increasing $Ra_0$ also enhances damage.  

Applying these scaling laws to planetary mantle convection, we demonstrate that increasing surface temperature slows plate speed and reduces heat flow dramatically, because the higher surface temperature increases the healing rate in the lithosphere and thus increases the viscosity of lithospheric shear zones.  This provides further support to the hypothesis that the lack of plate tectonics on Venus is due to the extremely hot climate there.  In addition, changing mantle temperature would have a similar effect, and could result in a non-conventional thermal evolution model for the Earth where plate speed decreases as mantle temperature increases in the past, as has been proposed previously.             

Contrary to many previous studies on plate generation, especially those employing the pseudoplastic rheology, we observe a large transitional regime between stagnant lid and fully mobile modes of convection.  The switch from stagnant lid convection to fully mobile convection does not occur as a sudden bifurcation, but instead is a gradual, continuous transition over a wide region of parameter space.  This means that most planets likely exist in a transitional regime, between stagnant lid convection and fully mobile convection.  Plate-tectonics occurs within this transitional regime, with plate-like convection happening over a broad range of surface mobility, covering most of the transitional regime.  In addition, transitional regime convection near the stagnant lid regime is characterized by ``sluggish subduction", where the high viscosity of lithospheric shear zones causes a slow, drip-like lithospheric foundering and low surface velocities.  This type of convection could be an explanation for why Venus shows subduction-like surface features yet lacks plate-tectonic style surface recycling.   

\section{Acknowledgments}  

This work was supported by NSF award EAR-1135382: Open Earth Systems, and by the facilities and staff of the Yale University Faculty of Arts and Sciences High Performance Computing Center.  We thank Adrian Lenardic for a thorough and thoughtful review that helped us significantly improve the manuscript.  

\clearpage



\clearpage

\begin{table}
\caption{Numerical Results for Temperature-Independent Viscosity and Healing. The grid resolution is listed as the number of grid points in the x-direction by the number of grid points in the z-direction; the aspect ratio can be determined by dividing the x-direction resolution by the z-direction resolution.  The same notation is used for Tables \ref{tab_stag}, \ref{tab_full}, \ref{tab_mp_data}, and \ref{tab_res}.}
\label{tab_simple}
\begin{tabular}{c c c c c c c c c c c c}
\hline
$D$ & $H$ & $Ra_0$ & $m$ & $p$ & $v_{rms}'$ & $v_l'$ & $Nu$ & $\tau_{xz}'$ & $\bar{A}'$ & $A_{max}'$ & Resolution \\ \hline
$10^{-7}$ & 100 & $10^6$ & 2  & 4 & 192 & 359 & 8.76 & 5848 & 0.30 & 0.37 & $512 \times 128$ \\
$10^{-6}$ & 100 & $10^6$ & 2  & 4 & 374 & 695 & 11.77 & 3631 & 0.57 & 0.74 & $512 \times 128$\\
$10^{-4}$ & 100 & $10^6$ & 2  & 4 & 1980 & 3800 & 30.35 & 1397 & 2.14 & 4.78 & $512 \times 128$\\
$10^{-2}$ & 100 & $10^6$ & 2  & 4 & 9272 & 17794 & 63.46 & 528 & 7.43 & 26.46 & $1024 \times 256$\\
$10^{-1}$ & 100 & $10^6$ & 2  & 4 & 23537 & 44585 & 93.30 & 328 & 14.25 & 57.98 & $1024 \times 256$\\
$10^{-6}$ & 100 & $10^5$ & 2  & 4 & 33 & 50 & 3.65 & 1446 & 0.22 & 0.28  & $512 \times 128$\\
$10^{-6}$ & 100 & $5 \times 10^5$ & 2  & 4 & 206 & 395 & 9.11 & 2966 & 0.44 & 0.61& $512 \times 128$ \\
$10^{-6}$ & 100 & $5 \times 10^6$ & 2  & 4 & 2497& 4528 & 34.69 & 6929 & 1.08 & 1.62 & $512 \times 128$\\
$10^{-6}$ & 100 & $10^7$ & 2  & 4 & 5705.6 & 10172 & 52.35 & 9002 & 1.41 & 2.28 & $1024 \times 256$\\
$10^{-6}$ & 100 & $5 \times 10^7$ & 2  & 4 & 27651 & 52452 & 105.45 & 16723 & 2.62 & 3.97 & $1024 \times 256$\\
\end{tabular}      
\end{table}

\begin{table}
\caption{Numerical Results for Temperature-Dependent Viscosity and Constant Healing}
\label{tab_stag}
\begin{tabular}{c c c c c c c c c c c c c}
\hline
$D$ & $H$ & $Ra_0$ & $m$ & $p$ & $E_v'$ & $T_s^*$ & $v_m'$ & $Nu$ & $\tau_{xz}'$ & $\bar{A}'$ & $A_{max}'$ & Resolution \\ \hline
$10^{-4}$ & 100 & $10^6$ & 2  & 4 & 23.03 & 1.0 & 326 & 2.55 & 1253 & 1.09 & 2.22 & $512 \times 128$\\
$10^{-3}$ & 100 & $10^6$ & 2  & 4 & 23.03 & 1.0 &  808 & 3.44 & 694 & 1.99 & 4.78 & $512 \times 128$\\
$10^{-2}$ & 100 & $10^6$ & 2  & 4 & 23.03 & 1.0 & 1915 & 5.12 & 412 & 3.71 & 10.76 & $512 \times 128$\\
$10^{-1}$ & 100 & $10^6$ & 2  & 4 & 23.03 & 1.0 &  4802 & 7.97 & 257 & 7.29 & 24.16 & $1024 \times 256$\\
$1$ & 100 & $10^6$ & 2  & 4 & 23.03 & 1.0 & 11256 & 12.10 & 165 & 14.51 & 51.73 & $1024 \times 256$\\
$10^{-3}$ & 100 & $3 \times 10^5$ & 2  & 4 & 23.03 & 1.0 & 188 & 2.48 & 417 & 1.41 & 2.58 & $512 \times 128$ \\
$10^{-3}$ & 100 & $3 \times 10^6$ & 2  & 4 & 23.03 & 1.0 &  2662 & 5.88 & 1042 & 3.06 & 8.24 & $512 \times 128$\\
$10^{-3}$ & 100 & $10^7$ & 2  & 4 & 23.03 & 1.0 & 9092 & 10.46 & 1664 & 4.87 & 13.85 & $512 \times 128$ \\
$10^{-3}$ & 100 & $2 \times 10^7$ & 2  & 4 & 23.03 & 1.0 &  19876 & 16.04 & 2177 & 6.51 & 20.27 & $1024 \times 256$\\
\end{tabular}      
\end{table}

\begin{table}
\caption{Numerical Results for Temperature-Dependent Viscosity and Healing}
\label{tab_full}
\begin{tabular}{c c c c c c c c c c c c c}
\hline
$D$ & $H$ & $Ra_0$ & $m$ & $p$ & $E_v'$ & $T_s^*$ & $E_h'$ & $v_l'$ & $v_m'$ & $Nu$ & $\bar{A}'$ & Resolution\\ \hline
$10^{-6}$	&	100	&	$10^6$	&	2	&	4	&	23.03	&	1	&	23.03	&	2.4	&	69	&	1.82	&	0.41	& $512 \times 128$\\
$10^{-5}$	&	100	&	$10^6$	&	2	&	4	&	23.03	&	1	&	23.03	&	26.4	&	306	&	3.73	&	0.94	& $512 \times 128$\\
$10^{-4}$	&	100	&	$10^6$ &	2	&	4	&	23.03	&	1	&	23.03	&	186	&	808	&	8.14	&	2.30	& $512 \times 128$\\
$10^{-3}$	&	100	&	$10^6$&	2	&	4	&	23.03	&	1	&	23.03	&	673	&	1823 	&	14.31	&	4.77	& $512 \times 128$\\
$10^{-2}$	&	100	&	$10^6$	&	2	&	4	&	23.03	&	1	&	23.03	&	2424 	&	4858	&	24.89	&	10.63	& $512 \times 128$\\
$10^{-1}$	&	100	&	$10^6$	&	2	&	4	&	23.03	&	1	&	23.03	&	7427 	&	12320	&	42.41	&	21.34	& $1024 \times 256$\\
$10^{-6}$	&	100	&	$10^6$	&	2	&	4	&	27.632	&	1	&	23.03	&	0.072	&	72	&	1.81	&	0.37	& $512 \times 128$\\
$10^{-5}$	&	100	&	$10^6$	&	2	&	4	&	27.632	&	1	&	23.03	&	2.2	&	191	&	2.67	&	0.77	& $512 \times 128$\\
$10^{-4}$	&	100	&	$10^6$	&	2	&	4	&	27.632	&	1	&	23.03	&	45.9	&	559	&	4.68	&	1.79	& $512 \times 128$\\
$10^{-3}$	&	100	&	$10^6$	&	2	&	4	&	27.632	&	1	&	23.03	&	246	&	1190 	&	9.44	&	4.11	& $512 \times 128$\\
$10^{-2}$	&	100	&	$10^6$ &	2	&	4	&	27.632	&	1	&	23.03	&	736	&	2737	&	15.55	&	8.34	& $512 \times 128$\\
$10^{-1}$	&	100	&	$10^6$	&	2	&	4	&	27.632	&	1	&	23.03	&	2555	 &	7657	&	27.42	&	17.10	& $512 \times 128$\\
$10^{-6}$		&	100	&	$10^6$	&	2	&	4	&	18.421	&	1	&	23.03	&	16.7	&	137	&	3.04	&	0.48	& $512 \times 128$\\
$10^{-4}$	&	100	&	$10^6$	&	2	&	4	&	18.421	&	1	&	23.03	&	446	&	1412 	&	11.88	&	2.68	& $512 \times 128$\\
$10^{-3}$	&	100	&	$10^6$	&	2	& 4	&	18.421	&	1	&	23.03	&	1807 	&	3703	 &	21.76	&	6.24	& $512 \times 128$\\
$1.0$	&	100	&	$10^6$	&	2	& 4	&	18.421	&	1	&	23.03	&	41917	&	44253 	&	93.66	&	55.50 &	$1024 \times 256$\\
$10^{-4}$	&	100	&	$10^6$	&	2	&	4	&	36.842	&	1	&	23.03	&	0.0044	&	286	&	2.36	&	1.26	& $512 \times 128$\\
$10^{-3}$	&	100	&	$10^6$	&	2	&	4	&	36.842	&	1	&	23.03	&	0.012	&	723	&	3.10	&	2.30	& $512 \times 128$\\
$10^{-3}$	&	100	&	$10^6$	&	2	&	4	&	21.64	&	1	&	23.03	&	899	&	2299 	&	16.09	&	5.12 & $512 \times 128$	\\
$10^{-5}$	&	100	&	$10^6$	&	2	&	4	&	32.237	&	1	&	23.03	&	0.037	&	161	&	2.28	&	0.71	& $512 \times 128$\\
$10^{-4}$	&	100	&	$10^6$&	2	&	4	&	32.237	&	1	&	23.03	&	0.178	&	348	&	2.97	&	1.38	& $512 \times 128$\\
$10^{-3}$	&	100	&	$10^6$	&	2	&	4	&	32.237	&	1	&	23.03	&	0.17	&	885	&	3.78	&	2.48 & $512 \times 128$	\\
$10^{-2}$	&	100	&	$10^6$	&	2	&	4	&	32.237	&	1	&	23.03	&	0.29	&	2259	&	5.49	&	4.72	& $512 \times 128$\\
$10^{-3}$	&	100	&	$8 \times 10^4$	&	2	&	4	&	23.03	&	1	&	23.03	&	46.4	&	133	&	4.26	&	2.31	& $512 \times 128$\\
$10^{-3}$	&	100	&	$10^5$ &	2	&	4	&	23.03	&	1	&	23.03	&	60.9	&	163	&	4.79	&	2.53	& $512 \times 128$\\
$10^{-3}$	&	100	&	$3 \times 10^5$	&	2	&	4	&	23.03	&	1	&	23.03	&	184	&	528	&	7.80	&	3.35	& $512 \times 128$\\
$10^{-3}$	&	100	&	$3 \times 10^6$	&	2	&	4	&	23.03	&	1	&	23.03	&	2770 	&	6318 	&	27.04	&	7.60	& $512 \times 128$\\
$10^{-3}$	&	100	&	$10^7$	&	2	&	4	&	23.03	&	1	&	23.03	&	14494	&	27758	&	59.98	&	12.92	& $1024 \times 256$\\
$10^{-3}$	&	100	&	$3 \times 10^5$	&	2	&	4	&	27.632	&	1	&	23.03	&	51.2	&	332	&	4.68	&	2.65	& $512 \times 128$\\
$10^{-3}$	&	100	&	$3 \times 10^6$	&	2	&	4	&	27.632	&	1	&	23.03	&	979	&	3889 	&	17.99	&	6.27	& $512 \times 128$\\
$10^{-3}$	&	100	&	$8 \times 10^6$	&	2	&	4	&	27.632	&	1	&	23.03	&	4341 	&	12694	&	35.70	&	9.44	& $512 \times 128$\\
$10^{-3}$	&	100	&	$10^5$	&	2	&	4	&	18.421	&	1	&	23.03	&	137	&	285	&	6.57	&	2.88	& $512 \times 128$\\
$10^{-3}$	&	100	&	$10^7$	&	2	&	4	&	32.237	&	1	&	23.03	&	3.5	&	9604 	&	11.50	&	6.08	& $512 \times 128$\\
$10^{-3}$	&	100	&	$3 \times 10^6$	&	2	&	4	&	36.842	&	1	&	23.03	&	0.015	&	2865	&	5.06	&	3.57	& $512 \times 128$\\
$10^{-3}$	&	100	&	$5 \times 10^5$	&	2	&	4	&	36.842	&	1	&	23.03	&	0.0034	&	320	&	2.35	&	1.82	& $512 \times 128$\\
$10^{-3}$	&	100	&	$10^6$&	2	&	4	&	23.03	&	1	&	23.03	&	847	&	2130 	&	13.52	&	4.62	& $768 \times 128$\\
$10^{-3}$	&	100	&	$10^6$&	2	&	4	&	23.03	&	1	&	23.03	&	1116	&	2632 	&	13.42	&	4.63	& $1024 \times 128$\\
$10^{-3}$	&	100	&	$10^6$&	2	&	4	&	23.03	&	1	&	23.03	&	1367	&	3401 	&	13.49	&	4.65	& $2048 \times 128$\\
\end{tabular}      
\end{table}

\begin{table}
\caption{Numerical Results for Temperature-Dependent Viscosity and Healing, Varying $m$ and $p$}
\label{tab_mp_data}
\begin{tabular}{c c c c c c c c c c c c c}
\hline
$D$ & $H$ & $Ra_0$ & $m$ & $p$ & $E_v'$ & $T_s^*$ & $E_h'$ & $v_l'$ & $v_m'$ & $Nu$ & $\bar{A}'$ & Resolution \\ \hline
$10^{-5}$	&	100	&	$10^6$	&	3	&	4	&	27.632	&	1	&	23.03	&	14	&	251	&	3.06	&	0.85	& $512 \times 128$\\
$10^{-4}$	&	100	&	$10^6$	&	3	&	4	&	27.632	&	1	&	23.03	&	168	&	1108	&	8.24	&	1.94	& $512 \times 128$\\
$2 \times 10^{-4}$	&	100	&	$10^6$	&	3	&	4	&	27.632	&	1	&	23.03	&	365	&	2049 	&	11.86	&	2.57	& $512 \times 128$\\
$10^{-4}$	&	100	&	$5\times10^5$	&	3	&	4	&	27.632	&	1	&	23.03	&	54	&	461	&	4.91	&	1.52	& $512 \times 128$\\
$10^{-4}$	&	100	&	$2\times10^6$	&	3	&	4	&	27.632	&	1	&	23.03	&	624	&	3595 	&	15.52	&	2.69	& $512 \times 128$\\
$10^{-5}$	&	100	&	$10^6$	&	3	&	4	&	23.03	&	1	&	23.03	&	63	&	508	&	5.22	&	1.08	& $512 \times 128$\\
$10^{-5}$	&	100	&	$10^6$ &	3	&	4	&	18.421	&	1	&	23.03	&	182	&	870	&	8.13	&	1.28	& $512 \times 128$\\
$10^{-5}$	&	100	&	$10^6$	&	3	&	5	&	27.632	&	1	&	23.03	&	5.5	&	207	&	2.83	&	0.83	& $512 \times 128$\\
$10^{-4}$	&	100	&	$10^6$	&	3	&	5	&	27.632	&	1	&	23.03	&	146	&	749	&	7.53	&	1.77 & $512 \times 128$	\\
$10^{-3}$	&	100	&	$10^6$	&	3	&	5	&	27.632	&	1	&	23.03	&	644	&	2081	&	14.55	&	3.33 & $512 \times 128$	\\
$10^{-4}$	&	100	&	$10^6$	&	3	&	5	&	23.03	&	1	&	23.03	&	382	&	1201	&	11.31	&	2.07	& $512 \times 128$\\
$10^{-4}$	&	100	&	$10^6$	&	3	&	5	&	18.421	&	1	&	23.03	&	880	&	2217 	&	16.02	&	2.35	& $512 \times 128$\\
$10^{-4}$	&	100	&	$5\times 10^5$	&	3	&	5	&	27.632	&	1	&	23.03	&	64	&	348 	&	5.21	&	1.50	& $512 \times 128$\\
$10^{-4}$	&	100	&	$3\times 10^6$ &	3	&	5	&	27.632	&	1	&	23.03	&	652	&	2794 	&	15.23	&	2.51 & $512 \times 128$	\\
$10^{-4}$	&	100	&	$10^6$	&	3	&	4	&	27.632	&	1	&	23.03	&	352	&	1967	&	8.67	&	1.94	& $1024\times 128$\\
$10^{-4}$	&	100	&	$10^6$	&	3	&	5	&	27.632	&	1	&	23.03	&	278	&	1004	&	7.53	&	1.73	& $1024 \times 128$\\

\end{tabular}      
\end{table}

\clearpage

\appendix
\section{Appendix}
\subsection{Influence of Dislocation Creep}
\label{sec:disl_creep}

\setcounter{equation}{0}
\renewcommand{\theequation}{A\arabic{equation}}

Our grain-damage formulation assumes diffusion creep is the dominant creep mechanism throughout the mantle under all conditions.  This is clearly a simplification, as under many conditions grainsize insensitive dislocation creep should prevail in the mantle.  Here we explain how to extend our scaling laws to include dislocation creep, and under what conditions we might expect dislocation creep to dominate in the mantle. However, as a more complete study of convection with grain-damage and a composite rheology, including comparisons with numerical experiments, is beyond the scope of this paper, the theory presented in this appendix should be considered preliminary.  In particular, we consider the case where the effective interior mantle viscosity, $\mu_{eff}$, is governed by dislocation creep as opposed to diffusion creep; we do not need to consider a dislocation creep lithosphere because the lithospheric shear zone viscosity is only relevant to convection when damage is effective (i.e. when there is considerable grainsize reduction and diffusion creep will be dominant).  We focus on the scaling laws for the problem of grain-damage with temperature-dependent viscosity and healing, as this is most applicable to convection in planetary mantles.  

Dislocation and diffusion creep are independent mechanisms, which operate simultaneously; in principle, this means that the total strain rate is the sum of the strain rates from each mechanism, $\dot{\varepsilon}_{tot} = \dot{\varepsilon}_{disl} + \dot{\varepsilon}_{diff}$ \citep[e.g.][]{karato1993,Karato2008}, and the creep mechanism which produces the larger strain rate will dominate. The flow laws for dislocation and diffusion creep are, respectively, \citep{karato1993}, 
\begin{equation}
\eqlbl{disl}
\dot{\varepsilon}_{disl} = a \tau^n \exp{\left(-\frac{E_{v,disl} + pV_{disl}}{RT} \right)}     
\end{equation}
and 
\begin{equation}
\eqlbl{diff}
\dot{\varepsilon}_{diff} = b A^m \tau \exp{\left(-\frac{E_{v} + pV}{RT} \right)}    
\end{equation}
where $\dot{\varepsilon}$ and $\tau$ are the second invariant of the strain-rate and stress tensors, respectively, $a$ and $b$ are constants, $n$ is the power law exponent and is typically 3-3.5, $E_{v,disl}$ is the activation energy for dislocation creep, and  $V$ and $V_{disl}$ are the activation volumes for diffusion and dislocation creep, respectively.  Given that $E_{v,disl}$ is nearly twice $E_v$ ($E_{v,disl} = 540$ kJ/mol and $E_{v} = 300$ kJ/mol \citep{karato1993}), inspection of \eqref{disl} and \eqref{diff} shows that diffusion creep is favored for large fineness, low stress, and low temperatures.    

If dislocation creep dominates in the mantle such that the effective interior mantle viscosity can be considered to be grainsize insensitive, our scaling laws for $Nu$ and $v_l'$ in the transitional regime (\eqref{Nu_trans_scale} and \eqref{vl_trans_scale}, respectively) will be unaffected because they only depend on the viscosity of damaged lithospheric shear zones, not the interior mantle viscosity.  Thus they are insensitive to the creep mechanism that prevails in the mantle.  However, our scaling laws for the basal mantle velocity, $v_m'$, mantle fineness, $A_i'$, and shear stress, $\tau_{xz}'$ would be significantly different.  The most important of these, $v_m'$, would follow the scaling law for non-Newtonian convection \citep{slava1995,Solomatov2000b} 
\begin{equation}
\eqlbl{vscale_disl}
v_m' \sim ( Ra_0(1-T_i') )^{\frac{2n}{n+2}} \left(\frac{\mu_{i,disl}}{\mu_m} \right)^{-\frac{2n}{n+2}}
\end{equation}
where $\mu_{i,disl}$, the reference viscosity for dislocation creep, is given by 
\begin{equation}
\mu_{i,disl} = a^{\frac{1}{n}} \left(\frac{\kappa}{d^2} \right)^{\frac{1-n}{n}} \exp{ \left( \frac{E_{v,disl}}{nRT_i} \right)} . 
\end{equation}
The mantle shear stress can be obtained from \eqref{vscale_disl}, and the mantle fineness would no longer be relevant.  

In both the fully stagnant lid regime and the fully mobile regime, our scaling laws based on diffusion creep would also not be applicable if dislocation creep dominates in the mantle.  Instead, the velocity would scale as 
\begin{equation}
\eqlbl{vscale_stag_disl}
v_m' \approx 0.1 \left(\frac{Ra_0}{\theta}\right)^{\frac{2n}{n+2}} \left(\frac{\mu_{i,disl}}{\mu_m} \right)^{-\frac{2n}{n+2}}
\end{equation}
in the fully stagnant lid regime and 
\begin{equation}
\eqlbl{vscale_mob_disl}
v_m' = v_l' \approx 0.1 Ra_0^{\frac{2n}{n+2}} \left(\frac{\mu_{i,disl}}{\mu_m} \right)^{-\frac{2n}{n+2}}
\end{equation}
in the fully mobile regime.  
The Nusselt number would scale as 
\begin{equation}
\eqlbl{Nu_mob_disl} 
Nu \approx 0.26 Ra_0^{\frac{n}{n+2}} \left(\frac{\mu_{i,disl}}{\mu_m} \right)^{-\frac{n}{n+2}}
\end{equation}
in the fully mobile regime and 
\begin{equation} 
\eqlbl{Nu_stag_disl}
Nu \approx 0.26 \theta^{\frac{2(n+1)}{n+2}} Ra_0^{\frac{n}{n+2}} \left(\frac{\mu_{i,disl}}{\mu_m} \right)^{-\frac{n}{n+2}}
\end{equation}
in the fully stagnant lid regime \citep{slava1995}.  As discussed in \S \ref{sec:stag_H0_scale}, there is some disagreement about the scaling for velocity in the stagnant lid regime; we give the result from \cite{slava1995} because it assumes bottom heating, as we use for our numerical experiments.  

Finally the regime diagram would be altered if the mantle is permanently in dislocation creep; the boundary betw,een the fully stagnant lid regime and transitional regime would now be defined by the intersection of \eqref{Nu_stag_disl} and \eqref{Nu_trans_scale}, and the boundary between the transitional regime and the fully mobile regime would be defined by the intersection of \eqref{Nu_trans_scale} and \eqref{Nu_mob_disl}.  The transitional regime would take up a smaller area in $\mu_l' - D/H$ space, but would still be a large, important regime for mantle convection. Moreover, the transitional regime in $Ra_0 - D/H$ space would be approximately the same size for the parameters used in this paper ($m=2$, $p=4$, and $n=3$); this is because the fully mobile and fully stagnant lid scaling laws have the same dependence on $Ra_0$ in both the diffusion creep dominated and dislocation creep dominated cases, as grain-damage produces an effectively non-Newtonian rheology with $n=3$, just like dislocation creep.  Writing the constitutive equation for grain-damage, 
\begin{equation}
\dot{\varepsilon} \sim \frac{\tau}{\mu} A^m
\end{equation}  
and substituting for the fineness as a function of stress (e.g. \eqref{steady_fine_stag2}), 
\begin{equation}
\dot{\varepsilon} \sim \tau^{\frac{p+m}{p-m}} \mu^{-\frac{p}{p-m}} \left(\frac{D}{H} \right)^{\frac{m}{p-m}},
\end{equation}  
which shows that $\dot{\varepsilon} \sim \tau^3$ for $m=2$ and $p=4$.  This demonstrates an important point; grain-damage causes convection to follow an effectively non-Newtonian rheology with a stress exponent similar to dislocation creep even when diffusion creep dominates.  Therefore the physics governing convection in the two end-member regimes is similar regardless of the creep mechanism that dominates in the mantle. 

\subsection{Resolution Tests}
\label{sec:res}

To ensure that the numerical models presented in the main text are sufficiently well resolved to constrain the scaling laws, we reran a subset of the numerical models (focusing on those with large $D$ and/or large $Ra_0$) at doubled resolution.  For all resolution tests (see Table \ref{tab_res}), the models were started from the same initial condition, and time averages taken over the same time window for both the lower resolution and higher resolution model.  The percent error in heat flow and plate speed by which the lower resolution model deviates  from the higher resolution model is shown in Table \ref{tab_res} as $Nu$ error and $v_l'$ error, respectively.  Higher resolution only changes the numerical results for $Nu$ by a maximum of 5.5\%, and only changes those for $v_l'$ by a maximum of $6.5$\%.  The close agreement between the lower resolution models and the test cases run at higher resolution, indicates that the numerical models used in the main text to constrain the scaling laws are sufficiently well resolved.    

\begin{table}
\caption{Resolution Tests}
\label{tab_res}
\begin{tabular}{c c c c c c c c c c c c c}
\hline
$D$ & $H$ & $Ra_0$ & $m$ & $p$ & $E_v'$ & $T_s^*$ & $E_h'$ & $v_l'$ & $Nu$ & Resolution & $Nu$ error & $v_l'$ error\\ 
\hline   
$10^{-4}$ & 100 & $10^6$ & 2 & 4 & 0 & - & 0 & 3518 & 29.6 & $512 \times 128$ & 1.6 & 3.5 \\  
$10^{-4}$ & 100 & $10^6$ & 2 & 4 & 0 & - & 0 & 3648 & 30.1 & $1024 \times 256$ & - & - \\ 
$10^{-1}$ & 100 & $10^6$ & 2 & 4 & 0 & - & 0 & 43402 & 93.2 & $1024 \times 256$ & 5.4 & 5.7 \\  
$10^{-1}$ & 100 & $10^6$ & 2 & 4 & 0 & - & 0 & 41048 & 98.6 & $2048 \times 512$ & - & - \\ 
$10^{-2}$ & 100 & $10^6$ & 2 & 4 & 23.03 & 1 & 23.03 & 2437 & 25.07 & $512 \times 128$ & 0.1 & 5.4 \\  
$10^{-2}$ & 100 & $10^6$ & 2 & 4 & 23.03 & 1 & 23.03 & 2313 & 25.05 & $1024 \times 256$ & - & - \\  
$10^{-3}$ & 100 & $3 \times 10^6$ & 2 & 4 & 23.03 & 1 & 23.03 & 2797 & 27.1 & $512 \times 128$ & 2.9& 5.5\\  
$10^{-3}$ & 100 & $3 \times 10^6$ & 2 & 4 & 23.03 & 1 & 23.03 & 2959 & 27.92 & $1024 \times 256$ & - & - \\  
$1$ & 100 & $10^6$ & 2 & 4 & 18.42 & 1 & 23.03 & 37749 & 93.99 & $1024 \times 256$ & 4.8 & 6.4 \\  
$1$ & 100 & $10^6$ & 2 & 4 & 18.42 & 1 & 23.03 & 40344 & 98.69 & $2048 \times 512$ & - & - \\ 
\end{tabular}      
\end{table}

\end{document}